\documentclass[11pt]{article}
\usepackage{graphicx}

[1999/03/29 PSNFSS v.7.2
Times font as default roman
: S Rahtz]

\newcommand \be{\begin{equation}}
\newcommand \ba{\begin{eqnarray}}
\newcommand \ee{\end{equation}}
\newcommand \ea{\end{eqnarray}}
\newcommand{\lp}{\left(}
\newcommand{\rp}{\right)}

\newcommand{\yo}{y_{1}}
\newcommand{\yt}{y_{2}}
\newcommand{\vy}{{\bf y}}
\newcommand{\vyz}{{\bf y}_{0}}
\newcommand{\bs}{\triangle b}
\newcommand{\Bs}{\triangle B}
\newcommand{\es}{\triangle e}
\newcommand{\tsb}{\triangle t_{b}}
\newcommand{\tsB}{\triangle t_{B}}
\newcommand{\tse}{\triangle t_{e}}

\newcommand{\Bo}{B_{1}}
\newcommand{\vpo}{{\bf p}}
\newcommand{\tauo}{t}
\newcommand{\tz}{t_{0}}
\newcommand{\tc}{t_{c}}
\newcommand{\sbtr}{_{\rm sing}}
\newcommand{\dytr}{\dot{y}_{2 {\rm sing}}}
\newcommand{\dyrv}{\dot{y}_{2 {\rm osc}}}
\newcommand{\sppm}{^{\pm}}
\newcommand{\spp}{^{+}}
\newcommand{\spm}{^{-}}
\newcommand{\tr}{\triangleright}
\newtheorem{theorem}{{\sffamily\bfseries\large Theorem}}[subsection]
\newenvironment{Theorem}[1]
         {\begin{theorem} {#1} \\ \makebox[0.5cm][l]{\ } \rm}{\end{theorem}}
\newtheorem{corollary}[theorem]{{\sffamily\bfseries\large Corollary}}
\newenvironment{Corollary}[1]
         {\begin{corollary} {#1} \\ \makebox[0.5cm][l]{\ } \rm}
         {\end{corollary}}
\newtheorem{lemma}[theorem]{{\sffamily\bfseries\large Lemma}}

\newtheorem{definition}[theorem]{{\sffamily\bfseries\large Definition}}
\newenvironment{Definition}[1]
         {\begin{definition} {#1} \\ \makebox[0.5cm][l]{\ } \rm}
         {\end{definition}}

\newtheorem{remark}[theorem]{{\sffamily\bfseries\large Remark}}
\newenvironment{Remark}{\begin{remark} \  \rm \begin{enumerate}}
         {\end{enumerate}\end{remark}}
\newenvironment{Remarks}[1]{\begin{remark} \rm \ \\
    \renewcommand{\theenumii}{\arabic{enumii}}
   {#1} \begin{enumerate}} {\end{enumerate}
    \renewcommand{\theenumii}{\alph{enumii}} \end{remark}}

\begin{document}

\begin{center}
\Huge Oscillatory Finite-Time Singularities\\
in Finance, Population and Rupture
\end{center}
\bigskip
\begin{center}
\Large Kayo Ide$^{\mbox{\ref{das}}}$ and Didier 
Sornette$^{\mbox{\ref{ess},\ref{lpec}}}$\\
\end{center}
\begin{center}
Institute of Geophysics and Planetary Physics\\ University of
California, Los Angeles\\ Los Angeles, CA 90095-1567\label{igpp}
\end{center}

\bigskip
\begin{enumerate}

\item Also at the Department of Atmospheric Sciences, UCLA\label{das}

\item Also at the Department of Earth and Space Sciences, UCLA\label{ess}

\item Also at the Laboratoire de Physique de la Mati\`ere Condens\'ee, CNRS UMR
6622 and
Universit\'e de Nice-Sophia Antipolis, 06108 Nice Cedex 2,
France\label{lpec}

\end{enumerate}

\vskip 2cm
\begin{abstract}

We present a simple two-dimensional dynamical system where two 
nonlinear terms, exerting respectively positive 
feedback and reversal, compete to create a singularity in finite time decorated by
accelerating oscillations.
The power law singularity results from the increasing growth rate.
The oscillations result from
the restoring mechanism. As a function of the order of 
the nonlinearity of the
growth rate and of the restoring term, a rich variety of behavior is
documented analytically and numerically.
The dynamical behavior is traced back fundamentally to
the self-similar spiral structure of trajectories in phase space
unfolding around an unstable spiral point at the origin. The interplay
between the restoring mechanism and the nonlinear growth rate leads
to approximately log-periodic oscillations with
remarkable scaling properties.
Three domains of applications are discussed: (1) the stock market with a
competition between nonlinear trend-followers and nonlinear value investors;
(2) the world human population with a competition between a
population-dependent
growth rate and a nonlinear dependence on a finite carrying capacity;
(3) the failure of a material
subjected to a time-varying stress with a competition between positive
geometrical feedback on the damage variable and nonlinear healing.

\end{abstract}

\vfill
\begin{flushleft}

\vskip 2cm
{\bf Acknowledgments}: We are grateful to Y. Malevergne and S. Roux for useful
discussions.
This work was partially supported by ONR N00014-99-1-0020 (KI) and by
NSF-DMR99-71475 and the James S. Mc Donnell Foundation 21st century 
scientist award/studying
complex system (DS).
\end{flushleft}

\pagebreak

\tableofcontents

\pagebreak

\section{Introduction}
\label{sec:intro}

The mathematics of singularities is applied routinely in the
physics of phase transitions to describe for instance the transformations
 from ice to water or from a magnet to a demagnetized state
when raising the temperature, as well as in many other condensed matter
systems. Such singularities characterize so-called critical phenomena.
In these problems, physical observables such as susceptibilities, specific
heat, etc., exhibit a singularity as the control parameter (temperature,
strength of the interaction) approaches a critical value.

Other classes of singularities occur in dynamical systems and are
spontaneously reached in finite time.
Spontaneous singularities in ordinary (ODE) and partial differential equations
(PDE) are quite common and have been found in many well-established models
of natural systems, either at special points in space such as in the Euler
equations of inviscid fluids \cite{Pumiersiggia,eulerstag},
in the surface curvature on the
free surface of a conducting fluid in an electric field \cite{Zubarev}, in
vortex collapse of systems of point vortices, in the equations of
General Relativity coupled to a mass field leading to the formation of
black holes \cite{Choptuik}, in models of micro-organisms aggregating to form
fruiting bodies \cite{Rascle}, or in the more prosaic rotating coin (Euler's
disk) \cite{Moffatt}. Some more complex examples
are models of rupture and material failure \cite{critruptcan,critrupt},
earthquakes \cite{earthquake} and stock market crashes \cite{crash,nasdaq}.

The normal form of a finite-time singularity is the equation
\be
{dp \over dt} = p^m~, ~~~~~~~{\rm with}~m>1~,    \label{nvnkal}
\ee
whose solution is
\be
p(t) = p(0)~\left({t_c-t \over t_c}\right)^{-1/(m-1)}~,  \label{bvnmz}
\ee
where the critical time $t_c=(m-1)/[p(0)]^{m-1}$ is determined by the 
initial condition $p(0)$.
The singularity results from the fact that the instantaneous growth
rate $d\ln p/dt = p^{m-1}$ is increasing with $p$ and thus
with time. This can be visualized by studying the doubling time,
defined at the time interval $\Delta t$ necessary for $p(t)$
to double, i.e., $p(t+\Delta t) = 2 p(t)$. When the growth rate of $p$ increases as
a power law of $p$, the doubling time decreases fast and the
sequence of doubling time intervals shrinks to zero sufficiently fast so that
its sum is a convergent geometrical series. The variable thus undergoes
an infinite number of doubling operations in a finite time, which the
essence of a finite-time singularity.

The power law solution (\ref{bvnmz})
possesses the symmetry of ``scale invariance'', namely a reduction
$t_c-t \to (t_c-t)/\lambda$ of the
distance $t_c-t$ from the singularity at $t_c$ by an arbitrary factor
$\lambda$ changes $p(t)$ to $\lambda^{1/(m-1)}~p(t)$, i.e., keeps
the same form of the solution up to a global rescaling.

This continuous scale invariance can be partially broken into a weaker
symmetry, called discrete scale invariance, according to which the 
self-similarity
holds only for integer powers of a specific factor $\lambda$ \cite{reviewsor}.
The hallmark of this discrete scale invariance is that the power law
(\ref{bvnmz}) transforms into an oscillatory singularity, with log-periodic
oscillations decorating the overall power law acceleration.
Such log-periodic power laws have been documented for many systems 
such as with a
built-in geometrical hierarchy, in
programming and number theory, for
Newcomb-Benford law of first digits and in the arithmetic system,
in diffusion in anisotropic quenched random lattices,
as the result of a cascade of ultra-violet instabilities in growth 
processes and rupture,
in deterministic dynamical systems
(cascades of sub-harmonic bifurcations in the transition to chaos,
two-coupled anharmonic oscillators, near-separatrix Hamiltonian 
chaotic dynamics,
kicked charged particle moving in a double-well potential giving a physical
realization of Mandelbrot and Julia sets, chaotic scattering),
in extension of percolation theory (so-called ``animals''),
in response functions of spin systems with quenched disorder, in
freely decaying 2D-turbulence,
in the gravitational collapse and black hole formation,
in spinodal decomposition of binary mixtures in uniform shear flow, etc.
(see \cite{reviewsor,BookSor} and references therein).

The novel interesting feature is the presence of a discrete hierarchy
of length and/or time scales in an otherwise scale-invariant system.
The presence of these scales may provide insight into the underlying
mechanisms. While there is a good general framework for the
description of discrete scale invariant systems
using renormalization group theory \cite{reviewsor}, a general understanding
of the possible physical mechanisms at its origin is still lacking. 
In particular,
dynamically generated discrete scale invariance is the most important problem,
as it might provide understanding in the origins of the ubiquitous
existence of hierarchies and cascades in natural and social systems.

Here, we introduce and study a
simple two-dimensional nonlinear
dynamical system with the minimal ingredients ensuring that it
exhibits both a finite-time singularity
(and its associated scale invariance) and
oscillatory behavior. The scale invariance is thus
partially broken by the existence of dynamically generated
length scales associated with the oscillations.  
We start from (\ref{nvnkal}) and enrich it by the minimal ingredient
to obtain what we believe is the simplest 
``normal form'' of an oscillatory finite-time singularity.
While the singularity emerges from the nonlinear growth law with 
positive feedback, the
hierarchy of length scales results from a nonlinear negative feedback.
The competition between the positive and negative nonlinear feedbacks create
an approximate self-similar oscillatory structures, which can be 
understood from
a spiral dynamics in phase space around a central unstable fixed point. Physically,
the self-similar oscillations result from the dependence of the local frequency
of the nonlinear oscillator on the amplitude.
This will be shown in phase space to result from the special role played by
the origin which is the unstable fixed point around which 
the spiral structures
of trajectories are organized.

This spiral structure of the dynamics
around the central unstable fixed point bears
a superficial resemblance to the
the Shilnikov's mechanism for chaos \cite{Guckenheimer}. However, both their 
dynamics and their behaviors are unrelated.
Shilnikov's systems are characterized by trajectories in phase space
spiraling towards the hyperbolic point along
the stable manifold and then blowing-up exponentially
along the unstable manifold of the hyperbolic point,
until they are reinjected again along the stable manifold.
In our system, trajectories in phase space spiral out slowly at first and
then accelerate until a singular point in finite-time is reached due to a
faster-than-exponential acceleration. Our system has thus a finite lifetime
while Shilnikov's systems are globally statistically stationary.

Our work is somewhat more related
to that of several authors who emphasized the possible role of
spiral structures in singular flows as a mechanism to promote the transfer of
energy from large scales to small scales \cite{eulerstag,Vassilicos,Vassilicos2}.
Kiehn \cite{Kiehn} has emphasized that vortex sheet evolution, governed by an 
integral form of the Biot-Savart law (known as the Rott-Birchoff equation)
leads to the production of discontinuities in finite time. Asymptotic
spiral type solutions in the vicinity of the singularity have been 
investigated both analytically and numerically (see \cite{Kiehn} and
references therein). Szydlowski et al. \cite{Szydlowski}
have analyzed a nonlinear second-order
ordinary differential equation, called the Kaldor-Kalecki business model in which
capital stock changes are caused by past investment decisions. Their study 
emphasizes the negative feedback connected with the lag-delay effect and thus lacks
the positive feedback trend effect discussed here. Canessa \cite{Canessa1,Canessa2} has
also a nonlinear second-order differential equation for the price but again
the emphasis is on the nonlinear feedback rather than on the possibility of 
explosive phases coupled with the oscillatory behavior.

Let us also mention another mechanism for log-periodicity:
scale invariant equations
which present an instability at finite wavevector decreasing with the field
amplitude may generate naturally a discretely scale-invariant
spectrum of internal scales \cite{ETC97sor}.

We first motivate the normal form studied here for
an oscillatory finite-time singularity by three physical examples, namely the
time evolution of a stock market price described in section 2,
the dynamics of human population described in section 3 and the coupled
evolution of a damage variable and the average stress leading to 
material rupture
given in section 4. We then present in section 5 an analysis of the effect of
each of the two components (the nonlinear
amplification and the nonlinear reversal term) of the dynamics taken 
separately.
Section 6 describes in a rather heuristic way
the fundamental characteristics of the overall dynamics
obtained when combining both terms. Section 7 provides a detailed dynamical
system approach giving  a complete
characterization of the dynamics in phase space and precise predictions
on the exponents of the scaling laws which are tested by numerical simulations.
Section 8 concludes.

\section{Stock market price dynamics}
\label{sec:market}
The importance of the interplay of two classes of investors, so-called
fundamental value investors and technical analysists (or trend followers),
has been stressed by several recent works \cite{Luxnature,farmerjoshi} to
be essential
in order to retrieve the important stylized facts of stock market 
price statistics.
We build on this insight and construct a simple model of price dynamics,
whose innovation is to put emphasis on the
fundamental {\it nonlinear} behavior of both classes of agents.

\subsection{Nonlinear value and trend-following strategies}
\label{sec:market-trend}
The variation of price of an asset on the stock market is controlled by
supply and demand, in other words by the net order size $\Omega$
through a market impact function \cite{Farmer}.
Assuming that the ratio ${\tilde p}/p$ of the price ${\tilde p}$ at 
which the orders are
executed over the previous quoted price $p$ is solely a function of $\Omega$
and using the condition
that it is impossible to make profits by repeatedly trading through a close
circuit (i.e. by buying and selling with final net position equal to zero),
Farmer \cite{Farmer} has shown that the logarithm of the price is given by the 
following equation written in discrete form
\be
\ln p(t+1) - \ln p(t)  = {\Omega(t) \over L} ~.
\label{mqmqmmq}
\ee
The so-called ``market depth'' $L$ is the typical number of 
outstanding stocks traded per unit time
and thus normalizes the impact of a given order size $\Omega(t)$ on the log-price
variations.
The net order size $\Omega$ summed over all traders is changing as a
function of time so as
to reflect the information flow in the market and the evolution of the
traders' opinions and moods. A zero net order size $\Omega=0$ corresponds
to exact balance between supply and demand.
Various derivations
have established a connection between
the price variation or the variation of the logarithm of the price
to factors that control the net order size itself 
\cite{Farmer,Boucont,PandeyStauffer}.
Two basic ingredients of $\Omega(t)$ are thought to be important in determining the price
dynamics: reversal to the fundamental value 
($\Omega_{\rm fund}(t)$) and trend following ($\Omega_{\rm trend}(t)$). Other factors,
such as risk aversion, may also play an important role.

We propose to describe the reversal to estimated fundamental value
by the contribution
\be
\Omega_{\rm fund}(t) = -c ~[\ln p(t) - \ln p_f]
~ |\ln p(t) - \ln p_f|^{n-1}~,   \label{fmaaak}
\ee
to the order size,
where $p_f$ is the estimated fundamental value and $n>0$ is an exponent
quantifying the nonlinear nature of reversion to $p_f$.
The strength of the reversion is measured by the coefficient $c>0$, which
reflects that the net order is negative (resp. positive) if the price is
above (resp. below) $p_f$. The nonlinear power law
$[\ln p(t) - \ln p_f]~ |\ln p(t) - \ln p_f|^{n-1}$ of order $n$ is chosen as the
simplest function capturing the following effect. In principle, the 
fundamental value
$p_f$ is determined by the discounted expected future dividends and is thus
dependent upon the forecast of their growth rate and of the risk-less 
interest rate,
both variables being very difficult to predict. The fundamental value is thus
extremely difficult to quantify with high precision and is often estimated
within relatively large bounds \cite{Malkiel,Chiang,Dow,Carbonara}:
all of the methods of determining intrinsic value rely on
assumptions that can turn out to be far off the mark.  For instance, 
several academic
studies have disputed the premise that a portfolio of sound, cheaply
bought stocks will, over time, outperform a portfolio selected by any
other method (see for instance \cite{Lamont}). As a consequence, a trader
trying to track fundamental value has no incentive to react when she feels
that the deviation is small since this deviation is more or less within
the noise. Only when the departure of price from fundamental value
becomes relatively large will the trader act. The relationship (\ref{fmaaak})
with an exponent $n>1$ precisely accounts for this effect: when $n$ 
is significantly
larger than $1$, $|x|^n$ remains small for $|x|<1$ and shoots up rapidly only when
it becomes larger than $1$, mimicking a smoothed threshold behavior.
The nonlinear dependence of $\Omega_{\rm fund}(t)$ on $\ln [p(t)/p_f]=
\ln p(t) - \ln p_f$ shown in 
(\ref{fmaaak}) is the first novel element of our model.
Usually, modelers reduce this term to the linear case $n=1$ while, as 
we shall show, generalizing to larger values $n>1$ will be a crucial feature of the 
price dynamics. In economic language, the exponent 
$n = d \ln \Omega_{\rm fund}/d \ln \left(\ln [p(t)/p_f]\right)$
is called the ``elasticity'' or ``sensitivity''
of the order size $\Omega_{\rm fund}$ with respect
to the (normalized) l0g-price $\ln [p(t)/p_f]$.

A related ``sensitivity'', that of the money demand to interest rate, has 
has been recently documented to be larger than $1$, similarly to our proposal
of taking $n>1$ in (\ref{fmaaak}).
Using a survey of roughly 2,700 households, 
Mulligan and Sala-i-Martin \cite{Mulligan} estimated the interest
elasticity of money demand (the sensitivity or log-derivative of money demand
to interest rate) to be very small at low interest rates. This is due to the fact
that few people decide to invest in interest-producing assets when rates are low,
due to ``shopping'' costs. In contrast, for large interest rates or for
those who own a significant bank account, the interest elasticity of money demand is 
significant. This is a clear-cut example of a threshold-like behavior characterized
by a strong nonlinear response. This can be captured by
$e \equiv d \ln M/ d\ln r = (r/r_{\rm infl})^n$ with $n>1$ such that the elasticity $e$
of money demand $M$ is negligible when the interest $r$ is 
not significantly larger than the inflation
rate $r_{\rm infl}$ and becomes large otherwise.

Trend following (in various elaborated forms)
was (and probably is still) one of the major strategy used by
so-called technical analysts (see \cite{Techanajorg} for a review and 
references therein).
More generally, it results naturally when investment strategies are 
positively related to
past price moves.
Trend following can be captured by the following expression of the order size
\be
\Omega_{\rm trend}(t) = a_1[\ln p(t) - \ln p(t-1)] +a_2 [\ln p(t) - \ln p(t-1)]
|\ln p(t) - \ln p(t-1)|^{m-1}~.   \label{jfala}
\ee
This expression corresponds to
driving the price up if the preceding move was up ($a_1>0$ and $a_2>0$).
The linear case $(a_1>0, a_2=0)$ is usually chosen by modelers. Here, we
generalize this model by adding the contribution proportional to $a_2>0$ 
from considerations
similar to those leading to the nonlinear expression
(\ref{fmaaak}) for the reversal term with an exponent $n>1$. We argue
that the dependence of the order size at time $t$ resulting from 
trend-following
strategies is a nonlinear function with exponent $m>1$
of the price change at previous time steps.
Indeed, a small price change from time $t-1$ to time $t$ may not be perceived
as a significant and strong market signal. Since many of the 
investment strategies
are nonlinear, it is natural to consider an average trend-following order size
which increases in an accelerated manner as the price change 
increases in amplitude.
Usually, trend-followers increase
the size of their order faster than just proportionally to the last 
trend. This is
reminiscent of the argument \cite{Techanajorg} that traders's psychology
is sensitive to a change of trend (acceleration or deceleration) and 
not simply to the
trend (velocity). The fact that trend-following
strategies have an impact on price proportional to the price change 
over the previous
period raised to the power $m>1$ means that trend-following strategies are
not linear when averaged over all of them: they tend to under-react 
for small price
changes and over-react for large ones.
The second term with coefficient $a_2$ captures this phenomenology.

\subsection{Nonlinear dynamical equation for stock market prices}
\label{sec:market-eq}

Introducing the notation
\be
x(t) = \ln [p(t)/p_f]~,  \label{nkkla}
\ee
and the time scale $\delta t$ corresponding to one time step,
and putting all the contributions (\ref{fmaaak}) and (\ref{jfala}) into
(\ref{mqmqmmq}), with $\Omega(t)=\Omega_{\rm fund}(t)+\Omega_{\rm trend}(t)$,
we get
\be
x(t+\delta t) - x(t) = {1 \over L} \left( a_1 ~[x(t) - x(t-\delta t)] + a_2 [x(t) - 
x(t-\delta t)]|x(t) - x(t-\delta t)|^{m-1}
- c ~x(t)  |x(t)|^{n-1} \right)~.   \label{ajkafa}
\ee
Expanding (\ref{ajkafa}) as a Taylor series in powers of $\delta t$, we get
\be
(\delta t)^2 {d^2 x \over dt^2} = - \left[1-{a_1 \over L}\right]~\delta t~ {dx \over dt} +
{a_2 (\delta t)^{m}  \over L} {dx \over dt}
|{dx \over dt}|^{m-1} - {c \over L} x(t)  |x(t)|^{n-1}~+~{\cal O}[(\delta t)^3]~, 
\label{kaklaklq}
\ee
where ${\cal O}[(\delta t)^3]$ represents a term of the order of $(\delta t)^3$.
Note the existence of the second order derivative,
which results from the fact that the price variation from present to
tomorrow is based on analysis of price change between yesterday and
present. Hence the existence of the three time lags leading to
inertia.
A special case of expression (\ref{ajkafa})
with a {\it linear} trend-following term $(a_2=0)$ and a {\it linear} reversal term
$(n=1)$ has been studied in \cite{Boucont,Farmer}, with the addition
of a risk-aversion term and a
noise term to account for all the other
effects not accounted for by the two terms (\ref{fmaaak}) and (\ref{jfala}). We shall
neglect risk-aversion as well as any other term and focus only on the
reversal and trend-following terms previously discussed to
explore the resulting price behaviors. Grassia has also studied a similar {\it linear}
second-order differential equation derived from market delay, positive feedback 
and including a mechanism for quenching runaway markets \cite{Grassia}.
Thurner \cite{Thurner} considers a three-dimensional system of three ordinary 
differential equations coupling price, ``friction'' and a state variable 
controlling friction, which can be mapped onto a third-order ordinary 
differential equation. The nonlinearity is on the friction term and not on
the trend term which is again assumed linear.

Expression (\ref{ajkafa}) is inspired by
the continuous mean-field limit of the model of
Pandey and Stauffer \cite{PandeyStauffer}, defined by starting from 
the percolation model of
market price dynamics
\cite{percocontbou,percocontbou2,percocontbou3} and
developed to account for the dynamics of the Nikkei and Russian market recessions
\cite{antilogperiod,antilogperiod2}. The generalization assumes
that trend-following and reversal to fundamental values are two forces that
influence the probability that a trader buys or sells the market. In addition,
Pandey and Stauffer \cite{PandeyStauffer} consider as we do here that the
dependence of the probability to enter the market is a nonlinear function
with exponent $n>1$ of
the deviation between market price and fundamental price. However, they do not
consider the possibility that $m>1$ and stick to the linear
trend-following case.
We shall see that the analytical control offered by our
continuous formulation allows us
to get a clear understanding of the different dynamical phases.

Among the four terms of equation (\ref{kaklaklq}), the first term of
the right-hand-side of (\ref{kaklaklq})
is the least interesting. For $a_1<L$, it corresponds to a
damping term which becomes
negligible compared to the second term
in the terminal phase of the growth close to the singularity
when $|dx/dt|$ becomes very large. For $a_1>L$, it corresponds
to a negative viscosity
but the instability it provides is again subdominant for $m>1$. The main
ingredients here are the interplay between the inertia provided by the second
derivative in the left-hand-side, the destabilizing nonlinear
trend-following term
with coefficient $a_2>0$ and the nonlinear reversal term. In order to
simplify the
notation and to simplify the analysis of the different regimes, we
shall neglect
the first term of the right-hand-side of (\ref{kaklaklq}), which amounts
to take the special value $a_1=L$. In a field theoretical
sense, our theory is tuned right at the ``critical point'' with a 
vanishing ``mass'' term.

Equation (\ref{kaklaklq}) can be viewed in two ways. It can be 
seen as a convenient short-hand notation for the intrinsically
discrete equation (\ref{ajkafa}), keeping the time step $\delta t$ small
but finite. In this interpretation, we pose
\ba
\alpha &=& a_2 (\delta t)^{m-2} /L ~,  \label{alphaeq1}    \\
\gamma &=& c /L(\delta t)^2~,   \label{gammaeq2} 
\ea
which depend explicitely on $\delta t$, 
to get
\be
{d^2 x \over dt^2} = \alpha {dx \over dt}
|{dx \over dt}|^{m-1} - \gamma x(t)  |x(t)|^{n-1}~.   \label{kakalaklq}
\ee
A second interpretation is to genuinely take the continuous limit 
$\delta t \to 0$ with the constraints $a_2/L \sim (\delta t)^{2-m}$
and $c/L \sim (\delta t)^2$. This allow us to define the now
$\delta t$-independent coefficients $\alpha$ and $\gamma$ according to 
(\ref{alphaeq1}) and (\ref{kakalaklq}) and obtain the 
truly continuous equation (\ref{kakalaklq}). This equation 
can also be written as
\ba
{d y_1 \over dt} &=& y_2~,  \label{dyn1bis}\\
{d y_2 \over dt} &=& \alpha y_2 |y_2|^{m-1} - \gamma y_1 |y_1|^{n-1}~.
\label{dyn2bis}
\ea
This is the system we are going to study for $m>1$ and $n > 1$. For
further discussions, we call
the term proportional to $\alpha$ (resp. $\gamma$) the trend or
positive feedback term
(resp. the reversal term). The richness of behaviors documented below results
from the competition between these two terms.

In defining the generalized dynamics (\ref{dyn1bis},\ref{dyn2bis}) for
the market price,
we aim at a fundamental dynamical understanding of the observed
interplay between
accelerating growth and accelerating (log-periodic) oscillations, that
have been
documented in speculative bubbles preceding large crashes
\cite{crash,antilogperiod2,nasdaq}.

We shall show below that 
the origin $(y_1=0, y_2=0)$ plays
a special role as the unstable fixed point around which spiral structures
of trajectories are organized in phase space $(y_1, y_2)$. It is 
particularly interesting that this point plays a special role since
$y_1=0$ means that the observed
price is equal to the fundamental price. If, in addition,
$y_2=0$, there is no trend, i.e., the
market ``does not know'' which direction to take. The fact that this is
the point of instability  around which the price trajectories organize 
themselves provides a fundamental understanding of the cause of
the complexity of market price time series based on the
instability of the fundamental price ``equilibrium''.

\section{Population dynamics}
\label{sec:pop}

As a standard model of population growth, Malthus' model assumes that the size
of a population increases by a fixed growth rate $\sigma$
independently of the size of the population and thus gives an exponential
growth:
\be
{d p \over dt} = \sigma p(t) ~. \label{ajdaak}
\ee
The logistic equation attempts to correct for the resulting unbounded
exponential growth by assuming a finite carrying capacity $K(t)$ such that the
population instead evolves according to
\be
{d p \over dt} =  \sigma_0 p(t) \left[K(t) - p(t)\right]~, \label{ajdak}
\ee
where $\sigma_{0}$ controls the amplitude of the nonlinear saturation term.
Applying this model to the human population on earth,
Cohen and others (see \cite{Cohenscience} and references therein) have put
forward idealized models taking into account interaction between the human
population $p(t)$ and the corresponding carrying capacity $K(t)$ by assuming
that $K(t)$ increases with $p(t)$ due to technological progress such as the
use of tools and fire, the development of agriculture, the use of fossil fuels,
fertilizers {\it etc.} as well an expansion into new habitats and the removal
of limiting factors by the development of vaccines, pesticides, antibiotics,
{\it etc.} If $K(t)$ grows faster than $p(t)$, then $p(t)$
explodes to infinity after a finite time creating a singularity
\cite{JohSorgreat}. In this case,
the limiting factor $-p(t)$ can be dropped out and, assuming a simple power
law relationship $K(t) \propto [p(t)]^{\delta}$ with $\delta >1$, (\ref{ajdak})
can be written as (\ref{ajdaak}) with an accelerating growth rate $\sigma$
replacing $\sigma_0$:
\be
\sigma = \sigma_0  [p(t)]^{\delta}~. \label{aafajdak}
\ee
The generic consequence of a power law acceleration in the growth rate is the
appearance of singularities in finite time:
\be \label{pow}
p(t) = p(0) \left({t_c - t \over t_c}\right)^{-{1 \over \delta}}, ~~~~\mbox{for $t$ close to $t_c$}~,
\ee
where
$t_c$ is determined by the constant of integration, {\it i.e.}, the initial
condition $p(0)$ as $t_c=[p(0)]^{-\delta}/\delta \sigma_0$.
Equation (\ref{aafajdak}) is said to have a ``spontaneous'' or ``movable''
singularity at the critical time $t_c$ \cite{benderorszag}, 

Note that, using (\ref{pow}), (\ref{aafajdak}) can be written
\be
{d\sigma \over dt} \propto \sigma^2~,   \label{bvnx}
\ee
showing that the finite-time singularity of the population $p(t)$
is the result of the finite-time singularity of its growth rate, resulting from
the quadratic growth equation (\ref{bvnx}).

We now generalize (\ref{bvnx}) as
  \be
{d\sigma \over dt} = \alpha \sigma |\sigma|^{m-1} - \gamma \ln (p/K_{\infty})
|\ln (p/K_{\infty})|^{n-1}~
  \label{bvnaax}
\ee
for the following reasons.
Apart from the absolute value, the first term in the r.h.s. of (\ref{bvnaax})
is the same as (\ref{bvnx})
with $m=2$. In addition, we allow the instantaneous growth rate 
$\sigma$ to be negative
and thus its growth has to be signed. The novel second term in the 
r.h.s. of (\ref{bvnaax})
takes into account a saturation or restoring effect such that by 
itself this term
attracts the population $p(t)$ to an asymptotic constant carrying 
capidity $K_{\infty}$. Using the
logarithm of the ratio $p/K_{\infty}$ is the natural choice for the 
dynamics of a growth rate
since $\ln (p/K_{\infty})$ is nothing but the effective cumulative growth rate.
For $n=1$,
$- \gamma \ln (p/K_{\infty}) |\ln (p/K_{\infty})|^{n-1} = - \gamma \ln (p/K_{\infty})$
corresponds to a linear (in $\ln (p/K_{\infty})$)
restoring term. A choice $n>1$ captures the following effect: the 
restoring term is
very weak when $p$ departs weakly from $K$ and then becomes rather 
suddenly stronger
when this deviation increases. This nonlinear feedback effect is 
intended to capture
the many nonlinear (often quasi-threshold) feedback mechanisms acting 
on population
dynamics. In the limit $n \to +\infty$, the reversal term acts as a threshold.
Note that the absolute values can be removed when the exponents
$m$ and $n$ are odd.

Expression (\ref{bvnaax}) generalizes (\ref{ajdak}) by putting together a
faster-than-exponential growth and an attraction to finite value. In contrast,
(\ref{ajdak}) puts together an exponential growth and an attraction to a finite
value.

Let us introduce the change of variable
\ba
y_1 &=& \ln (p/K_{\infty})~,  \label{nbvggd}\\
y_2 &=& \sigma ~.
\ea
The equations (\ref{ajdaak},\ref{bvnaax}) then retrieve (\ref{dyn1bis}) and 
(\ref{dyn2bis}).
For further discussions,
we shall refer to
the term proportional to $\alpha$ (resp. $\beta$) as the positive feedback or
acceleration term
(resp. the reversal term).

In defining the generalized dynamics (\ref{dyn1bis}) and 
(\ref{dyn2bis}) with (\ref{nbvggd}) for the population,
we aim at a fundamental dynamical understanding of the observed 
interplay between
accelerating growth and accelerating (log-periodic) oscillations, 
that have been
documented in \cite{Kapitza,Raan,Hanson,JohSorgreat}.

\section{Rupture of materials with competing damage and healing}
\label{sec:rapture}

Consider the problem of so-called creep or damage rupture \cite{Krajci} in which a rod
is subjected to uniaxial tension by a constant applied axial force $P$.
The intact cross section $A(t)$ of the rod is assumed to be
a function of time. The physical
picture is to envision myriads of microcracks damaging progressively the 
rod and decreasing its effective intact cross section that can sustain stress.
The problem is simplified by assuming that $A(t)$ is independent of the axial coordinate,
which eliminates necking as a possible mode of failure. The considered viscous
deformation is assumed to be isochoric, i.e., the rod volume remains constant
during the process. This provides a geometric relation between the 
rod cross-sectional
area and length $A_0 L_0 = A(t) L(t) =$ constant, which holds for all times.

The rate of creep strain $\epsilon_c$ can be defined as a function of 
geometry as
\be
{d \epsilon_c \over dt} = {1 \over L}{dL \over dt} = -{1 \over A}{dA 
\over dt}~,
\label{bvbncxs}
\ee
showing that
\be
\epsilon_c(t) = \ln {L(t) \over L_0} = - \ln {A(t) \over A_0}~,
\ee
where $L_0=L(t_0)$ and $A_0=A(t_0)$ correspond to the underformed state 
$\epsilon_c(t_0)=0$ at time $t_0$.

The rate of change of the creep strain $\epsilon_c(t)$ is assumed to follow the
rheological Norton's law, i.e.,
\be
{d \epsilon_c \over dt} = C \sigma^{\mu}~, ~~~~{\rm with}~\mu>0~, 
\label{nvbncv}
\ee
where the stress
\be
\sigma = P/A    \label{ajajs}
\ee
is the ratio of the applied force over the cross section of the rod.
Eliminating $d \epsilon_c / dt$ between (\ref{bvbncxs}) and 
(\ref{nvbncv}) and using
(\ref{ajajs})
leads to $A^{\mu - 1} dA/dt = - CP^{\mu}$, i.e., $A(t) = A(0)
\left({t_c-t \over t_c}\right)^{1/\mu}$, where
the critical failure time is given by $t_c=[A(0)/P]^{\mu}/(\mu C)$.
The rod cross section thus vanishes in a finite time $t_c$ and
as a consequence the stress diverges as the time $t$ goes to the 
critical time $t_c$ as
\be
\sigma = P/A  = {P \over A(0)} \left({t_c-t \over t_c}\right)^{-{1 \over \mu}}~.
\ee
Physically, the constant force is applied to a thinner cross
section, thus enhancing the stress, which in turn accelerate the 
creep strain rate,
which translates into an acceleration of the decrease of the rod 
cross section and so on.
In other words, the finite-time singularity results from the positive
feedback of the increasing stress on the thinner cross section and vice-versa.
This finite-time singularity for the stress can be reformulated as a
self-contained equation expressed only in terms of the stress:
\be
{d \sigma \over dt} = C \sigma^{\mu+1}~.   \label{nvbaancv}
\ee

Let us now generalize this model by allowing not only creep 
deformations leading to damage but
also recovery or healing as well as a strain-dependent loading. We 
thus propose to modify the expression (\ref{nvbaancv}) into
\be
{d \sigma \over dt} = \alpha \sigma |\sigma|^{m-1} - \gamma 
\epsilon_c |\epsilon_c|^{n-1}~,
\label{bvnklksz}
\ee
The first term in the right-hand-side of (\ref{bvnklksz}) is similar 
to (\ref{nvbaancv})
by redefining $\mu+1$ as $m$, and captures the accelerated growth of the
stress leading to a finite-time singularity. It embodies the positive
geometrical feedback of a reduced intact area on the effective stress
applied to whole system. The addition of 
the second term
in the right-hand-side of (\ref{bvnklksz}) implies a modification of 
Norton's law
which is no more specified by the exponent $\mu$ or $m$ and 
introduces the novel
physical ingredient that damage can be reversible. For convenience, we choose
a specific power law dependence $- \gamma \epsilon_c |\epsilon_c|^{n-1}$ to
capture the healing mechanism. This term alone tends to decrease the effective
stress and describes a recovery of the material 
since a reduction of the effective stress
is associated with an increase of carrying area of the intact material.
Alternatively, we can interpret
(\ref{bvnklksz}) as defining the loading, which becomes strain-dependent: a larger
strain implies less room for additional stress increase, as for instance occurs
in mechanical apparatus in the laboratory which are often limited to small 
deformations
and relax the applied stress beyond a given strain. The mechanism is also
attractive for describing the tectonic loading of faults which is occurring
with mixed stress and strain rates, rather than a pure imposed stress or
strain rate. 

Bringing the system out of
equilibrium and then releasing it, the equation (\ref{bvnklksz}) describes
how the system can either recover an equilibrium or rupture in finite-time
due to accumulating creep and damage in its dynamical attempt
to come back to equilibrium.
The novel second term in the r.h.s. of (\ref{bvnklksz})
takes into account a healing process or work-hardening, such that large
creep deformations hinder and may even reverse the stress increase.
By itself, this term
attracts the cross section $A(t)$ back to the equilibrium value $A_0$.
Since the cross-sectional area $A(t)$ can be alternatively interpreted as the surface
of intact material able to carry the stress, healing increases the area
of intact material and thus decreases the effective stress.

We close the model by assuming again Norton's law but with an exponent $\mu'$
different from $\mu$:
\be
{d \epsilon_c \over dt} =  C \sigma^{\mu'}~.
\ee
Incorporating the constant $C$ in a redefinition of time $C t \to t$ (with
suitable redefinitions of the coefficients $\alpha/C \to \alpha$ and
$\gamma/C \to \gamma$) and posing
\ba
y_1 &=& \epsilon_c = - \ln {A(t) \over A_0}~,  \label{mfmla}\\
y_2 &=& \sigma ~,
\ea
we retrieve the dynamical system (\ref{dyn1bis}) and (\ref{dyn2bis}) for the special 
choice $\mu'=1$, which we shall restrict to in the sequel.
We are going to study this system in the regime where $m>1$ and $n > 1$.
The first condition  $m>1$ ensures the existence of a finite-time singularity
describing a positive feedback between the stress increase and the
cross section  decrease.
The second condition $n >1$ ensures that the healing process is only active
for large deformations: the larger $n$ is, the more threshold-like is
this  effect
with respect to the amplitude of the creep strain.

In defining the generalized dynamics (\ref{dyn1bis},\ref{dyn2bis})
with (\ref{mfmla})
for the rupture dynamics,
we aim at a fundamental dynamical understanding of the observed
interplay between
accelerating growth and accelerating (log-periodic) oscillations, that
have been
documented in time-to-failure analysis of material rupture
\cite{thermalfuse,thermalfuse2,Anifrani,Lamai,Arbabi,Ray,
Ciliberto,Ciliberto2,Ciliberto3,critruptcan,critrupt}.

\section{Individual components of the dynamics}
\label{sec:element}

The system (\ref{dyn1bis},\ref{dyn2bis}) can also be written as
(\ref{kakalaklq}),
which we rewrite here for convenience, with uniform notation:
\be
{d^2 y_1 \over dt^2} = - \gamma y_1 |y_1|^{n-1}
  + \alpha {d y_1 \over dt} |{d y_1 \over dt}|^{m-1} ~.
\label{kaklakqwlq}
\ee
This autonomous expression has the following interpretation.
The left-hand side (l.h.s.) is the inertia for the variable $\yo$.
The right-hand side (r.h.s.) has two elements.
The first element in the r.h.s. describes a restoring force for $\gamma>0$
which will be the case studied here. Coupled
with inertia, this leads generally to an oscillatory behavior.
The second element is a positive trend
which can lead to singular behavior for $\alpha>0$ and $m>1$.

Our strategy is first to study these two components coupled to
the inertia as separate sub-dynamical systems. The interplay between these two
components will then help deepen our understanding of the overall
dynamics.

Each sub-dynamical system can be reduced
to a normalized model where one single (resp. pair of) orbit(s)
give(s) a template dynamics of the entire sub-dynamical system.
Because the system is autonomous, we describe
two complementary approaches to describe its dynamics:
(1) phase portrait using orbits graphed in the phase space $\vy=(\yo,\yt)$;
(2) time evolution of trajectories $\vy(t;\vyz,\tz)$
with initial condition $\vyz$ at time $\tz$.

\subsection{The restoring term: oscillations}
\label{sec:element-osc}

We first consider the restoring term and examine the nature of the 
corresponding
oscillations. Motivated by sections \ref{sec:market}, \ref{sec:pop} 
and \ref{sec:rapture}
describing the application to various physical processes,
we focus on the relevant case $\gamma>0$  and $n>1$.

\subsubsection{Model}
\label{sec:element-osc-model}

Keeping only the first term in the r.h.s. of (\ref{kaklakqwlq}) gives:
\be
{d^2 y_1 \over dt^2} = - \gamma~ y_1  |y_1|^{n-1}~.
\label{eq:osc-inertia}
\ee
This system can be written as a one-degree-of-freedom
Hamiltonian system
\ba
{d\over d t}
  \lp \begin{array}{l} \yo \\ \yt \end{array} \rp =
  \lp \begin{array}{cl} &{\partial\over\partial \yt}
                    \\ -&{\partial\over\partial \yo}  \end{array} \rp
   H(\vy;n,\gamma)~ =
  \lp \begin{array}{cl} \yt \\ -\gamma \yo|\yo|^{n-1} \end{array} \rp ~,
   \label{eq:osc-model-dydt}
\ea
where
\ba \label{eq:osc-model-H}
    H(\vy;n,\gamma)\equiv
   {\gamma\over n+1}(\yo^{2})^{{n+1\over 2}}+{1\over 2}\yt^{2}~.
\ea
An orbit of (\ref{eq:osc-model-dydt}) in the $\vy$ phase space
for fixed $(n,\gamma)$ can be given as a graph of constant
$H(\vy;n,\gamma)$.
In other words, a trajectory $\vy(t;\vyz,\tz)$ going through
$\vyz$  follows contours of constant $H(\vyz;n,\gamma)$ in time.

\subsubsection{Template dynamics in the normalized model}
\label{sec:element-osc-norm}
\begin{figure}
\begin{center}
\includegraphics[height=20cm]{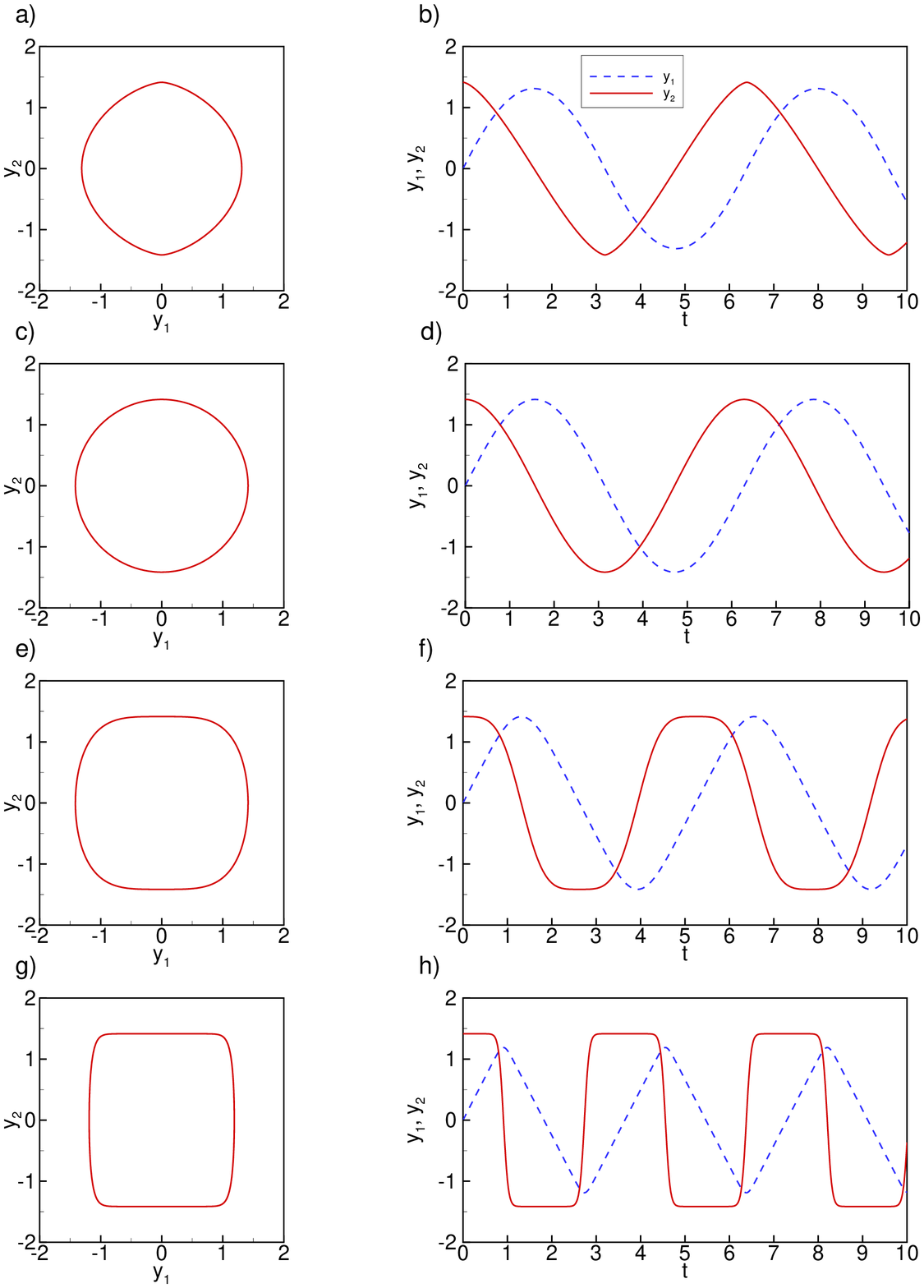}
\end{center}
\caption{Normalized model for the oscillatory term: phase space (left panels)
and time series (right panels) for a,b) $n=0.5$;
c,d) $n=1$; e,f) $n=3$; and g,h) $n=15$. The continuous (resp. dashed) line
is $y_2$ (resp. $y_1$).}
\label{fg:norm_osc}
\end{figure}

We now describe the template dynamics
using a normalized model.
We define the following normalized variables denoted by hat $\hat{\{\cdot\}}$:
\ba \label{eq:osc-norm-var}
  \lp \begin{array}{l}  \yo \\ \yt \\ t \end{array} \rp
  = \lp \begin{array}{llc}
   \gamma^{{-1\over n+1}} & H^{{1\over n+1}} & \hat{y}_{1} \\
   &  H^{{1\over 2}} & \hat{y}_{2} \\
   \gamma^{{-1\over n+1}} & H^{{1-n\over  2(n+1)}}~ & \hat{t} \end{array} \rp~~.
\ea
Then all $H(\vy;n,\gamma)$-contours in the original $\vy$ phase space
collapse onto a single closed curve in the
normalized  $\hat{\vy}$ phase space. The case
$H=0$ at $\vy=0$, which is an elliptic fixed point, is excluded from 
this analysis.
The closed curve corresponds to
the normalized orbit defined by:
   \ba
     1  & = &{1\over n+1}(\hat{y}_{1}^{2})^{{n+1\over 2}}
     + {1\over2} \hat{y}_{2}^{2}~.\label{eq:osc-norm-inv}
   \ea
Therefore, the range of oscillation for the
variable $\hat{y}_{1}$ and its velocity
$\hat{y}_{2}=\frac{d}{d\hat{t}}\yo$ is:
   \ba
     \hat{y}_{1}\in[-(n+1)^{{1\over n+1}},(n+1)^{{1\over n+1}}]~, ~~~
     \hat{y}_{2}\in[-\sqrt{2},\sqrt{2}]~, \label{eq:osc-norm-range}
   \ea
as shown in the left panels of Figure~\ref{fg:norm_osc}.
The original dynamics in $\vy$ can be recovered using 
(\ref{eq:osc-norm-var}).

The closed orbit in phase space represents an oscillatory
dynamics for any fixed $n>0$.
For $n=1$, the curve given by (\ref{eq:osc-norm-inv})
is a perfect circle with radius $\sqrt{2}$
corresponding to a linear harmonic oscillator
(Figure~\ref{fg:norm_osc}c).
For $n\neq 1$, the closed curve is a nonlinear generalization
of the circle.
As $n$ increases above $1$, $(\hat{y}_{1}^{2})^{n+1 \over 2}$ becomes
small for $|\hat{y}_{1}|<1$.
 From the condition (\ref{eq:osc-norm-inv}),
it follows that  $|\hat{y}_{2}|$ remains almost constant near
its maximum $\sqrt{2}$ for $|\hat{y}_{1}|<1$.
Furthermore from (\ref{eq:osc-norm-range}), the
range of $|\hat{y}_{1}|$ decreases monotonically to 1 as $n$ increases,
while the range of $\hat{y}_{2}$  remains unchanged for any $n$.
On the whole, the geometry of the closed orbit becomes closer and closer to
a square as $n$ increases,
as shown in Figures~\ref{fg:norm_osc}e and \ref{fg:norm_osc}g.

The normalized dynamical model can be obtained by substituting
(\ref{eq:osc-norm-var}) into (\ref{eq:osc-model-dydt}).
In addition, it can be decoupled into two independent first-order
nonlinear ordinary differential equations (ODEs) using the
condition (\ref{eq:osc-norm-inv}):
\ba
{d\over dt}
   \lp \begin{array}{c} \hat{y}_{1} \\ \hat{y}_{2} \end{array} \rp
  = \lp \begin{array}{c} \hat{y}_{2}
	\\ -\hat{y}_{1} |\hat{y}_{1}|^{n-1} \end{array} \rp
  = \lp \begin{array}{rll} {\rm sign}[\hat{y}_{2}]&
  \sqrt{2} & [1-{1\over n+1}(\hat{y}_{1}^{2})^{{n+1\over 2}}]^{{1\over 2}} \\
  - {\rm sign}[\hat{y}_{1}]
   &(n+1)^{{n\over n+1}} &[1-{1\over 2}\hat{y}_{2}^{2}]^{{n\over n+1}}
	\end{array} \rp~.
	\label{eq:osc-norm-dydt}
\ea
Therefore, the normalized trajectory goes around the
origin in a clockwise
direction along the closed curve given by (\ref{eq:osc-norm-inv}).

Given an initial condition $\hat{\vy}_{0}$ satisfying
(\ref{eq:osc-norm-inv}), the two ODEs in (\ref{eq:osc-norm-dydt})
can be solved separately and provide the trajectory
$\hat{\vy}(\hat{t};\hat{\vy}_{0},\hat{t}_{0})$.
However, once $\hat{y}_{1}(\hat{t};\hat{\vy}_{0},\hat{t}_{0})$ is
solved using the top ODE,
$\hat{y}_{2}(\hat{t};\hat{\vy}_{0},\hat{t}_{0})$ can be
algebraically computed using (\ref{eq:osc-norm-inv}) without solving
the bottom ODE for $\hat{y}_{2}$; and vice versa.
The period of oscillation can be obtained  by integrating
one of the two ODEs  over the range
defined by (\ref{eq:osc-norm-range}).

For some special $n$'s, the normalized model has explicit analytical
solutions. For $n=1$, corresponding to a harmonic oscillator, the
solution is:
  \ba
   \hat{\vy}(\hat{t};\hat{\vy}_{0},\hat{t}_{0})
   &=&\sqrt{2}(\sin (\hat{t}-\hat{t}_0+\theta),\cos(\hat{t}-\hat{t}_0+\theta))~,
	 \label{eq:osc-norm-n1}
  \ea
where the initial condition
$\hat{\vy}_{0}=\sqrt{2}(\sin\theta,\cos\theta)$
satisfies (\ref{eq:osc-norm-inv}) for any real number $\theta$.
Its period of oscillation is $2\pi$.

For $n=3$, the solution
for $\hat{y}_{1}(\hat{t};\hat{\vy}_{0},\hat{t}_{0})$ is \cite{Gradshtein}:
\be
F\left(\arccos {\hat{y}_{1}(\hat{t}) \over \hat{y}_{10}}, {1 \over 
\sqrt{2}}\right) =
\hat{y}_{10} ~ (\hat{t} - {\hat t}_0)~,  \label{eq:osc-norm-n3}
\ee
where $\hat{y}_{10} \equiv \hat{y}_{1}(\hat{t}=\hat{t}_0)$ and 
$\hat{y}_{2}(\hat{t}=\hat{t}_0)=0$
and
$F\left(u, k)\right)$ is the elliptic integral of the first kind
defined by
\be
F(u, k) = \int_0^u {dv \over \sqrt{1 - k^2 \sin^2 v}}~.\label{eq:osc-norm-Fuk}
\ee

The evolution of a trajectory with initial condition
$\hat{\vy}_{0}=(0,\sqrt{2})$ at $\hat{t}_{0}=0$ is shown in the
right panels of Figure~\ref{fg:norm_osc} as a function of $\hat{t}$.
For $n=1$, the normalized variable $\hat{y}_{1}$ is preceded by
its velocity $\hat{y}_{2}$ by a $\frac{\pi}{2}$-phase shift.
As $n$ increases, the amplitude of the normalized velocity $|\hat{y}_{2}|$
becomes nearly constant about $\sqrt{2}$ for $|\hat{y}_{1}|<1$
[see Figures~\ref{fg:norm_osc}e and \ref{fg:norm_osc}g as well as the
discussion concerning the geometry of the normalized orbit
given by (\ref{eq:osc-norm-inv})].
Accordingly, $\hat{y}_{1}$ evolves almost linearly in time
with a constant velocity $\hat{y}_{2}=\pm\sqrt{2}$ for
$|\hat{y}_{1}|<1$.
Thus, time series of $\hat{y}_{1}$ and $\hat{y}_{2}$
respectively have saw-teeth and step-function shapes.
The period of the oscillation becomes shorter as $n$ increases, because
the velocity amplitude $|\hat{y}_{2}|$ remains closer and closer to its
maximum as $n$ increases during each oscillation cycle.

\subsubsection{Global dynamics}
\label{sec:element-osc-global}

Having understood the template dynamics of the normalized model,
we describe the global dynamics of the oscillatory element
in the original model.
Two parameters, $n$ and $\gamma$, are used to define
the one-degree-of-freedom
Hamiltonian system given by (\ref{eq:osc-model-dydt}).
 From the normalization defined by (\ref{eq:osc-norm-var}) and the
resulting system given by (\ref{eq:osc-norm-dydt}),
we see that the exponent $n$ is the controlling parameter.
The coefficient $\gamma$ contributes only for the scaling of
$(\yo,t)$.

\begin{figure}
\begin{center}
\includegraphics[height=20cm]{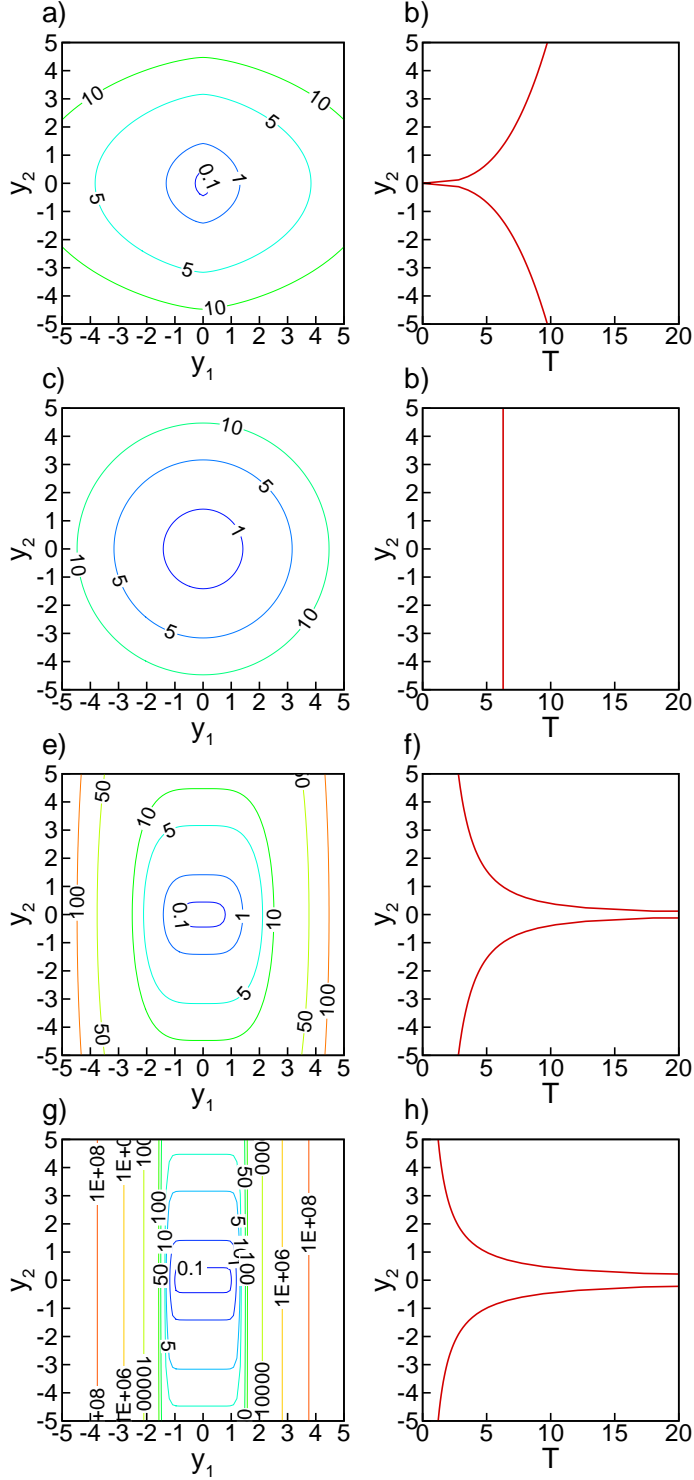}
\end{center}
\caption{Phase portrait (left panels) and period $T$ of the oscillations
(right panel) for the oscillatory term with $\gamma=1$ for all,
so that $H=1$ corresponds to the normalized model
(Figure~\ref{fg:norm_osc}):
a,b) $n=0.5$;
c,d) $n=1$; e,f) $n=3$; and g,h) $n=15$.
The period of the oscillation is on the abscissa as a 
function of the maximum equal to $(2 H)^{1\over 2}$ of $\yt$ on the ordinate.}
\label{fg:phase_osc}
\end{figure}

For fixed $(n,\gamma)$, the phase portrait in $\vy$ consists of a family of
closed orbits parameterized by $H$. Each orbit has a unique $H$
and oscillates in the clockwise direction around the
origin (left panels of Figure~\ref{fg:phase_osc}).
Given the properties of the normalized orbit in
Section~\ref{sec:element-osc-norm},
three main properties of individual orbit as function of $H$
follow.

First, $H$ measures the amplitude of oscillation.
Each orbit in $\vy$ ranges over:
\ba
   \yo\in[-\left(\frac{(n+1)H}{\gamma}\right)^{{1\over n+1}},
           \left(\frac{(n+1)H}{\gamma}\right)^{{1\over n+1}}],~~
   \yt\in[-(2H)^{1\over 2},(2H)^{1\over 2}]~ \label{eq:osc-global-range}
  \ea
 from (\ref{eq:osc-norm-var}) and (\ref{eq:osc-norm-range}).
The left panels of Figure~\ref{fg:phase_osc} show the
orbits in the  $\vy$ phase space as curves of constant $H$.
The higher $H$ is, the larger is the amplitude of the oscillations.

Second, $H$ determines the geometry of the orbits,
except for $n=1$ where all orbits are ellipses of the same
aspect ratio $\gamma^{{1\over n+1}}$ (Figure~\ref{fg:norm_osc}c).
For $n\neq 1$, we describe the geometry in terms of the
deformation from the normalized curve (\ref{eq:osc-norm-inv})
by comparing the two normalization coefficients for $\yo$ and $\yt$
in (\ref{eq:osc-global-range}), i.e.,  $H^{1\over n+1}$ and $H^{1\over 2}$,
respectively.
These two coefficients also govern the range of the oscillation
(\ref{eq:osc-norm-var}).
For $n<1$, the two coefficients have the following relation:
$H^{1\over n+1}<H^{1\over 2}<1$ for $H<1$, and
$H^{1\over n+1}>H^{1\over 2}>1$ for $H>1$.
Hence, an orbit with small amplitude ($H<1$) has a geometrical shape
stretched along the vertical direction
(Figure~\ref{fg:phase_osc}a), as deduced from the normalized closed curve
(Figure~\ref{fg:norm_osc}a). This is because $\yo$ is reduced more
than  $\yt$.
Similarly, an orbit with a large amplitude ($H>1$) has a geometrical shape
stretched horizontally.
For $n>1$, the relation is reversed:
$1>H^{1\over n+1}>H^{1\over 2}$ for $H<1$, and
$1<H^{1\over n+1}<H^{1\over 2}$ for $H>1$.
Accordingly, orbits with small ($H<1$) or large ($H>1$)
amplitudes respectively have horizontally or vertically stretched
geometries compared with the normalized closed curve
(Figure~\ref{fg:phase_osc}c,d).

Third, $H$ controls the speed of the oscillation, except
for the harmonic oscillator case $n=1$ which has
a constant period $2\pi/\sqrt{\gamma}$.
For $n\neq 1$, the period of the oscillations is obtained
using (\ref{eq:osc-norm-var}),  (\ref{eq:osc-norm-range}) and
(\ref{eq:osc-norm-dydt}):
  \ba
     T(H;n,\gamma) &=& C(n) ~ \gamma^{-1\over n+1} ~
	H^{{1-n\over 2(n+1)}}~,  \label{eq:osc-global-T}
  \ea
where $C(n)$ is a positive number given by:
  \be \label{eq:osc-global-C}
   C(n)\equiv {4\over (n+1)^{{n\over n+1}}}
  \int_{0}^{\sqrt{2}}
   {d\hat{y}_{2}  \over \lp 1-{1\over 2}\hat{y}_{2}^{2}\rp^{{n\over n+1}}}~~.
  \ee
Differentiating the expression of the period $T(H;n,\gamma)$ given by 
(\ref{eq:osc-global-T})
with respect to $H$ gives:
\ba \label{eq:osc-global-dTdH}
    {\partial\over\partial H}T(H;n,\gamma)&= &
    {1-n \over 2(n+1)}~ C(n)~
	\gamma^{-1\over n+1}~H^{-{3n+1\over 2(n+1)}} ~~.
  \ea
Therefore, ${\partial\over\partial H}T(H;n,\gamma)$
can be positive or negative depending on $n$.
For $n<1$ where $\frac{\partial}{\partial H}T>0$,
the period increases monotonically from $0$ to $\infty$
as $H$ increases. In contrast, for $n>1$,
the period decreases monotonically from $\infty$ to $0$
as $H$ increases.
The right panels of Figure~\ref{fg:phase_osc} show the
period of oscillation on the abscissa
as function of the maximum amplitude reached by  $y_{2}$
equal to $\sqrt{2 H}$, given
as a measure of the oscillation amplitude.

\subsection{The trend term: singular behavior}
\label{sec:element-sing}

We now consider the trend term and examine the nature of the singularity
which manifests itself in ``finite-time'' in the behavior of the
velocity $\yt$ and of the variable $\yo$.
Motivated by the applications to concrete physical processes discussed
in sections \ref{sec:market}, \ref{sec:pop} and \ref{sec:rapture},
we focus on the case $\alpha>0$  and $m>1$.

\subsubsection{Model}
\label{sec:element-sing-model}

Keeping only the second term in the r.h.s.~of (\ref{kaklakqwlq})
and re-writing it as a system of two-dimensional ODEs for the variable
$\yo$ and its velocity $\yt$ give:
\be
\frac{d}{dt} \lp \begin{array}{c} \yo \\ \yt \end{array} \rp
= \lp \begin{array}{c} \yt \\ \alpha \yt |\yt|^{m-1} \end{array} \rp~.
\label{eq:sing-model-dydt}
\ee
In this system,
\be
G(\vy; m,\alpha) \equiv \yo -
\left\{\begin{array}{ccl}
  {1\over \alpha (2-m)} & \yt |\yt|^{1-m} & ~~~\mbox{for $m\neq 2$}~, \\
  {1\over \alpha } & {\rm sign}[\yt] \log|\yt| & ~~~\mbox{for $m=2$}~
         \end{array} \right. \label{eq:sing-model-G}
\ee
is an invariant, i.e., ${d\over dt}G(\vy; m,\alpha)=0$.
This can be easily verified by differentiating (\ref{eq:sing-model-G})
with respect to $t$ and then substituting in (\ref{eq:sing-model-dydt}).
Therefore, like $H(\vy;n,\gamma)$ in the
one-degree-of-freedom Hamiltonian system discussed in the previous section
for the restoring term,
a curve of constant $G(\vy; m,\alpha)$
in the $\vy$ phase space corresponds to an orbit governed by
(\ref{eq:sing-model-dydt}).
A trajectory $\vy(t;\vyz,\tz)$ going through $\vyz$ satisfies
$G(\vy(t;\vyz,\tz);m,\alpha)=G(\vyz;m,\alpha)$ for any $t$.

\subsubsection{Template dynamics in the normalized model}
\label{sec:element-sing-norm}

To describe the template dynamics of the trend term,
we reduce the system (\ref{eq:sing-model-dydt}) to a normalized
model using the invariant $G$ (\ref{eq:sing-model-G}).
We define the following normalized variables denoted by
$\tilde{\{\cdot\}}$:
\ba \label{eq:sing-norm-var}
  \lp \begin{array}{c}
	\yo \\ \yt \\ t \end{array} \rp =
  \lp \begin{array}{clc} \alpha^{-1}&\tilde{y}_{1}&+~G\\
    &\tilde{y}_{2}& \\ \alpha^{-1}&\tilde{t}& \end{array} \rp~.
\ea
Substituting (\ref{eq:sing-norm-var}) into
(\ref{eq:sing-model-dydt}) gives the normalized dynamical model:
\ba
{d\over d \tilde{t}}
  \lp \begin{array}{l} \tilde{y}_{1} \\ \tilde{y}_{2} \end{array} \rp =
  \lp \begin{array}{c} \tilde{y}_{2} \\ \tilde{y}_{2}|\tilde{y}_{2}|^{m-1}
   \end{array} \rp~~.\label{eq:sing-norm-dydt}
\ea
Therefore, the normalized velocity $\tilde{y}_{2}$
undergoes an irreversible amplification if it starts from
$|\tilde{y}_{2}|\neq0$ and it pulls the variable $\tilde{y}_{1}$
along. The dividing point $\tilde{y}_{2}=0$ is a fixed point.
Furthermore, substituting (\ref{eq:sing-norm-var}) into (\ref{eq:sing-model-G})
gives an invariant condition for the normalized orbit:
\ba \label{eq:sing-norm-g}
  \tilde{y}_{1}=\tilde{g}(\tilde{y}_{2})~,
\ea
where
   \ba
     \tilde{g}(\tilde{y}_{2})=
\left\{\begin{array}{cll}
  {1\over (2-m)} & \tilde{y}_{2} |\tilde{y}_{2}|^{1-m}
	& ~~~\mbox{for $m\neq 2$} \\
  {\rm sign}[\tilde{y}_{2}] & \log|\tilde{y}_{2}|
	& ~~~\mbox{for $m=2$}~
         \end{array} \right.~. \nonumber
   \ea

\begin{figure}
\begin{center}
\includegraphics[height=15cm]{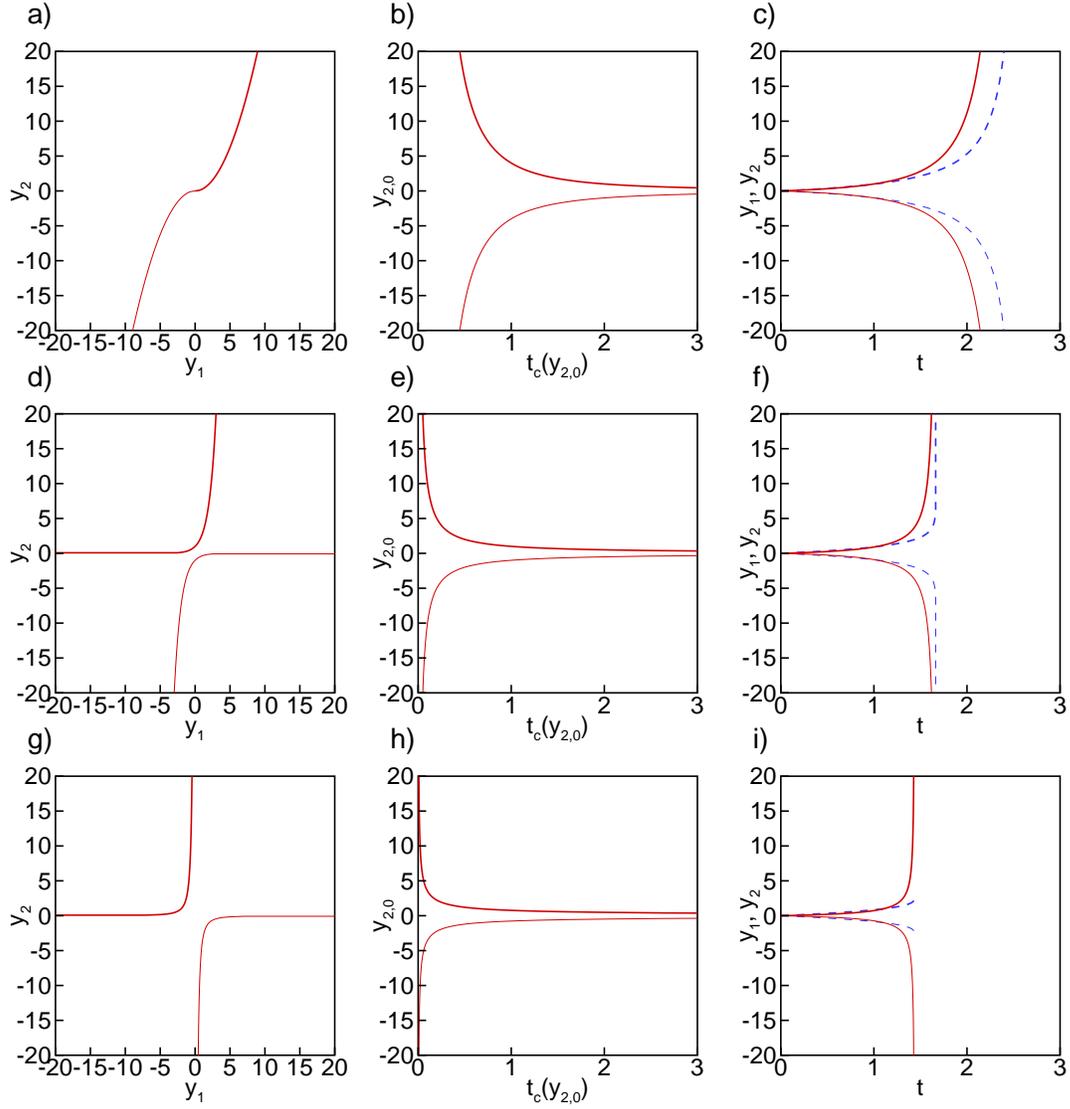}
\end{center}
\caption{
Normalized model for the singular term: phase space trajectories (left panels),
critical time $t_c$ (center panels) as a function of the initial
value $y_{2,0}$ and time series (right panels)
for a--c) $m=1.5$; d--f) $m=2$; g--h) $m=2.5$.
Thick and thin lines correspond to the pair of  normalized
orbits for $\yt>0$ and $\yt<0$, respectively, with initial 
condition $(y_{1,0}, y_{2,0})=(0, \pm 0.6)$. The continuous (resp. 
dashed) line
is $y_2$ (resp. $y_1$).}
\label{fg:norm_sng}
\end{figure}

The left panels of Figure~\ref{fg:norm_sng} show
graphs of (\ref{eq:sing-norm-g}) for $m=1.5$, 2 and 2.5.
Each graph consists of a pair of rotationally-symmetric orbits
for $\tilde{y}_{2}>0$ and $\tilde{y}_{2}<0$
with the symmetry given by $\tilde{\vy}\rightarrow-\tilde{\vy}$.
Through the normalization (\ref{eq:sing-norm-var}),
all orbits in the original $\vy$ phase space with $\yt>0$
collapse onto the single normalized orbit with
$\tilde{y}_{2}>0$ in $\tilde{\vy}$ phase space.
Similarly,
all orbits in the $\vy$ phase space with $\yt<0$ collapse onto
the single normalized orbit with
$\tilde{y}_{2}<0$ in the $\tilde{\vy}$ phase space.
Using (\ref{eq:sing-norm-dydt}), the slope of the graph is:
\ba
{d\tilde{y}_{2}\over d \tilde{y}_{1}} =
{d\tilde{y}_{2}\over d \tilde{t}}/
{d\tilde{y}_{1}\over d \tilde{t}} = |\tilde{y}_{2}|^{m-1}~.
  \label{eq:sing-norm-slope}
\ea
Integrating this equation retrieves the normalized invariant
condition given by (\ref{eq:sing-norm-g}).
For fixed $m$, the slope increases from 0 to $\infty$
as $|\tilde{y}_{2}|$ increases.
The higher $m$ is, the faster the slope increases.

The geometry of (the pair of)  symmetric normalized orbit(s)
depends on $m$ significantly (Figure~\ref{fg:norm_sng}).
This is due to a qualitatively different behavior
of $\tilde{y}_{1}=\tilde{g}(\tilde{y}_{2})$
as $\tilde{y}_{2}$ approaches towards either end of the
normalized orbit.  For $\tilde{y}_{2}>0$,
\ba \tilde{y}_{1}=\tilde{g}(\tilde{y}_{2})
   \in\left\{ \begin{array}{ll}
  (0,\infty)  & \mbox{for $1<m<2$} \\
  (-\infty,\infty)  & \mbox{for $m=2$} \\
  (-\infty,0)  & \mbox{for $m>2$}  \end{array} \right.,~
  \tilde{y}_{2}\in(0,\infty)~.
  \label{eq:trend-norm-lmt}
\ea
In other words, the normalized model undergoes a bifurcation at $m=2$.
Precisely for $m=2$, $\tilde{g}(\tilde{y}_2)$ spans the whole
interval $(-\infty,\infty)$ and thus interpolates
between the cases for $m<2$ and $m>2$ for which only one-half of the interval
is covered by the dynamics.

Accordingly, the dynamical behavior of the normalized trajectory changes
qualitatively when $1<m<2$, $m=2$ or $m>2$.
Any initial condition
$\tilde{\vy}_{0}\equiv(\tilde{y}_{1,0},\tilde{y}_{2,0})$
of the normalized trajectory
$\tilde{\vy}(\tilde{t};\tilde{\vy}_{0},\tilde{t}_{0})$
must satisfy  (\ref{eq:sing-norm-g}), i.e.,
$\tilde{y}_{1,0}=\tilde{g}(\tilde{y}_{2,0})$.
Using this initial condition, (\ref{eq:sing-norm-dydt})
can be solved analytically:
\ba
  \tilde{y}_{1}(\tilde{t};\tilde{\vy}_{0},\tilde{t}_{0})
    &= & \tilde{g}(\tilde{y}_{2,0})
    + {\rm sign}[\tilde{y}_{2,0}]~  \nonumber \\
        & & \times\left\{\begin{array}{ll}
          (m-1)^{{m-2\over m-1}}~{1\over 2-m}~
         [ (\tilde{t}_{c}(\tilde{y}_{2,0})-\tilde{t})^{{m-2\over m-1}} -
           (\tilde{t}_{c}(\tilde{y}_{2,0})-\tilde{t}_{0})^{{m-2\over m-1}} ]
                & \mbox{for $m\neq 2$} \\
          \log
         \lp{\tilde{t}_{c}-\tilde{t}_{0} \over
	\tilde{t}_{c}(\tilde{y}_{2,0})-\tilde{t}}\rp
                & \mbox{for $m=2$}
        \end{array} \right. ~, \nonumber \\
  \tilde{y}_{2}(\tilde{t};\tilde{\vy}_{0},\tilde{t}_{0})
    &=& {\rm sign}[\tilde{y}_{2,0}]~
    (m-1)^{-{1\over m-1}}~(\tilde{t}_{c}(\tilde{y}_{2,0})
		-\tilde{t})^{-{1\over m-1}} ~.
	\label{eq:sing-norm-range}
\ea
The solution is valid only for a semi-infinite time interval
$\tilde{t}\in(-\infty,\tilde{t}_{c}(\tilde{y}_{2,0}))$ up
to the normalized ``critical time:''
\be\label{eq:sing-norm-tc}
    \tilde{t}_{c}(\tilde{y}_{2,0})=\tilde{t}_{0}+
          {1\over m-1}|\tilde{y}_{2,0}|^{1-m},
\ee
for any arbitrary $\tilde{t}_{0}$.
The center panels of Figure~\ref{fg:norm_sng} shows
$\tilde{t}_{c}$ on the abscissa as a function
of $\tilde{y}_{2,0}$ on the ordinate with $\tilde{t}_{0}=0$.
The solution becomes singular as $\tilde{y}_{2}$ approaches
$\pm\infty$,  and
no solution exists for $\tilde{t}>\tilde{t}_{c}(\tilde{y}_{2,0})$.

This is the ``finite-time'' singular behavior for the velocity amplitude
$|\tilde{y}_{2}|$, which occurs for any $m>1$.
It is driven by the nonlinear positive feedback of the trend term
producing a faster than exponential growth rate, leading to a infinite growth
of $|\tilde{y}_{2}|$ in finite time.
For a fixed $m$, the initial normalized velocity $\tilde{y}_{2,0}$
solely determines the behavior of the trajectory
given by (\ref{eq:sing-norm-range}) and (\ref{eq:sing-norm-tc}),
as a consequence of the fact that the dynamics 
(\ref{eq:sing-norm-dydt}) is solely
determined by $\tilde{\vy}_{2}$.
The evolution of $\tilde{\vy}_{2}(\tilde{t};\tilde{\vy}_{0},\tilde{t}_{0})$
in time with a pair of initial conditions
$\tilde{\vy}_{0}=(0,\pm0.6)$ at $\tilde{t}_{0}=0$
is shown in the right panels of Figure~\ref{fg:norm_sng}.

We now examine separately the behavior for forward $(\tilde{t}>\tilde{t}_{0})$
and backward $(\tilde{t}<\tilde{t}_{0})$ time intervals.
Using the parity symmetry, we focus on
the dynamics described by the normalized orbit with
$\tilde{y}_{2,0}>0$.
Similar results hold for $\tilde{y}_{2,0}<0$ using
$\tilde{\vy}\rightarrow-\tilde{\vy}$.

Over a forward finite time interval
$\tilde{t}\in[\tilde{t}_{0},\tilde{t}_{c}(\tilde{y}_{2,0}))$
up to $\tilde{t}_{c}(\tilde{y}_{2,0})$,
the normalized trajectory with initial condition $\tilde{y}_{2,0}>0$
ranges over:
\ba
\tilde{y}_{1}\in
\left\{ \begin{array}{ll}
    \left[\tilde{g}(\tilde{y}_{2,0}),\infty)\right. & \mbox{for $1<m\leq2$} \\
    \left[0, \tilde{g}(\tilde{y}_{2,0})) \right.& \mbox{for $m>2$}
   \end{array}\right.~,~~
  \tilde{y}_{2}\in [\tilde{y}_{2,0},\infty)~,
  \label{eq:sing-norm-y1f}
\ea
as shown in the left and right panels of Figure~\ref{fg:norm_sng}.
During this finite time interval up to $\tilde{t}_{0}(\tilde{y}_{2,0})$,
$\tilde{y}_{2}$ grows from $\tilde{y}_{2,0}$ to $\infty$.
For $1<m<2$, $\tilde{y}_{1}$ also blows up to infinity,
dragged along by $\tilde{y}_{2}$. On the contrary for $m>2$,
$\tilde{y}_{1}$ culminates at a finite value, given by 
$\tilde{g}(\tilde{y}_{2,0})$.
This finiteness of $\tilde{y}_{1}$
occurs only for $m>2$ for which the
relative growth rate of $\tilde{y}_{2}$ compared to
$\tilde{y}_{1}$ also becomes singular,
as given by the slope of the graph
$\frac{d \tilde{y}_{2}}{d \tilde{y}_{1}}=
\frac{d \tilde{y}_{2}}{d\tilde{t}}/\frac{d
\tilde{y}_{1}}{d\tilde{t}}$ in (\ref{eq:sing-norm-slope}).
The finite final value $\tilde{g}(\tilde{y}_{2,0})$ of $\tilde{y}_{1}$
can be observed in Figure~\ref{fg:norm_sng}g
using the graph of the normalized orbit (\ref{eq:sing-norm-g})
by replacing $(\tilde{y}_{1},\tilde{y}_{2})$ by
$(\tilde{g}(\tilde{y}_{2,0}),\tilde{y}_{2,0})$.

In contrast, over a backward semi-infinite time interval
$\tilde{t}<\tilde{t}_{0}$, the normalized trajectory
ranges over:
\ba
\tilde{y}_{1}\in
\left\{ \begin{array}{rl}
   (0,\tilde{g}(\tilde{y}_{2,0})] & \mbox{for $1<m<2$} \\
   (-\infty,\tilde{g}(\tilde{y}_{2,0})]
	& \mbox{for $m\geq2$} \end{array}
\right.,~~~
\tilde{y}_{2}\in (0,\tilde{y}_{2,0}]~~.
  \label{eq:sing-norm-y1b}
\ea
The normalized velocity $\tilde{y}_{2}$ hence shrinks to $0$, resulting in
a finite increment equal $\tilde{y}_{2,0}$ over the whole time interval.
For $1<m<2$, the increment $\tilde{g}(\tilde{y}_{2,0})$ in $\tilde{y}_{1}$ is
also finite.
On the contrary for $m>2$, the increment in $\tilde{y}_{1}$ is
infinite because the relative growth rate given by
slope $\frac{d \tilde{y}_{1}}{d \tilde{y}_{2}}$
diverges for $\tilde{y}_{2}\ll 1$.

\subsubsection{Global dynamics}
\label{sec:element-trend-observation}

We now derive the global dynamics from the template dynamics of the
normalized model constructed with the trend term.
Two parameters, $m$ and $\alpha$, are used to define the two-dimensional
representation of the trend term given by (\ref{eq:sing-model-dydt}) and
(\ref{eq:sing-model-G}).
For fixed $(m,\alpha)$, the model has the parity
symmetry: $\vy\rightarrow-\vy$.
We refer to the pair of symmetric orbits
corresponding to the graph of $G(\vy;m,\alpha)=0$ as the ``reference orbits''.
 From the normalization (\ref{eq:sing-norm-var}),
we see that the dynamics along the pair of reference orbits is
related to the dynamics along the normalized
orbits (\ref{eq:sing-norm-g})--(\ref{eq:sing-norm-y1b})
through scaling of $(\tilde{y}_{1},\tilde{t})$ by $\alpha^{-1}$.
We also see that all orbits in the $\vy$ phase space collapse onto
the corresponding reference orbit through linear translations in the $\yo$
phase space given by
the distance $G(\vy;m,\alpha)$.
Therefore, the dynamics along any orbit is exactly the same as
the one along the corresponding reference orbit, except for a 
translation in the
$\yo$ phase space.

\begin{figure}
\begin{center}
\includegraphics[height=20cm]{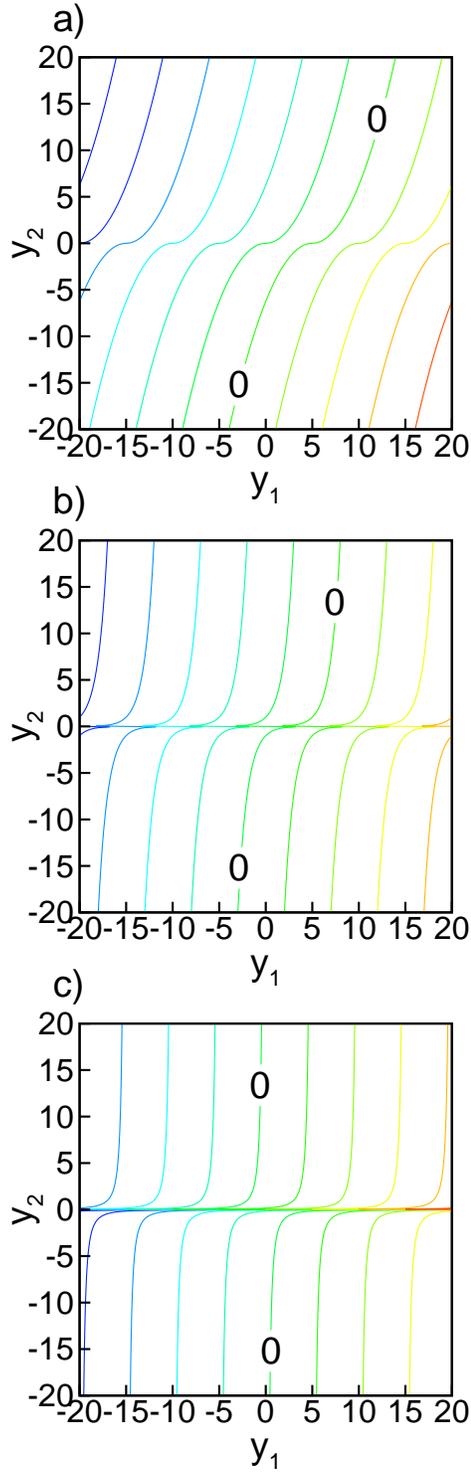}
\end{center}
\caption{
Phase portrait for the singular term with $\alpha=1$ for all cases,
so that the reference orbits with $G=0$ (labeled by 0)
corresponds to the normalized model (Figure~\ref{fg:norm_sng}):
a) $m=1.5$; b) $m=2$; c) $m=2.5$.}
\label{fg:phase_sng}
\end{figure}

Accordingly, the phase portrait for any $m$ consists in a pair of
symmetric families of open orbits.
Each family is parameterized by $G$
(Figure~\ref{fg:phase_sng}) where the reference orbits are labeled
by ``0'' [see also the left panels of Figure~\ref{fg:norm_sng}
for comparison with the normalized orbits].
On $\yt=0$, the dynamics is at rest.
Therefore, the $\vy$ phase space can be divided into
dynamically distinct regions as follows.

\begin{figure}
\begin{center}
\includegraphics[height=20cm,width=16.cm]{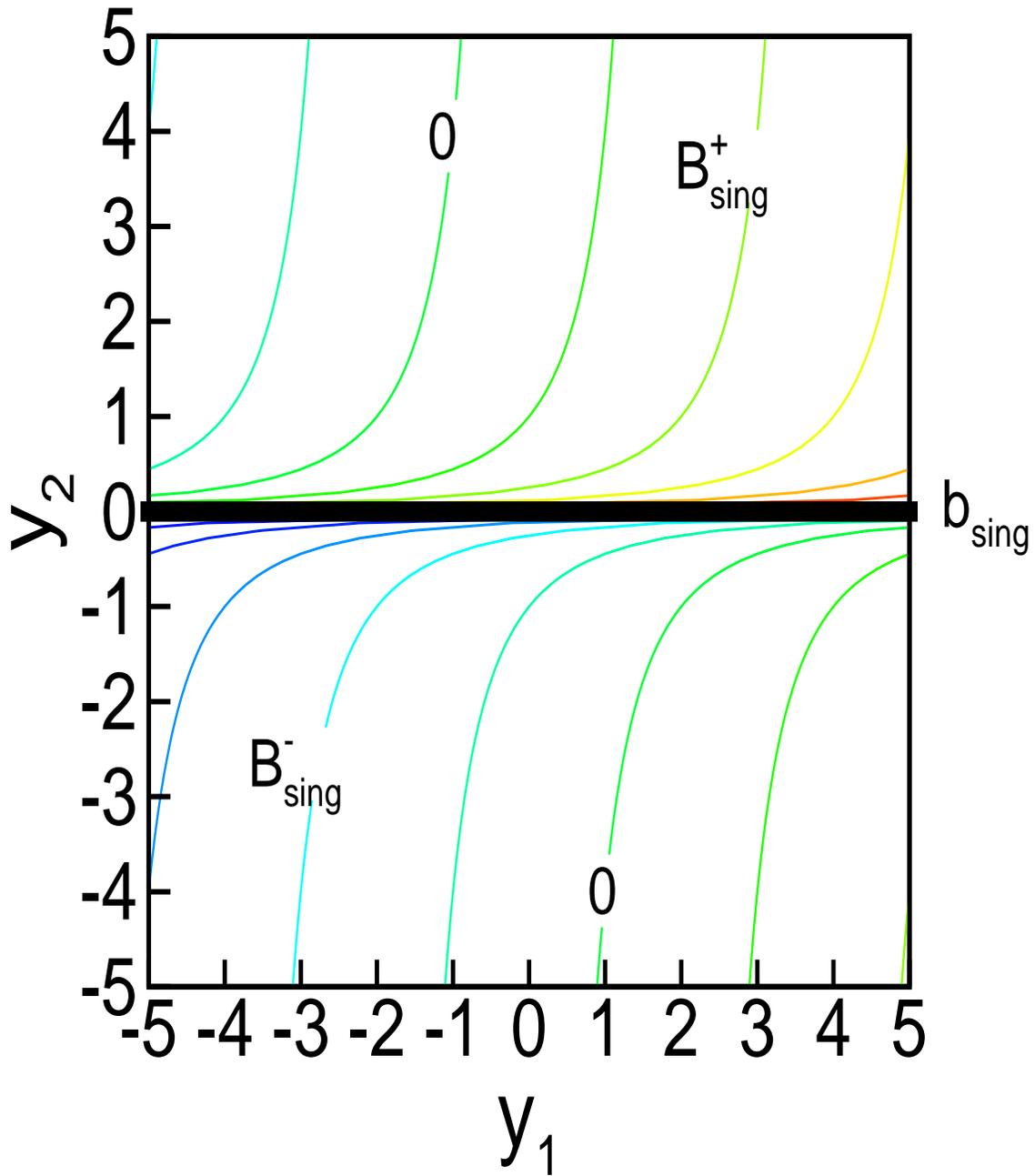}
\end{center}
\caption{
Two singular basins and the boundary between them
for the singular term with  $(m,\alpha)=(2.5,1)$
(see also Figures~\ref{fg:norm_sng}g and \ref{fg:phase_sng}c
for the normalized model)
in the $\vy$ phase space.
The reference orbits are labeled by 0.}
\label{fg:basin_sng}
\end{figure}

\begin{Definition}{Singular basins $B\sbtr\spp$ and
	$B\sbtr\spm$ and boundary $b\sbtr$}
         \label{def:sing-B}
	We define a pair of singular basins:
        \ba
         B\sbtr\spp&\equiv&\{\vy|~\yt>0, ~\mbox{where }
         {d\over dt}\yt>0 \} \nonumber \\
         B\sbtr\spm&\equiv&\{\vy|~\yt<0, ~\mbox{where }
         {d\over dt}\yt<0\}~~.\label{eq:trend-B}
         \ea
	The boundary which separate the two basins
	is defined by:
         \ba \label{def:sing-b}
         b\sbtr &\equiv&\{\vy|\yt=0, ~\mbox{where } {d\over dt}\yt = 0\}~.
         \ea
Any point $\vyz$ in the phase space belongs to either
	one of $B\sbtr\spp$, $B\sbtr\spm$ or $b\sbtr$, as shown in figure
	\ref{fg:basin_sng}.

\end{Definition}

For any $m>1$, the individual trajectory satisfies the following.

\begin{Corollary}{ \ }
    \label{cor:sing-B}
	Consider a trajectory $\vy(t;\vyz,\tz)$ with initial
	condition $\vyz=(y_{1,0},y_{2,0})$ at time $\tz$.
	If it starts in $B\sbtr\sppm$, then it will remain
	 within the  basin without never leaving it. It will reach the
	corresponding finite-time singularity, i.e.,
           \ba
          && \mbox{if $\vyz\in B\sbtr\sppm$}, \qquad
           \mbox{then  $\vy(t;\vyz,\tz)\in B\sbtr\sppm$,} \nonumber \\
          && \qquad \mbox{for $t\in(-\infty,\tc(y_{2,0}))$ ~~~
           with $\yt(-\infty;\vyz,\tz)=\pm0$ ~
	and ~$\yt(\tc(y_{2,0});\vyz,\tz)=\pm \infty$}.
	\label{eq:sing-corB}
           \ea
The critical time
$\tc(y_{2,0})={\alpha}^{-1}\tilde{t}_{c}(\tilde{y}_{2,0})$
is a function of the initial velocity $y_{2,0}$ only.
In the case where  $\vy(t;\vyz,\tz)$ starts on $b\sbtr$,
it will remain on the boundary
basin without never leaving it for a bi-infinite time interval, i.e.,
           \ba
           \mbox{if $\vyz\in b\sbtr$},&  \qquad
           \mbox{then} & \vy(t;\vyz,\tz)=\vyz\in b\sbtr , \qquad
           \mbox{for $t\in(-\infty,-\infty)$} ~.\nonumber
           \ea
\end{Corollary}

\begin{Definition}{Source strips $S\sbtr\spp$ and $S\sbtr\spm$}
         \label{def:sing-S}
         Right next to the boundary $b\sbtr$ in
         each basin  $B\sbtr\sppm$, we define
         a thin vertical strip of constant width:
         \ba
         S\sbtr\sppm&\equiv&
                 \{\vy\in B\sbtr\sppm~|~|\yt|<1,
	~\mbox{where $|\frac{d}{dt}\yt|=|\yt|^{m+1}\ll 1$}\} ~,
         \ea
where  $\vy(t;\vyz,\tz)$ moves away from $b\sbtr$ extremely slowly.  Once it
leaves $S\sbtr\sppm$, it then reaches very quickly the finite-time singularity.
For any $\vyz$, integrating with backward time, any $\vy(t;\vyz,\tz)$
will eventually enter $S\sbtr\sppm$.
Therefore, $S\sbtr\sppm$ can be considered as source regions
of the finite-time singularity.
\end{Definition}

However, the qualitative behavior of each individual orbit in each
singular basin depends on the specific value of the exponent $m$,
as discussed in Section~\ref{sec:element-sing-norm}.

\section{Overall dynamics: Fundamental characteristics}
\label{sec:all}

\subsection{Normalized model}
\label{sec:all-model}
We now consider the overall dynamics obtained by
combining the restoring and trend terms which have been analyzed
separately in Section \ref{sec:element}.
We use the following normalized variables:
\ba \label{eq:all-model-var}
  \lp \begin{array}{l} \bar{y}_{1} \\ \bar{y}_{2}
                    \\ \bar{t} \\ \bar{\gamma} \end{array} \rp =
  \lp \begin{array}{lc}
   \alpha&\yo \\ &\yt \\ \alpha&t \\ \alpha^{-(n+1)}&\gamma
  \end{array} \rp~~
\ea
to minimize the number of parameters by removing the coefficient $\alpha$
of the positive feedback trend term.
For simplicity in the notations, we drop the bar $\bar{\{\cdot\}}$
 from here on.
Then, the overall dynamical systems is written as:
\ba
{d \over d t}
  \lp \begin{array}{l} \yo \\ \yt \end{array} \rp &=&
  \lp \begin{array}{l} \yt \\
     \dot{y}_{2~{\rm osc}}
   + \dot{y}_{2~{\rm sing}} \end{array} \rp \label{eq:all-model-dydt}
\ea
where
\ba
  \dot{y}_{2~{\rm osc}} &=& - \gamma \yo |\yo|^{n-1} \\
  \dot{y}_{2~{\rm sing}}    &=&   \yt | \yt|^{m}~,
   \label{eq:all-model-dytr}
\ea
are the oscillatory ($\dot{y}_{2~{\rm osc}}$) and
singular ($\dot{y}_{2~{\rm sing}}$) source terms for the equation on the
acceleration $\frac{d}{dt}\yt$ (inertia) of $\yo$,
as discussed separately in Section~\ref{sec:element}.

\subsection{Heuristic discussion: Time evolution}
\label{sec:overall-heuristic}

The interplay between the
two previously documented regimes of oscillatory and singular behaviors
results into oscillatory finite-time singularities.
As a result of the nonlinearity of
the restoring term ($n>1$),
the oscillations have local frequencies modulated by the amplitude of $y_1$.
We stress again that the solution $y_1(t; {\bf y}_0, t_0)$ 
is controlled by the initial condition $({\bf y}_0, t_0)$. In this heuristic
discussion, we shall use the simplified notation $y_1$ for 
$y_1(t; {\bf y}_0, t_0)$.

A naive and approximate way to understanding the origin of the frequency modulation is that
the expression $- \left(\gamma~ |y_1|^{n-1}\right)~ y_1$ defines
a local frequency proportional to $\sqrt{\gamma} |y_1|^{n-1 \over 2}$:
the local frequency of the oscillations increases with the amplitude 
of $y_1$. It turns out that this naive guess is correct, as shown 
by the expressions (\ref{dejbglal}) with (\ref{eqexp1}) of section \ref{scalitheo}.
We thus expect the local frequency to
accelerate as the singular time $t_c$ is approached. if the amplitude
$|y_1(t)|$ grows like $(t^*-t)^{-z}$ (see the derivation leading to 
(\ref{ghahfdjgas}), then the
local period, corresponding to the distance between successive
peaks of the oscillations, will be modulated and proportional to
$\left(t_c-t\right)^{-{z(n-1) \over 2}}$.

\subsubsection{Case $m=1$}

We study
\be
{d^2 y_1 \over dt^2} = \alpha {dy_1 \over dt}
- \gamma~ {\rm sign}[y_1]  |y_1|^n~,   \label{kaklakqwwwlq}
\ee
and re-introduce the parameter $\alpha$ to allow us investigating the effect
of its sign.

\begin{figure}
\begin{center}
\includegraphics{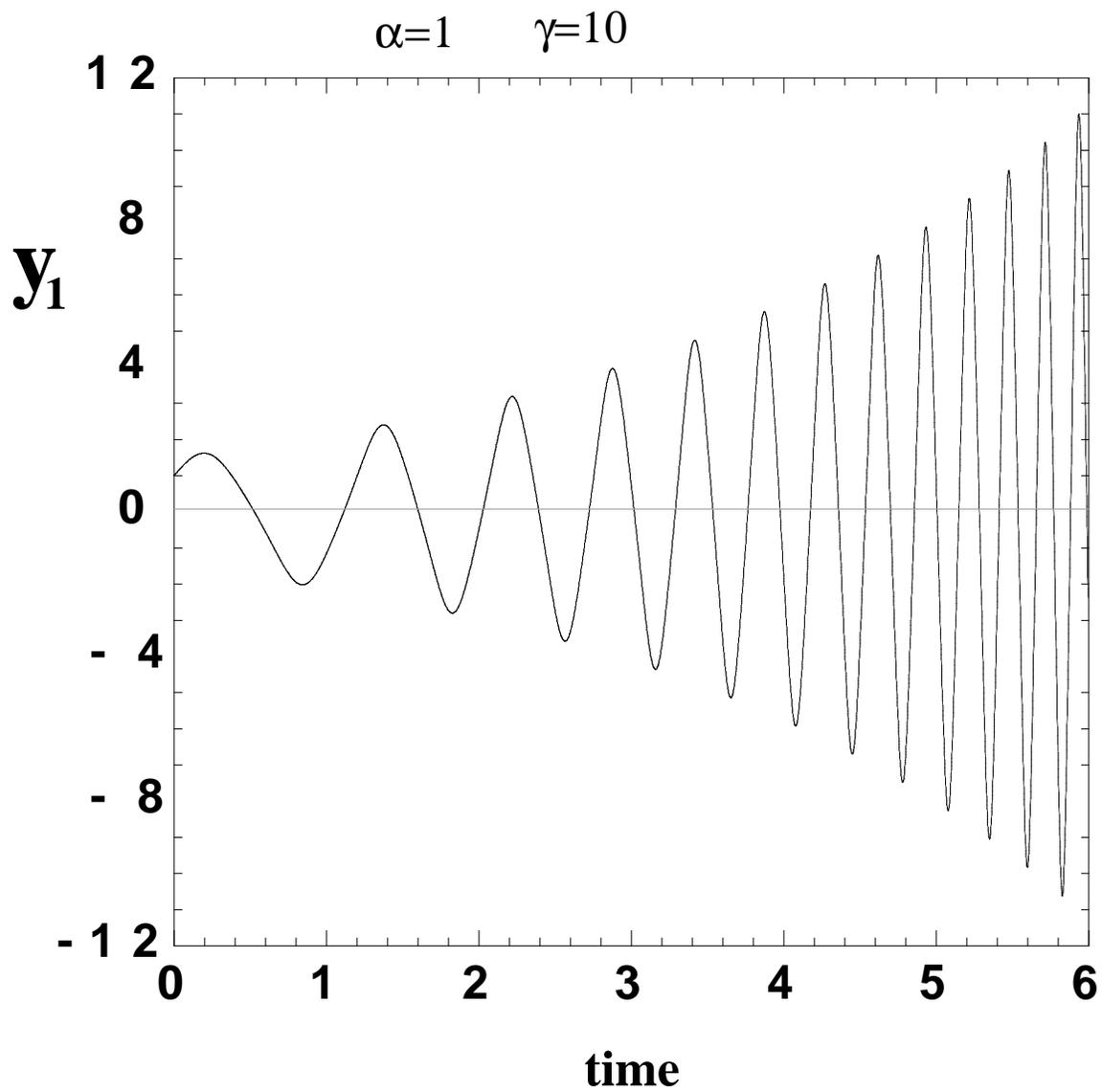}
\end{center}
\caption{Solution of (\ref{kaklakqwwwlq}) for the parameters
$m=1$, $n=3$, $\alpha=1$, $\gamma=10$, $y_{10} \equiv y_1(t=0)=1$ and
$y_{20} \equiv dy_1/dt|_{t=0}=5$.
The amplitude of $y_1(t)$ grows exponentially and
the accelerating oscillations have their frequency increasing also
approximately exponentially with time.}
\label{fig8}
\end{figure}

The interplay between the
l.h.s. and the first term of the r.h.s. of (\ref{kaklakqwwwlq}) leads to
an exponentially growing trend ${dy_1 \over dt}$ and thus an 
exponentially growing
typical amplitude of $y_1(t)$. If $y_1(t) \sim e^{\alpha t}$, both
${d^2 y_1 \over dt^2}$ and ${dy_1 \over dt}$ are of the same order 
while the reversal term
is of order $y_1^n \sim - e^{n \alpha t}$, showing that the 
oscillations will be a dominating
feature of the solution. This is indeed what we observe in figure
\ref{fig8} which shows the solution of (\ref{kaklakqwwwlq}) for the parameters
$\alpha=1$, $m=1, n=3$, $\gamma=-10$, $y_1(t=0)=1$ and $y_{2}(t=0)=5$.
The amplitude of $y_1(t)$ grows exponentially and
the accelerating oscillations have their frequency increasing also 
approximately
exponentially with time, in agreement with our qualitative argument.

\subsubsection{Case $m>1$ \label{xsectionsaa}}

\paragraph{Case $\alpha>0$ and $1 < m < 2$: $|y_1(t \to t_c)|=|y_2(t \to t_c)| \to +\infty$}

\begin{figure}
\begin{center}
\includegraphics{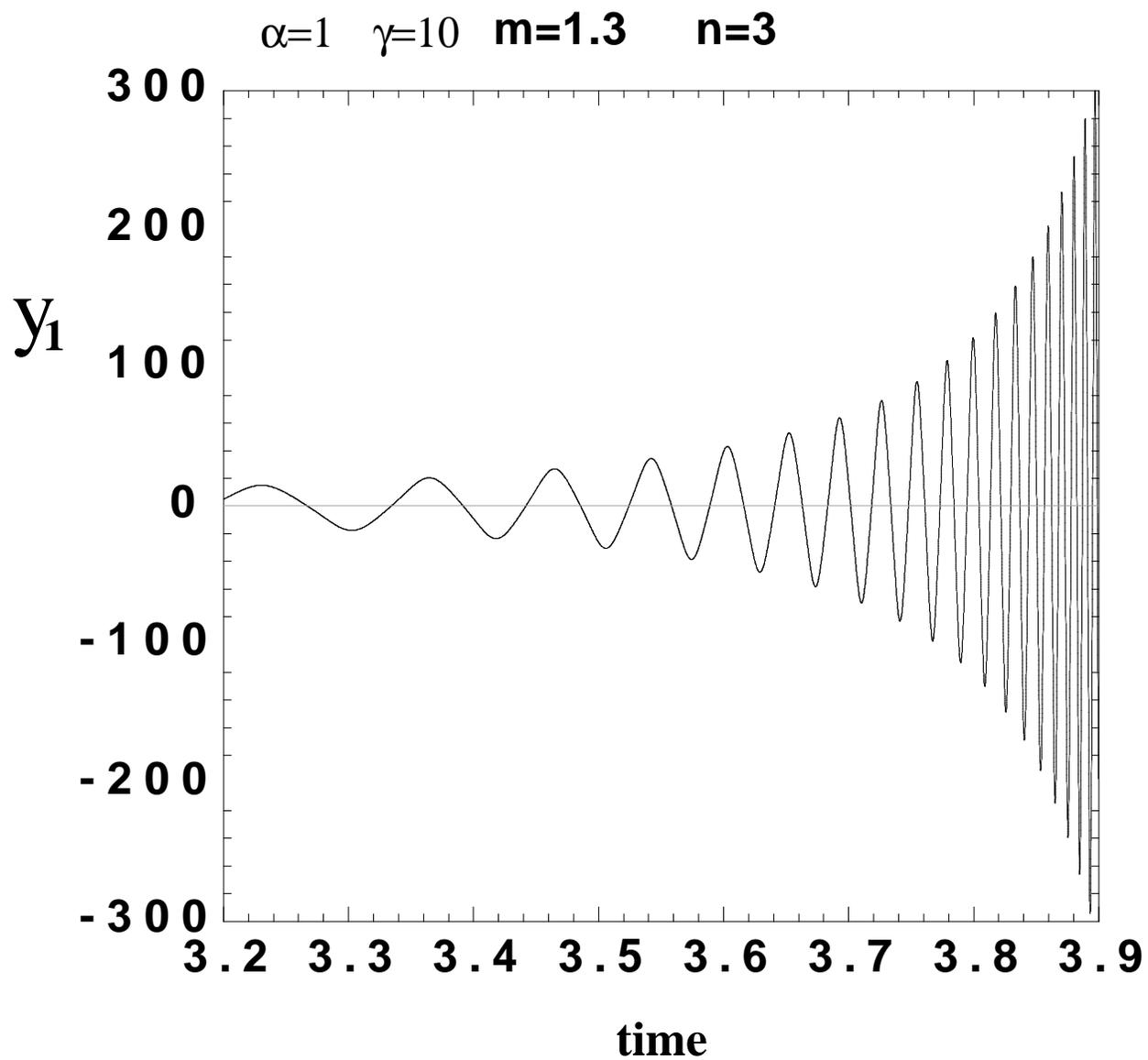}
\end{center}
\caption{Solution of (\ref{kaklakqwlq}) for the parameters
$m=1.3$, $n=3$, $\alpha=1$, $\gamma=10$ and $y_{20} \equiv dy_1/dt|_{t=0}=1$.
The envelop of $y_1(t)$ grows faster than
exponential and approximately as $(t_c -t)^{-1.5}$ where $t_c \approx 4$.
}
\label{fig9}
\end{figure}

\begin{figure}
\begin{center}
\includegraphics{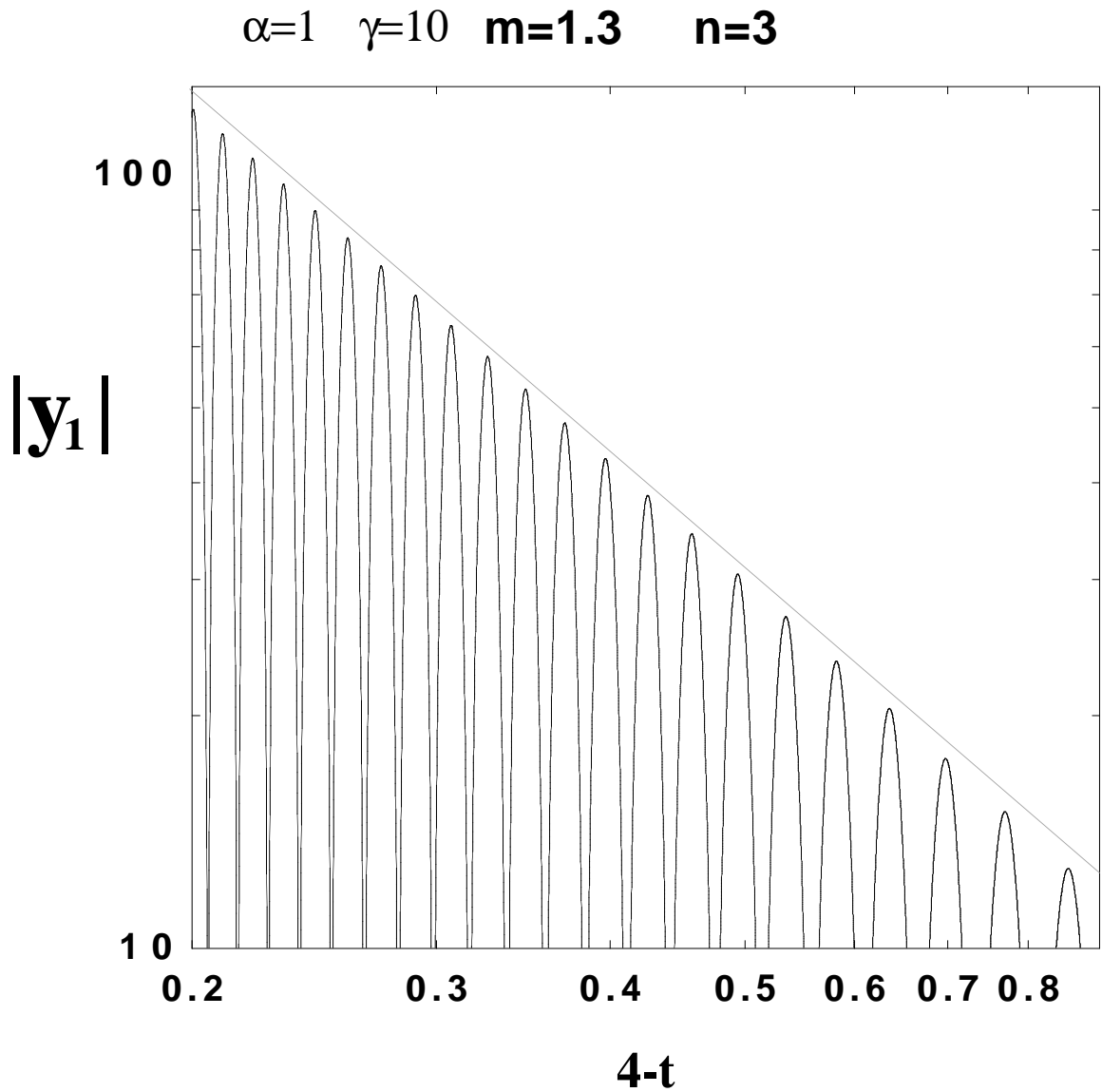}
\end{center}
\caption{Same data as in figure \ref{fig9}: the absolute value
$|y_1(t)|$ is shown as a function of $t_c - t$ where $t_c=4$ in log-log
coordinates, such that a linear envelop qualifies the power law divergence
$(t_c - t)^{-1.5}$. The slope of the line is $-1.5$. Notice also that the
oscillations are approximately equidistant in the variable $\ln (t_c - t)$
resembling a log-periodic behavior of accelerating oscillations on
the approach to the singularity.}
\label{fig10}
\end{figure}

In this regime,  $y_1(t)$
diverges on the approach of $t_c$ as an inverse power of 
$(t_c - t)$.
The accelerating oscillations are shown in figures \ref{fig9} and 
\ref{fig10} for
the parameters
$m=1.3$, $n=3$, $\alpha=1$, $\gamma=10$, $y_1(t=0)=1$ and $y_2(t=0)=1$.
We observe that the envelop of $y_1(t)$ grows faster than
exponential and approximately as $(t_c - t)^{-1.5}$ where $t_c \approx 4$.
In figure \ref{fig10},
$|y_1(t)|$ is represented as a function of $t_c - t$ where $t_c=4$. A double
logarithmic coordinate is used such that a linear envelop qualifies 
the power law divergence. The slope of the line shown on the figure
gives the exponent $-1.5$ which is significantly different from the
prediction $-(2-m)/(m-1)=-2.33$ given by (\ref{eq:sing-norm-range}) on the
basis of the trend term only, i.e., by neglecting the reversal
oscillatory term.
The reversal term has the effect of ``renormalizing'' the 
exponent downward. Notice also that the
oscillations are approximately equidistant in the variable $\ln (t_c - t)$
resembling a log-periodic behavior of accelerating oscillations on 
the approach to the singularity. Here, we shall not dwell more on this regime
which gives divergent $y_1$ and $y_2$ and concentrate rather on the rest of the
paper (except for the next subsection) on the case $m \geq 2$.

\paragraph{Case $\alpha<0$ and $1 < m < 2$: power law decay}

Equation (\ref{kaklakqwlq}) obeys the symmetry of scale invariance for
special choices of the two exponents $m$ and $n$. Consider indeed the
following transformation where $t$ is changed into $\lambda t'$ and
$y_1$ is changed into $\mu y_1'$. Inserting these two changes of variables
in (\ref{kaklakqwlq}) gives
\be
\mu \lambda^2 {d^2 y_1' \over dt'^2} =
\alpha {\mu^m \over \lambda^m} {d y_1' \over dt'} |{d y_1' \over dt'}|^{m-1} -
\gamma  \mu^n y_1' |y_1'|^{n-1}~.
\label{kaaxklakqwfflq}
\ee
We see that (\ref{kaaxklakqwfflq}) is the same equation as 
(\ref{kaklakqwlq}) if
\be
n= 1 + 2 {m-1 \over 2-m}~, \label{jgflal}
\ee
for which we also have
\be
\mu = \lambda^{-{2 \over n-1}}~.  \label{giqoq}
\ee
The condition (\ref{jgflal}) holds for instance with $m=1.5$ and $n=3$.
When the relationship (\ref{jgflal}) is true, the two
equations (\ref{kaaxklakqwfflq}) and (\ref{kaklakqwlq}) are identical
and their solutions are thus also identical for the
same initial conditions: $y_1(t) = y_1'(t')$.
This implies that the solution of  (\ref{kaklakqwlq}) obeys
the following exact renormalization group equation in the limit of
large times when the effect of the initial conditions have been damped out:
\be
y_1(t) = {1 \over \mu} y_1(\lambda t)~,
\ee
where $\lambda$ is an arbitrary positive number and $\mu(\lambda)$ is given
by (\ref{giqoq}). Looking for a solution of the form $y_1 \sim t^{\beta}$,
we get
\be
\beta = - {2 \over n-1}~.
\ee
This exact solution, describing the asymptotic regime $t \to +\infty$,
corresponds to the decaying regime obtained when $\alpha$ is negative
and will not be further explored in the sequel which focus on the singular
case $\alpha>0$ and $m >1$.

\paragraph{Case $m> 2$: $|y_1(t \to t_c)|<+\infty$ and $|y_2(t \to t_c)| \to +\infty$}

In this case with $\alpha>0$, the solution of (\ref{eq:sing-model-dydt})
gives a singularity in finite time with divergence
as ${dy_1 \over dt} \sim (t_c - t)^{-1/(m-1)}$.
Since $1/(m-1) < 1$, $y_1(t)$ remains finite
with a singularity in finite time of the type
\be
y_1(t) \sim y_c - A (t_c - t)^{m-2 \over m-1}    \label{jfakka}
\ee
  with infinite slope
but finite value $Y$ at the critical time $t_c$ since $(m-2)/(m-1) >0$.

The consequence is that there can be only at most a finite number of 
oscillations. Indeed,
since $y_1(t)$ goes to a finite constant, it becomes
negligible compared to its first and second derivatives which both 
diverge close to $t_c$.
Therefore, the two first terms in (\ref{kaklakqwlq}) dominate close 
to the singularity and
the oscillations, which are controlled by the last term, finally disappear and
the solution  becomes a pure power law (\ref{jfakka}) asymptotically 
close to $t_c$.

\begin{figure}
\begin{center}
\includegraphics{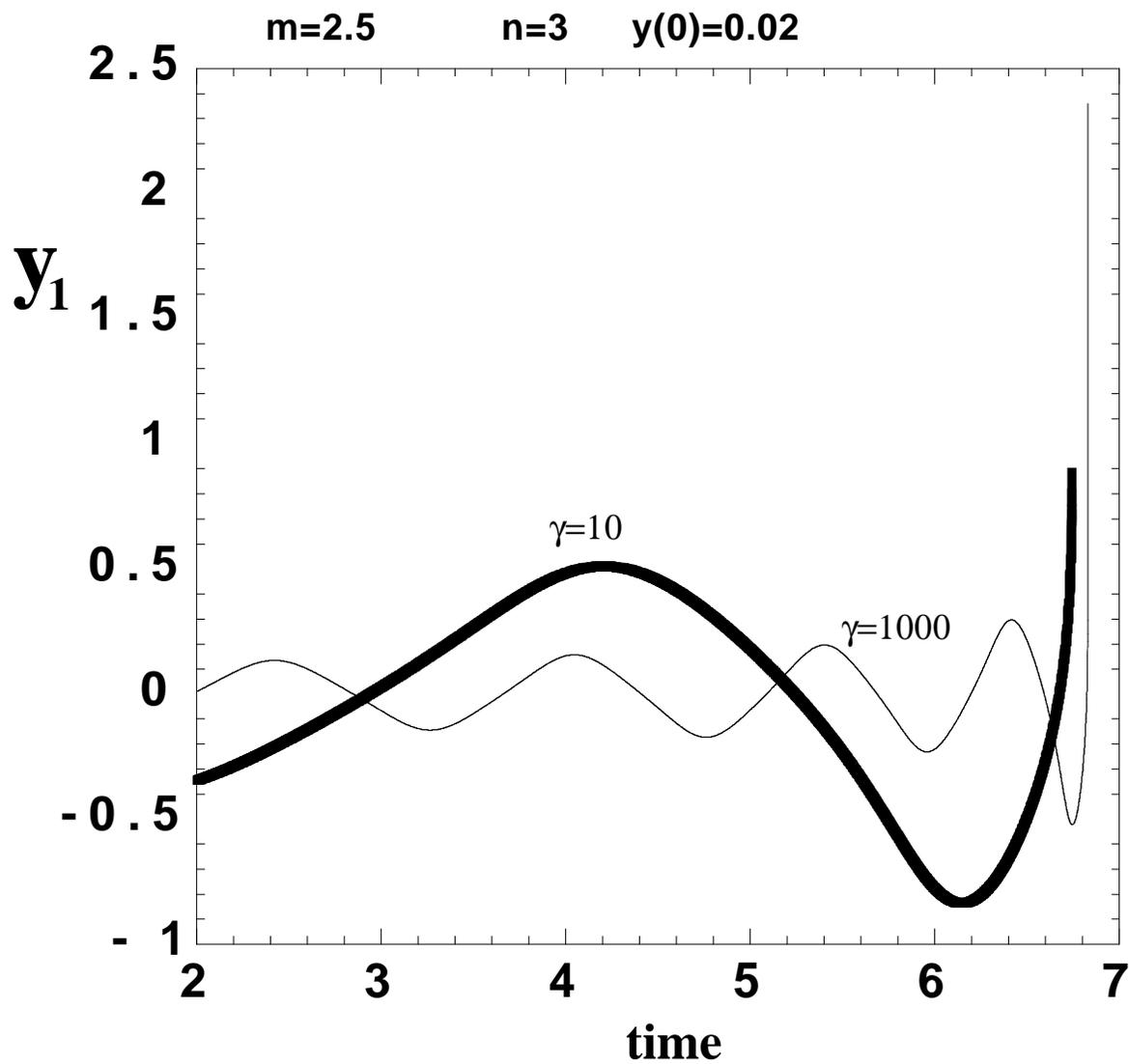}
\end{center}
\caption{Solutions obtained from a numerical integration of
(\ref{kaklakqwlq}) with $m=2.5$ yielding the exponent
${m-2 \over m-1}=1/3$ for the terminal singular behavior of $y_1
\sim y_{1c} - A (t_c -t)^{m-2 \over m-1}$ close to $t_c$, for
$n=3$, $\alpha=1$ and initial value  $y_{10}=0.02$ and derivative
$y_{20}={dy_1 \over dt}|_0=-0.3$ and for two amplitudes $\gamma=10$ and
$\gamma=1000$ of the  reversal term.}
\label{Figm1.5n3}
\end{figure}

Figure \ref{Figm1.5n3} shows the solutions obtained from a numerical 
integration of
(\ref{kaklakqwlq}) with $m=2.5$ yielding the exponent ${m-2 \over m-1}=1/3$,
for $n=3$ and initial value  $y_1(t=0)=0.02$ and derivative ${dy_1 
\over dt}|_{t=0} \equiv y_{2,0}=-0.3$ for
two amplitudes $\gamma=10$ and $\gamma=1000$ of the reversal term.
Notice the existence of a finite number of oscillations and the 
upward divergence of the slope.
As expected, the stronger the reversal term, the larger is the number 
of oscillations
before the pure power law singularity sets in.
The number of oscillations is very strongly controlled
by the initial value of the slope ${dy_1 \over dt}|_0$.
For $m=2.5, n=3$ and initial value  $y_1(0)=0.02$ with $\gamma=10$, 
for instance increasing
the slope in absolute
value to ${dy_1 \over dt}|_0=-0.7$ gives a single
dip followed by a power law acceleration. Intuitively, the number
of oscillations is controlled by the proximity of this initial starting 
point to the unstable fixed
point $(0,0)$, the closest to it, the larger is the number of oscillations.

These properties are formalized into
a systematic dynamical system approach in section 7.

\subsection{Heuristic discussion: Phase space}
\label{sec:all-phase}

\subsubsection{Properties in the phase space}
\label{sec:all-phase-prop}

Having understood the dynamics of the two elements separately
(Section~\ref{sec:element}) and with the qualitative insight provided
by the previous examples, we pose the natural question:
  \begin{itemize}
  \item[Q:] Can oscillatory and/or singular dynamics persist
in the presence of their interaction?
  \end{itemize}
The direct numerical integration of the equations of motion 
in Section \ref{sec:overall-heuristic} suggests
a positive answer.
In the next section \ref{sec:overall-contfrac},
we address this question in a formal way 
and construct a precise phase portrait of the overall
dynamics  for given $(n,m,\gamma)$.
Here we articulate the problem by identifying
fundamental properties of a trajectory
$\vy(t;\vyz,\tz)$ in the phase space.
First, we recall that the full dynamical equation is invariant under parity symmetry 
in phase space:
$(\vy,\frac{d}{dt}\vy)\to(-\vy,-\frac{d}{dt}\vy)$.
The origin $\vy=(0,0)$ is a fixed point:
\be
   {d\over dt}\vy\big|_{(0,0)} = 0~~.
\ee
More precisely, it is a clockwise unstable nonlinear focus 
in the $\vy$ phase space because the flow is divergent everywhere:
\be
  \nabla\cdot {d\over dt}\vy= {\partial\over \partial \yt} \dyrv  =
   m |\yt|^{m-1}\geq0~~. \label{eq:all-phase-div}
\ee
Therefore, starting extremely
near the origin  $|\vyz|\ll 1$, $\vy(t;\vyz,\tz)$ undergoes a 
clockwise oscillation with
increasing amplitude.

Next, we examine the properties of the oscillations.
When only the oscillatory term $\dyrv$ is present (i.e.,
$\dytr=0$), 
Section~\ref{sec:element-osc-global} showed
that the amplitude and the period of the oscillations can be 
determined by $H$ alone.
When the singular term $\dytr$ is added, $H$ is no longer conserved
along any trajectory but increases instead:
\ba
     {d\over dt}H (\vy;n,m,\gamma) &=&
        \lp {d\over dt}\yo\rp~\dytr = |\yt|^{m+1}\geq 0~.
\label{eq:all-phase-dHdt}
\ea
The growth rate of $H$ also increases as $H$ increases,
because a higher $H$ corresponds to a wider range of $\yt$ during
an oscillation cycle, as seen from (\ref{eq:osc-global-range}).
The higher $H$ becomes, the more effective is the impact of $\dytr$,
especially once $|\yt(t;\vyz,\tz)|$ reaches
${\cal O}(1)$.
The region where $|\yt(t;\vyz,\tz)|<1$ corresponds to the source
strips $S\sbtr\sppm$ (Definition \ref{def:sing-S})
for the case with only the singular term $\dytr$.

  \begin{Remarks}{For $n>1$, $\vy(t;\vyz,\tz)$ has the following
characteristics:}
\label{rmk:all-y}
\item  From (\ref{eq:osc-global-dTdH}), the frequency of oscillation
increases from $0$ to $\infty$ as $H$ increases. 
As a consequence, for $|\yt|\ll 1$ where ${d\over dt}H$ is very small,
$\vy(t;\vyz,\tz)$ undergoes extremely slowly divergent oscillation
with increasing frequency.
\item  From (\ref{eq:osc-global-range}), the amplitude  
of the oscillations  monotonically
grows as $H$ grows to $\infty$ at an increasing rate.
As a consequence, $|\vy(t;\vyz,\tz)|$ eventually diverges to $\infty$.
\item From (\ref{eq:all-phase-dHdt}),
$\vy(t;\vyz,\tz)$ starting from any $\vyz$ approaches the 
origin backward in time.
\item Therefore, $|\vy(t;\vyz,\tz)|$ starting from
any $\vyz$ connects the origin $|\vy(t;\vyz,\tz)|\rightarrow 0$ in
backward time and infinity $|\vy(t;\vyz,\tz)|\rightarrow \infty$
in forward time.  \end{Remarks}

We now formulate mathematically the notion of an oscillation for
the overall dynamics.
When $\vy(t;\vyz,\tz)$ undergoes a sequence of oscillation cycle
 in the $\vy$ phase space
(see for example Figure~\ref{fg:norm_osc}),
$\yo$ and $\yt$ change their direction of motion in succession.

\begin{Definition}{Turn of a trajectory}
   	\label{def:all-phase-turn}
   We say that $\vy(t;\vyz,\tz)$ makes a turn at
   $t=t'$ if the variable $\yo$ changes its direction of motion
   at $t=t'$, i.e.,
   \be\label{eq:all-phase-turn}
   {d\over dt}\yo\big|_{\vy(t';\vyz,\tz)}=\yt=0.
   \ee
\end{Definition}

Each complete oscillation cycle requires two turns of
$\vy(t;\vyz,\tz)$.
During a time interval between two adjacent turns,
$\yt$ changes directions (i.e., it achieves $\frac{d}{dt}\yt=0$)
if the oscillation is around the origin $\vy=(0,0)$;
see for example Figure~\ref{fg:norm_osc}.

\begin{Definition}{Zero velocity curves $F^{(1)}$ and $F^{(2)}$}
   \label{def:all-f}
  We define the two zero-velocity curves
     in the phase space
with respect to $\yo$ and $\yt$:
     \ba
      F^{(1)} &\equiv & \{\vy\big|~~ {d\over dt} \yo=0~,
	~~\mbox{i.e., $\yt=0$} \}~,   \label{eq:all-phase-F1} \\
      F^{(2)} &\equiv&\{\vy \big|~~ {d\over dt} \yt=0~,
	~~\mbox{i.e., $\gamma\yo|\yo|^{n-1}=\yt|\yt|^{m-1}$} \}~.
     \label{eq:all-phase-F2}
    \ea
$F^{(1)}$ is nothing but the $y_1$-axis. On the
curve $F^{(2)}$, $\yo$ can be expressed as a monotonic
function of   $\yt$:
   \be
    \yt\equiv f^{(2)}(\yo)=\gamma^{{1\over m}}~\yo |\yo|^{{n\over m}-1}~,
	\label{eq:all-gt}
   \ee
where $(\yo,f^{(2)}(\yo))$ is on $F^{(2)}$.
An alternative way for obtaining (\ref{eq:all-phase-F1}) and
(\ref{eq:all-phase-F2})
is to use the slope of the trajectory $\vy(t;\vyz,\tz)$ in phase space:
      \ba\label{eq:all-phase-cf2}
      {d \yt\over d\yo} \big|_{\vy(t;\vyz,\tz)} &=&
      {d \yt\over dt} /{d \yo\over dt}\big|_{\vy(t;\vyz,\tz)} =
 \frac{-\gamma\yo|\yo|^{n-1} + \yt |\yt|}{\yt}~,
    \ea
where ${d \yt\over d\yo}=\pm\infty$ results in $F^{(1)}$ and
${d \yt\over d\yo}=0$ gives $F^{(2)}$.
  \end{Definition}

\begin{Corollary}{Complete oscillation cycle} 
\label{cor:all-phase-osc}
Staring from a point on $F^{(1)}$ where $\vy(t;\vyz,\tz)$ makes a turn
(i.e., $\vyz\in F^{(1)}$),
one complete oscillation cycle requires a set of
four conditions to be satisfied in sequence:
[$\frac{d}{dt}\yo=0$
$\mapsto$ $\frac{d}{dt}\yt=0$ $\mapsto$ $\frac{d}{dt}\yo=0$
$\mapsto$ $\frac{d}{dt}\yt=0$].
Accordingly, in phase space, $\vy(t;\vyz,\tz)$ cuts across
[$F^{(1)}$ $\mapsto$ $F^{(2)}$ $\mapsto$ $F^{(1)}$ $\mapsto$ $F^{(2)}$]
in succession.
  \end{Corollary}

\begin{Corollary}{Transition to non-oscillatory motion}
\label{cor:all-phase-trans}
If $\vy(t;\vyz,\tz)$ ceases to reach $F^{(1)}$ or $F^{(2)}$,
then it can no longer oscillate around the origin.
  \end{Corollary}



\subsubsection{Schematic dynamics in phase space}
\label{sec:all-phase-sch}

Having Corollaries \ref{cor:all-phase-osc} and
\ref{cor:all-phase-trans} in hands,
we rephrase the question in Section~\ref{sec:all-phase-prop} into more
specific ones: 
  \begin{itemize}
  \item[Q1:] Under what conditions
  and how far does the clockwise oscillatory motion owing to
    $\dyrv$  persist away from the origin?
  \item[Q2:] Under what conditions does the finite-time singular
  behavior persist, and the two singular basins resulting from
   $\dytr$ exist?
  \end{itemize}

\begin{figure}
\begin{center}
\includegraphics{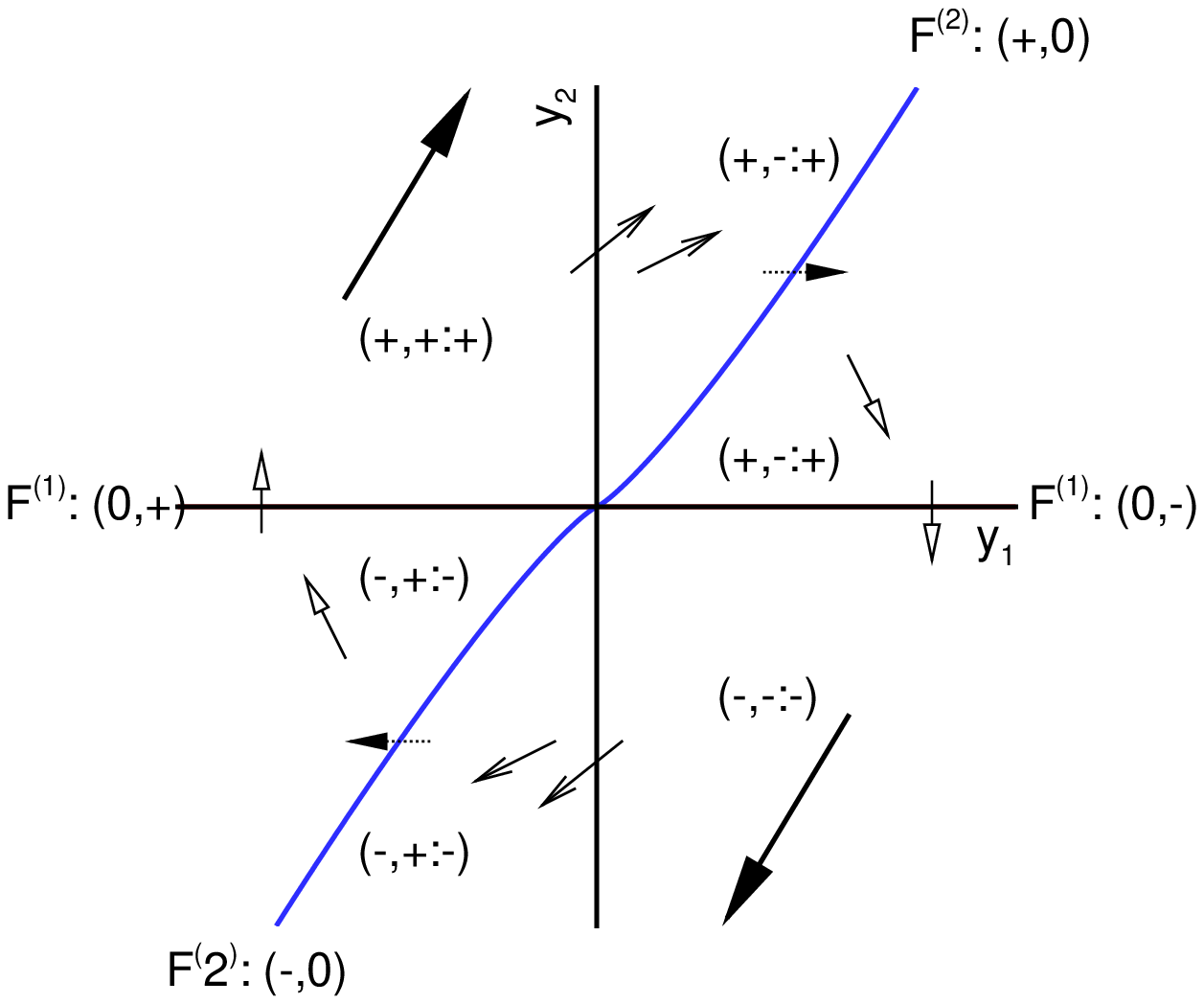}
\end{center}
\caption{Schematic velocity field ${d\over dt}\vy$
indicated by arrows.
The phase space is divided into six regions
by $F^{(1)}$, $F^{(2)}$ and $y_{1}=0$.
The effect of the reversions and trend terms in each region are shown 
by the sign of
$\dyrv$ and $\dytr$ in $({d\over dt}\yo,\dyrv:\dytr)$.
$\dyrv$ and $\dytr$ may enhance each other (long thick arrows);
$\dytr$ dominates $\dyrv$ (plain arrowhead);
$\dyrv$ dominates $\dytr$ (hollow arrowhead).
On $F^{(2)}$, they balance (dotted line).}
\label{fg:all_F}
\end{figure}

For an intuitive grasp of issues associated with these
questions, we schematically summarize in Figure~\ref{fg:all_F}
the dynamical properties
along a trajectory $\vy(t;\vyz,\tz)$ due respectively to
$\dyrv$ and $\dytr$.
The total velocity vector $(dy_1/dt, dy_2/dt)$ is indicated by the arrows.
The sign of the two contributing terms $\dyrv$ and $\dytr$ are given
in the triplet $({d\over dt}\yo,\dyrv:\dytr)$.
The two terms $\dyrv$ and $\dytr$ can enhance or oppose each other
depending on their relative signs.

Furthermore, we make the following observations.


\begin{Remark} \label{rmk:all-phase-dyn}
\item The phase space is divided into six domains by $F^{(1)}$, $F^{(2)}$ and
$y_{1}=0$.
On $F^{(1)}$ and  $F^{(2)}$, the components of the velocity vector,
${d\over dt}\yo$ and ${d\over dt}\yt$, respectively change their sign.
On $y_{1}=0$ and $F^{(1)}$, ${d\over dt}\yo$ and $\dytr$ change sign.
\item
In the second or fourth quadrant ($\yo\yt<0$)
defined by  $F^{(1)}$ and $\yo=0$,
both $\dytr$ and  $\dyrv$ have the same sign and hence
so does $\frac{d}{dt}\yt$.
In Figure \ref{fg:all_F}, it is indicated by long thick arrow
with the velocity triplet $({d \over dt}\yo,\dyrv:\dytr)=(-,-:-)$
in the second quadrant or $(+,+:+)$ in the fourth quadrant.
It can be thought that $\dyrv$ enhances $\dytr$
in a way that $\vy(t;\vyz,\tz)$  flows towards one of the two
($+$ or $-$) singular directions in $\yt$
 from the following reason.
\begin{itemize}
   \item Because $\dytr$ and $\dyrv$ have the same sign,
   $|{d\over dt}\yt|=|\dytr|+|\dyrv|$ is larger than  $|\dytr|$.
   Therefore, if $\vy(t;\vyz,\tz)$ remains in the second or fourth
   quadrant without ever leaving, it must be driven to a finite-time
   singularity with the same sign as in the case where only $\dytr$
   operates, but at a faster rate.
   Therefore, two singular basins inevitably exist:
   \begin{enumerate}
    \item  In the fourth quadrant where $\yt>0$ and $\yo<0$:
	$\yt\to +\infty$ with ${d\over dt}\yt>0$ and ${d\over dt}\yo>0$,
    \item  In the second quadrant where $\yt<0$ and $\yo>0$:
	$\yt\to -\infty$ with ${d\over dt}\yt<0$ and ${d\over dt}\yo<0$.
   \end{enumerate}
   These two basins are the generalization of the two basins
   $B\sbtr\sppm$ defined  previously for $\dytr$ only
	(Definition \ref{def:sing-B}).
  This partially answers Q2 above.
   \item The only way for $\vy(t;\vyz,\tz)$ to escape from the
     singular behavior is to leave the second or fourth
    quadrant respectively for the third or first quadrant
     by cutting $\yo=0$ while keeping the sign of $\yt$.
   \end{itemize}
\item  In the first or third quadrant ($\yo\yt>0$)
defined by $\yo=0$ and $F^{(1)}$ with $F^{(2)}$ inside,
$\dyrv$ and $\dytr$ have  opposite signs and the total velocity
vector cannot be determined unambiguously
as indicated by the velocity triplet 
$({d \over dt}\yo,\dyrv:\dytr)=(+,-:+)$
in the first quadrant or $(-,+:-)$ in the third quadrant.
When $\vy(t;\vyz,\tz)$ enters into the first of third quadrant
according to clockwise motion around the origin,
it respectively comes from the fourth or second quadrant
where $\dytr$ enhances $\dytr$ 
as indicated by plain arrowheads in Figure \ref{fg:all_F}.
Therefore, $\dytr$ is dominant first. Then, the effect of
$\dyrv$ gradually kicks in as  $|\yo|$ increases towards
$F^{(2)}$ where $\dyrv$ balances $\dytr$ as indicated by dotted
line in Figure \ref{fg:all_F}.
\begin{itemize}
  \item If $\dytr$ remains dominant and $\vy(t;\vyz,\tz)$ never reaches
    $F^{(2)}$,
        then ${d\over dt}\yt$ does not change sign and
    $\yt$ keeps growing.
    Eventually, $\dytr$ may
     completely dominate $\dyrv$, and
     $\vy(t;\vyz,\tz)$ moves quickly towards the terminal singularity,
  \begin{enumerate}
    \item In the first quadrant where $\yo>0$ and $\yt>0$ above $F^{(2)}$:
    $\yt\to +\infty$  with ${d\over dt}\yt>0$ and ${d\over dt}\yo>0$,
    \item In the third quadrant where $\yo<0$ and $\yt<0$ below $F^{(2)}$:
    $\yt\to -\infty$  with ${d\over dt}\yt<0$ and ${d\over dt}\yo<0$.
  \end{enumerate}
  The sign of $\yt$ towards the singularity is consistent for all quadrants.
  \item If $\vy(t;\vy,\tz)$ reaches $F^{(2)}$, $\dyrv$ overcomes $\dytr$ and
  $\frac{d}{dt}\yt$ changes sign as indicated by hollow arrowhead in
  Figure \ref{fg:all_F}.
  Eventually $\vy(t;\vy,\tz)$ has to exit the quadrant by
  passing  $F^{(1)}$.
  It follows that, if $\vy(t;\vy,\tz)$ reaches $F^{(2)}$,
  then it also reaches $F^{(1)}$ and makes a turn of an oscillation
(Definition~\ref{def:all-phase-turn}).
\end{itemize}
\end{Remark}
In summary, the following two conditions sequentially determine
whether or not $\vy(t;\vy_0,\tz)$ can make another turn
 starting from a turning point $\vyz\in F^{(1)}$:
\begin{enumerate}
\item In the fourth or second quadrant: whether or not it reaches
       $\yo=0$.
\item In the first or third quadrant if it reaches $\yo=0$: 
	whether or not it reaches $F^{(2)}$.
\end{enumerate}

For fixed $(n,m,\gamma)$,
the global dynamics can be completely described by
the phase portrait because this is a system of
two-dimensional autonomous ODEs.
However, this geometrical structure of the phase portrait may
bifurcate as the value of the exponents $n$ and $m$ vary
(Section \ref{sec:element-osc} and
\ref{sec:element-sing}).
In the following section, we will examine the structure of the
global dynamics when both elements have high nonlinearity,
i.e., $n>1$ and $m>2$.


\section{Overall dynamics for $n>1$ and $m>2$ with $\alpha>0$: $|y_1(t \to t_c)|<+\infty$ 
(except for isolated initial conditions)
and $|y_2(t \to t_c)|=+\infty$}
\label{sec:overall-contfrac}

Recall from Section \ref{sec:element-osc} 
for the sub-dynamical system with only the oscillatory element
that the case $n>1$ corresponds to highly nonlinear oscillations with a
monotonically decreasing period as the amplitude of the
oscillations increases (Figure \ref{fg:phase_osc}). 
From Section \ref{sec:element-sing} 
for the sub-dynamical system with only the singular element,
the case $m>2$ corresponds to finite-time
singularity with finite increment in $\yo$ and
infinite increment in $\yt$ (Figure \ref{fg:phase_sng}).
Furthermore, Section \ref{sec:all-phase} on the phase space 
of the full dynamical system showed the following results:
\begin{enumerate}
\item any trajectory $\vy(t;\vyz,\tz)$ 
starting away from the origin connects the origin in backward time and 
$|\yt|\rightarrow\infty$ in forward time;
\item the oscillations may persist especially near the origin;
\item the finite-time singular behavior should persist;
\item the $F^{(2)}$-curve 
is critical in determining whether or not a trajectory transits from
oscillatory to singular behavior.
\end{enumerate}

As we demonstrate below, the most striking dynamical feature for the case
$n>1$ and $m>2$ is the finite-time oscillatory singularity.
In Section \ref{sec:case-phase}, we heuristically describe the global 
dynamics by identifying the boundaries and basins in the phase space
using two examples.
The mathematical definitions of the boundaries and basins
are given in Section \ref{sec:case-Bb}.
Using the template maps,
we describe the global dynamics of 
the boundaries in Section \ref{sec:case-dynb} 
and the basins in Section \ref{sec:case-dynB}.
Finally, we study the scale-invariant properties associated with the finite-time
oscillatory singularity in  Section \ref{sec:case-fractal}.

\subsection{Phase space description} 
\label{sec:case-phase}

\begin{figure}
\begin{center}
\includegraphics[width=16cm]{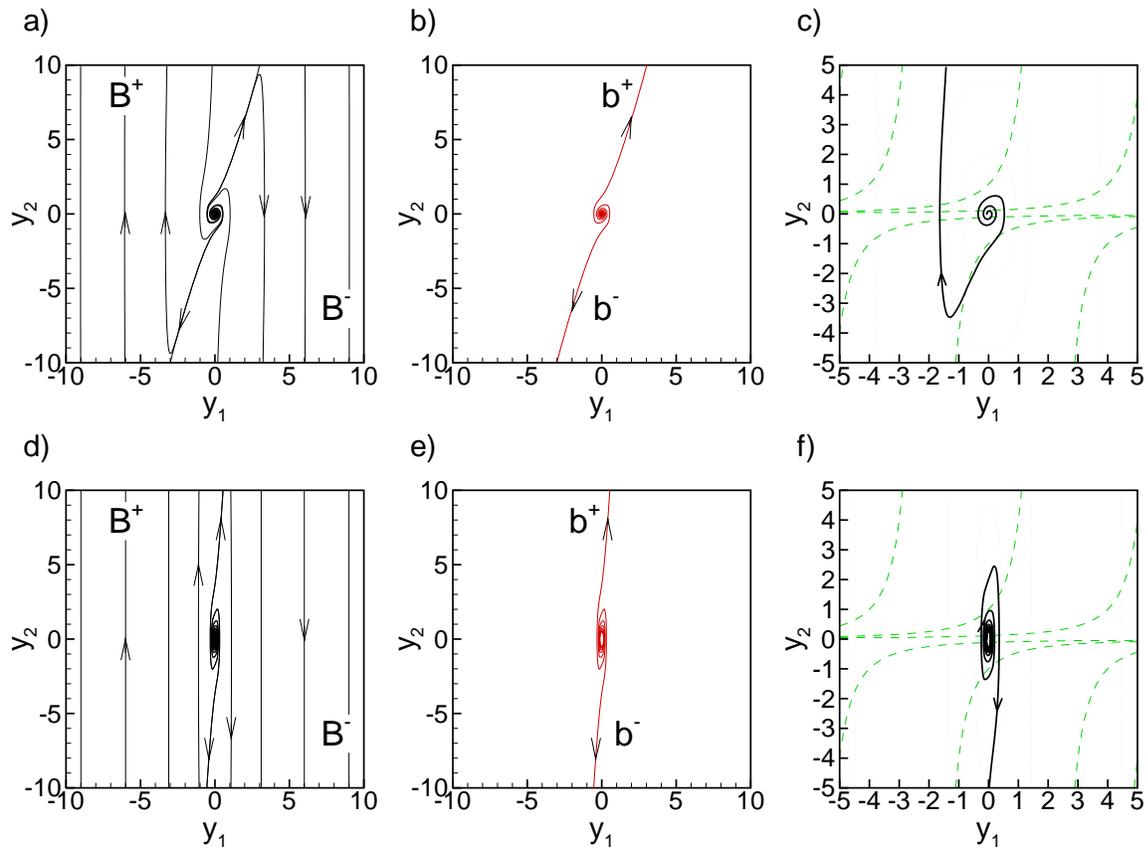}
\end{center}
\caption{Global dynamics in the phase space with 
$(n,m)=(3,2.5)$ for $\gamma=10$ (a-c)  and $\gamma=1000$ (d--f):
a,d) phase portrait as a collection of trajectories; 
b,e) singular boundary $b\sppm$, and
c,f) a trajectory starting $\vyz=(-0.06,0)$ with  contours of $H$
(dotted lines) and $G$ (dashed lines) for reference 
(see also Figures \ref{fg:phase_osc} and \ref{fg:phase_sng},
respectively).
Arrows along trajectories indicate the forward direction of time.}
\label{fg:phase_all}
\end{figure}

Two examples of phase portraits are shown in 
Figure~\ref{fg:phase_all} as a collection of trajectories 
with $(n,m)=(3,2.5)$  for $\gamma=10$ (Figure~\ref{fg:phase_all}a--c) 
and $\gamma=1000$ (Figure~\ref{fg:phase_all}d--f),
where arrows  along the individual trajectories indicate the direction of forward time.
We observe in both examples  that there are two basins 
(labeled by  $B\spp$ and $B\spm$) in the phase space.
The superscript of individual basins correspond to the sign of the
terminal direction $\frac{d}{dt}\yt$ as $|\yt|\rightarrow\infty$.
The boundary between the basins is kinematically defined by  the
special trajectories spirally out of the origin to reach 
$|\yo|\rightarrow\infty$ as well as $|\yt|\rightarrow\infty$.
Any other trajectories result in a finite terminal value of
$\yo$ as $|\yt|\rightarrow\infty$.
These boundary trajectories are singled out in 
Figures~\ref{fg:phase_all}b and e (labeled by $b\spp$ and $b\spm$).
A typical trajectory $\vy(t;\vyz,\tz)$ starting from $\vyz=(-0.06,0)$
are also shown in 
Figures~\ref{fg:phase_all}c and f.

These phase portraits confirm that oscillations indeed persist 
and are confined near the origin about $|\vy|<1$ and hence 
$H<1$ from (\ref{eq:osc-model-H}).
The amplitude of the oscillations continuously grows along $\vy(t;\vyz,\tz)$
in forward time as seen from (\ref{eq:all-phase-dHdt}).
For $|\yt|\ll 1$ starting near the origin $|\vyz|\ll 1$, 
$\vy(t;\vyz,\tz)$  is nearly vertical 
because of $|\frac{d}{dt}\yt|\gg |\frac{d}{dt}\yo|$ and
follows a constant $H$-curve closely (Section
\ref{sec:element-osc-global}, see also Figure \ref{fg:phase_osc}).
For $|\yt|$ no more much smaller than $1$, $H$ increases
more efficiently by growing further in $\yt$ 
as observed in Figures~\ref{fg:phase_all}c and f.
The period of the oscillations decreases continuously, because 
$\frac{d}{dt}T(H;n,\gamma) 
	=\frac{\partial T}{\partial H} \frac{dH}{dt}\leq 0$
along $\vy(t;\vyz,\tz)$ for $n>1$.

Once the oscillation reaches  $|\vy|\approx 1$ and hence 
$H\approx 1$, the singular element $\dytr$  works on
$\frac{d}{dt}\yt$ more effectively.
As a result, $\vy(t;\vyz,\tz)$ starts to grow rapidly, especially when 
it is moving vertically in the phase space.

Recall also that closed $H$ contours for $H>1$ are stretched out
vertically for $n>1$ (Section \ref{sec:element-osc-global}).
Therefore,  the oscillatory element $\dyrv$ can enhance or suppress the
singular behavior of $\vy(t;\vyz,\tz)$ significantly when 
it moves vertically.
Such stretching effect of $H$ is more prominent for larger $\gamma$
(compare Figures~\ref{fg:phase_all}c to f).

For $|\yt|>1$, the dynamics is extremely singular in the second and fourth
quadrants where $\dyrv$ enhances $\dytr$ (Remark
\ref{rmk:all-phase-dyn}).
The terminal increment of $\yo$ is nearly zero as $\yt$ approaches singularity in
finite time as indicated by the almost vertical trajectories.

In the first and third quadrants where $\frac{d}{dt}\yt$ 
changes sign with respect to $\dyrv$ on $F^{(2)}$,
the  boundaries $b\spp$ and $b\spm$ divide the phase space into
$B\spp$ and $B\spm$.
Any trajectory starting near $b\spp$ and $b\spm$ with $|\yt|>1$
accelerates extremely fast 
into $B\spp$ or $B\spm$, indicating that the dynamics is
extremely sensitive near $b\spp$ and $b\spm$ for $|y_2|>1$.
Any trajectory that moves away from $b\spp$ or $b\spm$ for $|y_2|>1$
does so almost vertically due to the stretched structure of $H$
for $H>1$.
Near vertical trajectories away from $b\spp$ or $b\spm$ 
indicate that increment of  $\yo$ is finite in the first and
third quadrants like in the second and fourth quadrants.

\subsection{Singular basins and boundary} \label{sec:case-Bb}

When the dynamics has only the trend element
(Section \ref{sec:element-sing}), there exist two singular
basins $B\sbtr\spp$ and $B\sbtr\spm$ separated by a boundary
$b\sbtr$ determined by a collection of stagnation points
where the velocity is identically zero (Definition \ref{def:sing-B}).
In the full dynamical system, we define the two boundaries 
$b\spp$ and $b\spm$ kinematically 
as observed in Figure \ref{fg:phase_all}.
The definition of the two basins $B\spp$ and $B\spm$ follow in a natural way.
For simplicity and economy in notation,
    	we use $\pm$ and $\mp$ to represent two distinct
	cases by choosing them consistently in order.
	For example, by ``$\pm A$ with $\mp B$ for  $\pm C$,''
	we mean: i) ``$+A$ with $-B$ for  $+C$,'' and
	ii) ``$-A$ with $+B$ for  $-C$.''
Use of  $\pm$ and $\mp$ is possible because 
the full dynamical equation is invariant under parity symmetry.

\begin{figure}
\begin{center}
\includegraphics{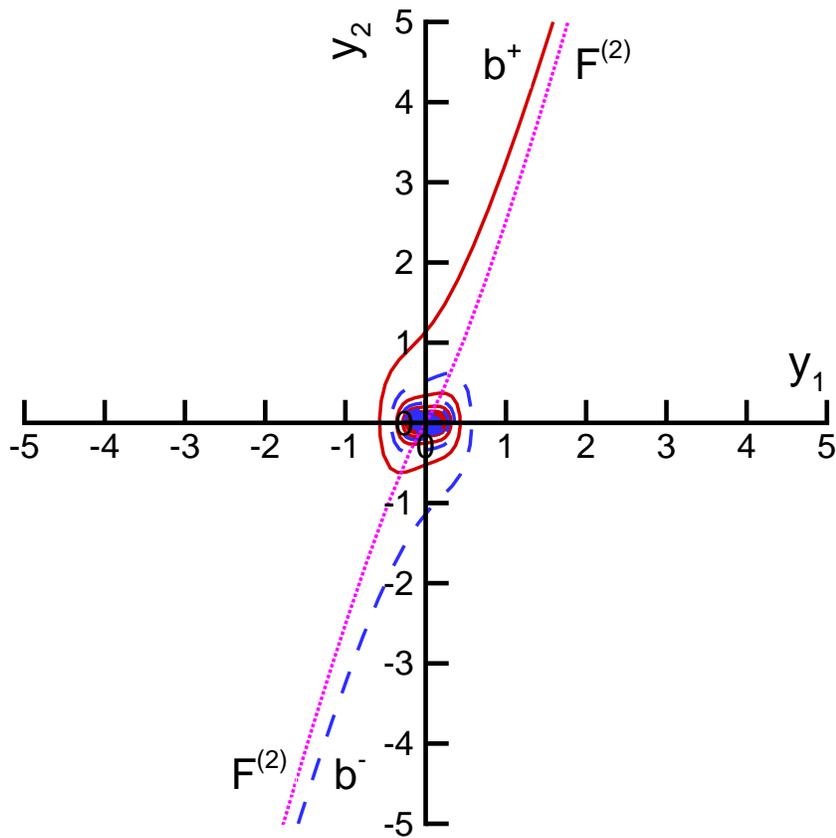}
\end{center}
\caption{Geometrical relation between $b\sppm$ and
$F^{(2)}$ for $(n,m)=(3, 2.5)$ and $\gamma=10$
as in Fig.~\ref{fg:phase_all}.}
\label{fg:basin_crv_n3m25g10}
\end{figure}

\begin{Definition}{Boundaries $b\spp$ and $b\spm$.
  {\rm [see Definition~\ref{def:sing-B}]}}
         \label{def:all-Bb}
   We define the boundary $b\sppm$ by the special
  trajectories that connect the origin $\vy=(0,0)$ 
  to  $(+\infty,+\infty)$ and to $(-\infty,-\infty)$
 (see Figures \ref{fg:phase_all} and \ref{fg:basin_crv_n3m25g10}).
  Any trajectory $\vy(t;\vyz,\tz)$ starting on $\vyz\in b\sppm$ will
 remain   on it in forward and backward time:
   \ba
   \mbox{if $\vyz\in b\sppm$}, \qquad
   &\mbox{then}& \vy(t;\vyz,\tz)\in b\sppm \qquad \qquad
   \mbox{for $t\in(-\infty,\infty)$} \nonumber \\
   &\mbox{with}& \vy(-\infty;\vyz,\tz)=(0,0), \qquad
   \vy(\infty;\vyz,\tz)=(\pm\infty,\pm\infty).
   \ea
   \end{Definition}

   \begin{Corollary}{Asymptotic behavior of $b\spp$ and $b\spm$.}
         \label{crlly:all-Bb}
      As $|\yt|\to +\infty$, $b\sppm$ asymptotically approaches
         $F^{(2)}$ where ${d\over dt}\yt=0$
  (Figure \ref{fg:basin_crv_n3m25g10}).
  	A trajectory  $\vy(t;\vyz,\tz)$ with 
 $\vyz\in b\sppm$ has the following properties.
	\begin{itemize}
	\item It can never reach $F^{(2)}$, because if it does,
    	it will have to exit the region where $\yo\yt>0$ and this
	leads to a contradiction
	(Remark \ref{rmk:all-phase-dyn}, Item 3),
	\item It must stay near $F^{(2)}$ so that the near zero
	velocity ${d\over dt}\yt\sim 0$ keeps the trajectory from
         growing rapidly to a singularity in a finite time.
        This also
         leads to a contradiction (Theorem \ref{def:all-Bb}). 
	\end{itemize}
  \end{Corollary}

\noindent
  Because the boundary is kinematically defined,
 we have the following theorem for the basins.

   \begin{Theorem}{Basins $B\spp$ and $B\spm$.
  {\rm [see Definition~\ref{def:sing-B}]}} \label{thm:all-B}
 There exist two distinct basins $B\spp$ and $B\spm$ kinematically
  divided by $b\spp$ and $b\spm$.
  Any trajectory $\vy(t;\vyz,\tz)$ with $\vyz\in B\sppm$ 
	will remain within it and reaches a finite-time
 singularity, i.e.,
   \ba
   \mbox{if $\vyz\in B\sppm$}, \qquad
   &\mbox{then}& \vy(t;\vyz,\tz)\in B\sppm \qquad \qquad
    \mbox{for $t\in(-\infty,\tc(\vyz)+\tz)$} \nonumber \\
   &\mbox{with} &\lim_{t\to -\infty}\vy(t;\vyz,\tz)=(0,0),
     \nonumber \\
   & & \lim_{t\to \tc(\vyz)+\tz}\vy(t;\vyz,\tz)=\vy_{c}(\vyz), \qquad
     \vy_{c}(\vyz)=(y_{1c}(\vyz),\pm\infty).
   \ea
where $\vy_{c}(\vyz)$ and $\tc(\vyz)$ are the finite-time singularity
and finite-time singular interval.
They depend only on the initial condition $\vyz$ because the system is autonomous.
 Any other trajectory not starting in $B\sppm$ initially
 can never enter into $B\sppm$ by
   cutting across $b\sppm$ due to the
    uniqueness of the solution.
   \end{Theorem}

\noindent
As seen on the two examples in  Figure \ref{fg:phase_all}, 
in the presence of both the trend and the reversal terms,
the oscillatory
behavior persists near the origin.
Technically, whether or not the oscillations persist depends on the 
competition between the oscillatory and
singular elements with respect to the time-scales.
In Section \ref{sec:case-fractal}, we will examine this issue
in details using scaling arguments.
Here, we proceed with our discussion assuming
that such oscillations do exist, as observed in figure \ref{fg:phase_all}.

\begin{remark}\label{rmk:all-nested} \ \\
 \renewcommand{\theenumii}{\arabic{enumii}}
{\rm {For $|\vy|\gg 1$ outside of the oscillatory region,
two basins $B\sppm$ are clearly visible
(see Figures \ref{fg:phase_all} and \ref{fg:basin_crv_n3m25g10}):
$B\spp$ lies ``above'' $b\spp$, and $B\spm$  lies ``below'' $b\spm$.
This description is carried into the oscillatory region
using the direction of the flow as follows.}
\begin{enumerate}
\item $B\spp$ basin: ``above'' the boundaries $b\sppm$, i.e.,
to the left of $b\spp$ and to the right of $b\spm$   
with respect to the forward direction in the flow
(see also in Figure \ref{fg:phase_all}).
  Any trajectory $\vy(t;\vyz,\tz)$ with $\vyz\in B\spp$ 
 goes to $\yt(\tc;\vyz,\tz) \to +\infty$.
\item $B\spm$ basin: ``below'' the boundaries $b\sppm$, i.e.,
to the left of $b\spm$ and to the right of $b\spp$
with respect to the forward direction in the flow.
  Any $\vy(t;\vyz,\tz)$ with $\vyz\in B\spm$ 
 goes to $\yt(\tc;\vyz,\tz) \to -\infty$.
\end{enumerate}
{In other words, $b\sppm$ lies to the right of $B\sppm$ 
with respect to the forward direction of the flow.}}
 \renewcommand{\theenumii}{\alph{enumii}}
	\end{remark}

\subsection{Global dynamics}
\label{sec:case-global}
\subsubsection{Dynamical properties along the boundaries}
\label{sec:case-dynb}
 
The structure of the phase space is completely governed by 
the boundary $b\sppm$.
Therefore, we first study the dynamical properties along $b\sppm$.

\begin{figure}
\begin{center}
\includegraphics{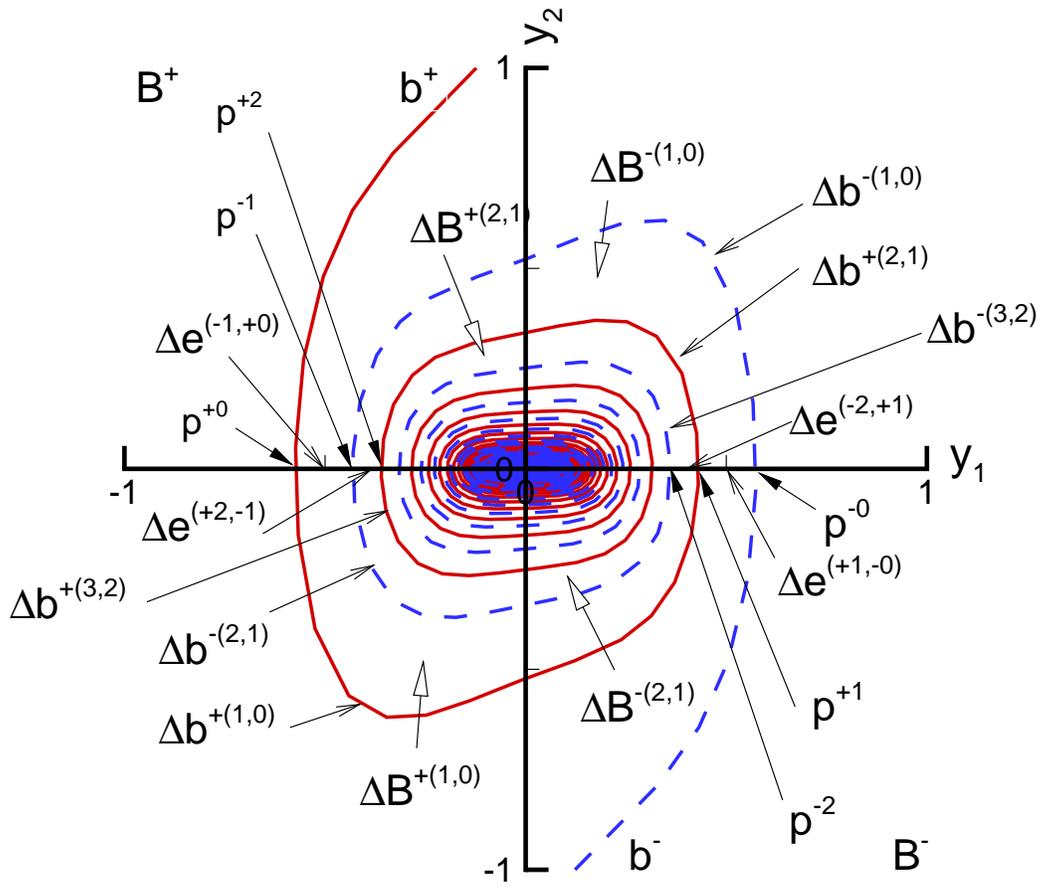}
\end{center}
\caption{Geometry of boundaries $b\sppm$ and basins $B\sppm$ for
$(n,m)=(3,2.5)$ and $\gamma=10$. 
See text for details.}
\label{fg:basin_bB_n3m25g10}
\end{figure}

   \begin{Definition}{Exit turn $\vpo^{\pm 0}$ as the intersection of $b\sppm$ 
with the $\yo$-axis}\label{def:po}
    We define a point $\vpo^{\pm 0}$ 
 as the out-most intersection of $b\sppm$ with the $\yo$-axis
 as shown in Figure \ref{fg:basin_bB_n3m25g10}.
 Therefore,
 $\vy(t;\vpo^{\pm 0},\tz)$ makes no further turn for $t>\tz$
 (Definition \ref{def:all-phase-turn}).
  We call $\vpo^{\pm 0}$ the exit turn point.
 The transition from oscillatory to singular behaviors
 occurs at  $\vpo^{\pm 0}$.
   \end{Definition}

  \begin{Definition}{Reference trajectory $\vy\sppm(t)$, turn points
   $\vpo^{\pm k}$ and  exit time $\tauo^{\pm k}$}
    \label{def:all-flightt}
    We define $\vy\sppm(t)$ as the reference trajectory on
    $b\sppm$ which goes through the exit turn point 
	$\vpo^{\pm 0}$ at time $t=0$, i.e., 
   \be \label{eq:all-vypm}
   \vy\sppm(t)\equiv\vy(t;\vpo^{\pm 0},0)~.
   \ee
 In backward time,  $\vy\sppm(t)$ makes a turn
 by intersecting the $\yo$-axis (Definition \ref{def:all-phase-turn}).
 At time $\tauo^{\pm k}(<0)$, $\vy\sppm(t)$ makes the
 $k$-th (backward) turn:
   \be \label{eq:all-taopm}
   \vy\sppm(\tauo^{\pm k})=\vpo^{\pm k},
   \ee
 where
   \be 
	\vpo^{\pm k}\equiv(\yo^{\pm k},0)~\in b\sppm~~
   \ee
 is defined as the $k$-th turn points.
 It is located at an intersection of 
 $b\sppm$ and $\yo$ axis (Figure \ref{fg:basin_bB_n3m25g10}).
 By construction, a trajectory $\vy(t;\vpo^{\pm k},\tauo^{\pm k})$ 
 is on the reference trajectory.
 Starting from $\vpo^{\pm k}$, the trajectory makes $k$ turns before reaching the exit
 turn point  $\vpo^{\pm 0}$ at time 0 after a time interval
$-\tauo^{\pm k}(>0)$.
 It does not make any more turns for $t>0$.
 We call $\tauo^{\pm k}(<0)$ the $k$-th exit time.
   \end{Definition}

  \begin{Definition}{Template map for the dynamics associated with
   $\vpo^{\pm k}$ along $b\sppm$}
   \label{crlly:n3m25-map}
   We define the template map of the dynamics along each boundary
   $b\sppm$ using the sequence of turn points
   $\vpo^{\pm k}$:
  \ba
     \ldots \tr \vpo^{\pm k+1} & \tr & \vpo^{\pm k}
 	\tr \ldots \tr \vpo^{\pm 0}~.
  \ea
  By construction, there is no other turn points between any
  $\vpo^{\pm k+1}$ and $\vpo^{\pm k}$ along $b\sppm$.
 \end{Definition}

   \begin{Remarks}{As shown in Figure \ref{fg:basin_bB_n3m25g10},}
 \label{rmk:all-bipo}
\item Along $b\sppm$ in forward time, 
the turn points $\vpo^{\pm k}$ jump between  $\yo>0$  and $\yo<0$  
  as $\vy^{\pm}(t)$ oscillates around the origin:
  \begin{itemize}
  \item along $b\spp$: $\yo^{+ (2l)}<0$ and $\yo^{+ (2l+1)}>0$,
  \item along $b\spm$:  $\yo^{- (2l+1)}<0$ and $\yo^{- (2l)}>0$.
  \end{itemize}
\item 
On the $\yo$-axis, the turn points  alternate between  $b\spp$ and $b\spm$:
\be  \label{eq:all-byo}
\yo^{+0}~<~\yo^{-1}~<~\yo^{+2}~<~\yo^{-3}<~\yo^{+4}~<~\ldots~
<~\yo^{-4}~<~\yo^{+3}~<~\yo^{-2}~<~\yo^{+1~}<~\yo^{-0}~,
\ee 
\end{Remarks}

  \begin{remark} \label{rmk:all-po} \ \\
{\rm 
 {Three main dynamical
  properties associated with $\vpo^{\pm k}$ are as follows:}
\begin{enumerate}
  \item the alternating signs of $\yo^{\pm k}$, i.e., 
 $(-1)^{k}\yo^{+k}<0$ along $b\spp$ and
   $(-1)^{k}\yo^{-k}>0$ along $b\spm$
  (Corollary \ref{rmk:all-bipo});
  \item the finite number of turns $N_{turns}=k$ in forward time before
	exiting from the oscillatory region (Corollary 
\ref{def:all-flightt});
  \item the exit time $\tauo^{\pm k}$ 
	(Definition \ref{def:all-flightt}).
\end{enumerate}
These properties are summarized in  Table \ref{tbl:all-mapp} 
(see also Figure \ref{fg:basin_bB_n3m25g10})
using the template map  ($\tr$) to show the dynamics in forward time.
}  \end{remark}

\begin{table}[!ht]
\begin{center}
\begin{tabular}{|c||ccccccccccccc|} \hline
   $\vpo^{\pm k}$ & $\ldots$ & $\tr$ & $\vpo^{\pm (k+1)}$ & $\tr$
& $\vpo^{\pm (k)}$ & $\tr$ & $\ldots$  & $\tr$
& $\vpo^{\pm 1}$ & $\tr$ & $\vpo^{\pm 0}$  & $\tr$
& [$(\pm\infty,\pm\infty)$] \\ \hline \hline
   sign of $\yo^{\pm k}$ & $\ldots$ & $\tr$ & $(-1)^{(k+1)}\mp$ & $\tr$
& $(-1)^{k}\mp$ & $\tr$ & $\ldots$  & $\tr$
& $(-1)^{1}\mp$ & $\tr$ & $\mp$  & ($\tr$
& [$\pm$]) \\ \hline
   $N_{turn}$ & $\ldots$ & $\tr$ & $k+1$ & $\tr$
& $k$ & $\tr$ & $\ldots$  & $\tr$
& $1$ & $\tr$ & $0$  & 
&  \\ \hline 
   $\tauo^{\pm k}$  & $\ldots$ & $\tr$ & $\tauo^{\pm (k+1)}$ & $\tr$
& $\tauo^{\pm k}$ & $\tr$ & $\ldots$  & $\tr$
& $\tauo^{\pm 1}$ & $\tr$ & $0$  &
&   \\ \hline
\end{tabular}
\caption{Dynamical properties associated with the template map
of $\vpo^{\pm k}$ along the singular boundary $b\sppm$ 
(Remarks \ref{rmk:all-po}).}
\label{tbl:all-mapp}
\end{center}
\end{table}

\noindent
We now present the dynamical properties of arbitrary trajectories
along $b\sppm$ using the template map.
To do so, we partition $b\sppm$ into boundary segments bounded by the points
$\vpo^{\pm (k)}$.

  \begin{Definition}{Boundary segment}\label{def:all-bsk}
   We define the boundary segment $\bs^{\pm(k+1,k)}$
   of $b\sppm$ to be a segment between two adjacent turn
   points $\vpo^{\pm (k+1)}$ and  $\vpo^{\pm k}$
   (both exclusive) :
  \ba \label{eq:all-bs}
  \bs^{\pm(k+1,k)}\equiv\{\vy\sppm(t),~
   \mbox{for $t\in (\tauo^{\pm k+1},\tauo^{\pm k})$}\}~~,
  \ea
(see also Figure \ref{fg:basin_bB_n3m25g10}). 
The symbol ``$($'' and  ``$)$'' in the superscript 
are used to indicate that both the left and right endpoints
are outside of the interval.
  \end{Definition}

\begin{Remark}\label{rmk:all-bprop}
\item
The semi-infinite unions of the boundary segments together with the
 semi-infinite unions of turn points forms the boundary:
   \be
    b\sppm =\cup_{k=0}^{\infty} \bs^{\pm(k+1,k)}
   ~\cup~\bs^{\pm(\cdot,0)}~\cup_{k=0}^{\infty} \vpo^{\pm k}~~,
   \ee
   where
  \ba \label{eq:all-bsz}
  \bs^{\pm(0,\cdot)}\equiv\{\vy=\vy\sppm(t),~
   \mbox{for $t\geq 0$}\}~.
  \ea
\item 
By construction, $\vy\sppm(t)$ makes no turn over any boundary segment
$\bs^{\pm(k+1,k)}$.
\end{Remark}

\noindent
The complete description of the dynamical properties along  $b\sppm$ follows.

 \begin{Corollary}{Dynamical properties of a trajectory 
 $\vy(t;\vyz,\tz)$ on $b\sppm$}
    \label{crlly:all-exitt}
Let us consider a trajectory $\vy(t;\vyz,\tz)$ on $b\sppm$  starting
 from  a point $\vyz$ in the boundary segment 
 $\vyz\in \bs^{\pm(k+1,k)}$ at an arbitrary time $\tz$.
 It reaches the turn point $\vpo^{\pm k}$ in forward time
 after a time interval $-\tsb^{\pm(k+1,k)}(>0)$ without making any
turn, where
\ba
\tauo^{\pm (k+1)}-\tauo^{\pm k}<\tsb^{\pm(k+1,k)}<0~,
\ea
because $\tauo^{\pm k}-\tauo^{\pm (k+1)}$ is the total time of flight
over $\bs^{\pm(k+1,k)}$.
Accordingly, $\vy(t;\vyz,\tz)$  with $\vyz\in \bs^{\pm(k+1,k)}$ 
has the same dynamical properties as $\vy^{\pm}(t)$ 
starting from $\vpo^{\pm k}$ (Table \ref{tbl:all-mapp}),
except that the exit time is extended from
$\tauo^{\pm k}$ to $\tauo^{\pm k}+\tsb^{\pm(k+1,k)}$.
It can also be expressed using the reference trajectory
\be
	\vy(t;\vyz,\tz)=\vy^{\pm}(t-\tz+\tauo^{\pm k}+\tsb^{\pm(k+1,k)})~,
\ee
because $\vy(\tz;\vyz,\tz)=\vy^{\pm}(\tauo^{\pm k}+\tsb^{\pm(k+1,k)})$.
   \end{Corollary}

\subsubsection{Dynamical properties in the basins}
\label{sec:case-dynB}

As the dynamical properties along the boundaries $b\sppm$ can be 
described by a template map of turn points $\vpo^{\pm k}$, the
dynamical properties in the basin $B\sppm$ can also 
described by a template map of turn segments as follows.

  \begin{Definition}{Turn segments $\es^{(\pm (k+1),\mp k)}$}
   \label{def:all-es}
Any trajectory $\vy(t;\vyz,\tz)$ can make a turn only on the
$\yo$-axis (Definition \ref{def:all-phase-turn}).
  We define a turn segment $\es^{(\pm (k+1),\mp k)}$ as being
bounded by two adjacent 
 turn points $\vpo^{\pm (k+1)}$  and  $\vpo^{\mp k}$
  on the $\yo$-axis (Figure \ref{fg:basin_bB_n3m25g10}):
 \begin{itemize}
 \item\makebox[3.cm][l]{for $\yo<0$}
  \makebox[3.0cm][l]{$\es^{(+0,-\infty)}$}
	\makebox[8.0cm][l]
	{$\equiv\{\vy~|~-\infty<\yo<\yo^{+ 0}<0, \yt=0\}$}{$\in B\spp$}; \\
	\makebox[3.cm][l]{\ }
  \makebox[3.0cm][l]{$\es^{(-(2l+1),+(2l))}$}
	\makebox[8.0cm][l]
	{$\equiv\{\vy~|~p^{+(2l)}<\yo<p^{-(2l+1)}<0, \yt=0\}$}{$\in B\spm$};  \\
	\makebox[3.cm][l]{\ }
  \makebox[3.0cm][l]{$\es^{(+(2l),-(2l-1))}$}
	\makebox[8.0cm][l]{$\equiv\{\vy~|~p^{- (2l-1)}<\yo<p^{+(2l)}<0,~
	\yt=0\}$}{$\in B\spp$}; 
 \item\makebox[3.cm][l]{for $\yo>0$}
  \makebox[3.0cm][l]{$\es^{(-(2l),+(2l-1))}$}
	\makebox[8.0cm][l]{$\equiv\{\vy~|~0<p^{-(2l)}<\yo<p^{+(2l-1)},~
	\yt=0\}$}{$\in B\spm$};  \\
	\makebox[3.cm][l]{\ }
  \makebox[3.0cm][l]{$\es^{(+(2l+1),-(2l))}$}
	\makebox[8.0cm][l]{$\equiv\{\vy~|~0<p^{+(2l+1)}<\yo<p^{-(2l)},~
	\yt=0\}$}{$\in B\spp$};  \\
	\makebox[3.cm][l]{\ }
  \makebox[3.0cm][l]{$\es^{(-0,+\infty)}$}
 	\makebox[8.0cm][l]{$\equiv\{\vy~|~0<p^{- 0}<\yo<\infty, \yt=0\}$}{$\in B\spm$}.
 \end{itemize}
The notational convention for the superscript of the turn
segment $\es^{(\pm (k+1),\mp k)}$ is that the left and 
right indices, $\pm (k+1)$ and $\mp k$,  respectively
correspond to the superscript of the turn points, 
$\vpo^{\pm (k+1)}$ and $\vpo^{\mp k}$,
which are  respectively at the left and right ends of the segment
with respect to forward direction of the flow
(see Remark \ref{rmk:all-nested}).
The symbols ``('' and  ``)'' in the superscript mean that 
these intersection points are not included in the segment.
\end{Definition}

 \begin{Remarks}{In comparison with Remark \ref{rmk:all-bipo}:}\label{rmk:all-es} 
\item Along the flow in forward time, the turn segments jump between $\yo<0$
and $\yo>0$ as a trajectory in $B\sppm$ oscillates around the origin:
  \begin{itemize}
  \item in $B\spp$: $\yo<0$ for $\es^{(+(2l+1),-(2l))}$
  and $\yo>0$ for $\es^{(+(2l),-(2l-1)}$;
  \item in $B\spm$: $\yo<0$ for $\es^{(-(2l),+(2l-1))}$
  and $\yo>0$ for $\es^{(-(2l+1),+(2l)}$;
  \end{itemize}
\item Moreover, the
$\yo$-axis consists of the union of all intersection points and segments;
 \ba
 \mbox{$\yo$-axis } & = &\es^{(+0,-\infty)}\cup\vpo^{+0}
 \cup\es^{(-1,+0)}\cup\vpo^{-1} \cup\es^{(+2,-1)}\cup\vpo^{+2}\cup
 \nonumber \\ 
 & & ~~~\ldots \cup (0,0) \cup \ldots
 \cup\vpo^{-2}\cup\es^{(-2,+1)}\cup\es^{(+1,-0)}\cup
  \vpo^{-0} \cup\es^{(-0,+\infty)}~.
 \ea
 Note that the superscript of $\es^{(\pm (k+1),\mp k)}$
 for $\yo<0$ is not in sequence with the subscript of
 $\vpo^{\pm (k+1)}$ as in the case for $\yo>0$.  
 This is because $\vpo^{\pm (k+1)}$ and $\vpo^{\mp k}$
 are located at the right and left ends of $\es^{(\pm (k+1),\mp k)}$  
 with respect to the forward direction of the flow
 (Definition \ref{def:all-es}), but geometrically
 at the left  and right ends of the segment
  on $\yo$-axis (Figure \ref{fg:basin_bB_n3m25g10}). 
 \end{Remarks}

\noindent
To describe the dynamics in the basins,
we first define the oscillatory source near the origin and separate it
from the singular region outside.
We show that the transition from the oscillatory to the singular behavior in $B\sppm$
occurs at the exit turn segment, associated with the fact that the
transition along $b\sppm$
occurs at the exit turn point 
(Definitions \ref{def:po} and \ref{def:all-flightt}).
The global dynamics in  $B\sppm$ is structured into two regimes 
separated by the exit turn segment which determines the
singular behavior in forward time and the oscillatory behavior in
backward time, as follows.

\begin{Definition}{Exit turn segments $\es^{(\pm 1,\mp 0)}\in B\sppm$}
\label{def:all_es}
We call   $\es^{(\pm 1,\mp 0)}\in B\sppm$
the exit turn segments.
\end{Definition}   

\begin{Definition}{Oscillatory Source $S$}
   \label{def:all-S}
  The exit turn segments and  the out-most boundary segments,
  along with the  turn points at the end of these segments,
  formally define the oscillatory source region $S$:
  \ba \begin{array}{cllll}
   S=\{\vy|~\mbox{ region surrounded by } &
     \vpo^{-1}, & \bs^{-(1,0)}, &
	\vpo^{-0}, & \es^{(+1,-0)},  \\
   & \vpo^{+1}, & \bs^{+(1,0)}, & 
	\vpo^{+0}, &\es^{(-1,+0)} \}~ \end{array}
  \ea
 (see Figure \ref{fg:basin_bB_n3m25g10};
 as well as Definition \ref{def:sing-S} for the case with only the
 singular element).
  \end{Definition}

\begin{Corollary}{Transition from oscillatory to singular behavior}
\label{crlly:all-transite}
In the sequel, we note $\vpo^{(\pm 1,\mp 0)}$ for short to include one of the
four points $\vpo^{\mp 0}, \vpo^{\pm 1}$.
Let us consider a trajectory $\vy(t;\vyz,0)$ starting
from a point $\vyz=\vpo^{(\pm 1,\mp 0)}$ on the
exit turn segment $\es^{(\pm 1,\mp 0)}\in B\sppm$.
In forward time, $\vy(t;\vyz,0)$ will make only one turn at a point 
in $\es^{(\pm 0,\mp\infty)}$  but never completes an 
oscillation cycle  (Corollary \ref{cor:all-phase-osc}).
Therefore, the exit segment
$\es^{(\pm 1,\mp 0)}$ defines the transition from oscillatory to
singular behavior.
\end{Corollary}

\begin{Corollary}{Dynamics outside $S$ in the singular region.}
\label{crlly:all-es-f}
Let us consider a trajectory $\vy(t;\vyz,0)$ 
starting from the exit turn segment in forward time with 
$\vyz=\vpo^{(\pm 1,\mp0)}\in \es^{(\pm 1,\mp 0)}$ in $B\sppm$.
After making the final turn in  $\es^{(\pm 0,\mp\infty)}$,
it reaches the corresponding singularity:
\ba
\vy_{c}(\vpo^{(\pm 1,\mp0)})
  =\vy(\tc(\vpo^{(\pm 1,\mp 0)});\vpo^{(\pm 1,\mp 0)},0)~\in B\sppm
\ea
which depends only on the initial condition  $\vpo^{(\pm 1,\mp 0)}$
(see Theorem \ref{thm:all-B}).
Here $\tc(\vpo^{(\pm 1,\mp 0)})$ is the finite singular time.

By Definition \ref{def:all-es}, the left and right end points of 
$\es^{(\pm 1,\mp 0)}$ are next  to $\vpo^{\pm 1}$ and $\vpo^{\mp 0}$.
Therefore, the terminal value of $\yo$ at the singularity ranges over:
\ba\label{eq:all_y1cp}
 y_{1c}(\vpo^{(\pm 1,\mp 0)})=\langle\pm\infty,\mp\infty\rangle~,
\ea
where $\langle a,b\rangle$ denotes that it is 
$a$ or $b$ asymptotically  if $\vpo^{(\pm 1,\mp 0)}$ is  respectively
at the left or right end point of $\es^{(\pm 1,\mp 0)}$ 
with respect to the forward direction of the flow.
  \end{Corollary}

  \begin{Remarks}{Two main dynamical properties associated with a point 
  $\vpo^{(\pm 1,\mp 0)}$ on the exit turn segment in the singular
 region outside $S$  are as follows:}
\label{rmk:all-poes0}
  \item finite singular time, $\tc(\vpo^{(\pm 1,\mp 0)})$;
  \item terminal $\yo$ value, $y_{1c}(\vpo^{(\pm 1,\mp 0)})$. \\
  \end{Remarks}

\begin{Corollary}{Dynamics inside $S$ in the oscillatory region.}
\label{crllry:all-y}
Let us consider a trajectory $\vy(t;\vyz,\tz)$ 
starting from the turn segment $\es^{(\pm k,\mp (k-1))}\in B\sppm$ in
forward time .
Including the starting point as the first turn, the trajectory
makes the $l$-th turn ($1\leq l\leq k$) at a point in
$\es^{(+k-l+1,-(k-l))}$ and the sign of $\yo$ alternates between $+$ 
and $-$ at each turn (Remark \ref{rmk:all-es}).
The trajectory reaches $\vpo^{(\pm 1,\mp 0)}$ of an exit turn segment
$\es^{(\pm 1,\mp 0)}$ to make the $k$-th and final turn at time
$\tz+\tse^{(\pm k,\mp (k-1))}$, with
\ba
\tse^{(\pm k,\mp (k-1))} \in
\langle\tauo^{\pm k},\tauo^{\mp (k-1)}\rangle_{t}~
\ea
where $\tauo^{\pm k}$ are defined by (\ref{eq:all-taopm}) (see also 
table \ref{tbl:all-mapp}.
$\langle a,b\rangle_{t}$ denotes that it is 
$a$ or $b$ asymptotically  if $\vyz$ is respectively at the left
or right end  of $\es^{(\pm(k+1),\mp k)}$
with respect to the forward direction of the flow, 
like $\langle a,b\rangle$ in (\ref{eq:all_y1cp}) for $y_{1c}$.
\end{Corollary}

  \begin{Definition}{Template map for the dynamics associated with
 $\es^{(\pm (k+1),\mp k}$}
   \label{crlly:n3m25-mapB}
   We define the template map of the dynamics in
   $B\sppm$ using the sequence of intersection segments
   $\es^{(\pm (k+1),\mp k)}$ on the $\yo$-axis:
  \ba
     \ldots \tr \es^{(\pm (k+1),\mp k)} &  \tr & \es^{(\pm k,\mp (k-1)}
 	\tr \ldots \tr \es^{(\pm 1,\mp 0)} \tr \es^{(\pm 0,\mp \infty)}
  \ea
  By construction, there is no other turn segments between 
  $\es^{(\pm (k+1),\mp k}$ and $\es^{(\pm k,\mp (k-1))}$ in  $B\sppm$
 (compare with Definition \ref{crlly:n3m25-map}).
 \end{Definition}

\begin{remark}\label{rmk:all-poe} \ \\
{\rm 
{Three main dynamical properties associated with a point 
  $\vyz\in \es^{(\pm(k+1),\mp k)}$ in the oscillatory source $S$
are summarized in table \ref{tbl:all-mapB} (compare with Remark \ref{rmk:all-po}). They
comprise:}
\begin{enumerate}
  \item the signs of $\yo$ (Definition \ref{def:all-es});
  \item the finite number of turns $N_{turn}$ to transit from $S$ into the
  final region where the monotonous singular behavior occurs
(Corollary \ref{crllry:all-y});
  \item the exit time interval $\tse^{(\pm (k+1),\mp k)}$
        to reach $\es^{(\pm 1,\mp 0)}$ (Corollary \ref{crllry:all-y}).
 \end{enumerate}
}
 \end{remark}

\begin{table}[!ht]
\begin{center}
\begin{tabular}{|c|c||ccccccccccc|} \hline
   \multicolumn{2}{|c||}{$\es^{(\pm (k+1),\mp k)}$} 
& $\ldots$ & $\tr$ & $\es^{(\pm (k+1),\mp k)}$ & $\tr$
& $\es^{(\pm k,\mp (k-1))}$ & $\tr$ & $\ldots$  & $\tr$
& $\es^{(\pm 1,\mp 0)}$ & $\tr$ & $\es^{(\pm 0,\mp\infty)}$  
 \\ \hline \hline
inside &  sign of $\yo$ & $\ldots$ & $\tr$ & $(-1)^{(k+1)}\mp$ & $\tr$
& $(-1)^{k}\mp$ & $\tr$ & $\ldots$  & $\tr$
& $(-1)^{1}\mp$ & $\tr$ & $\mp$   \\ \cline{2-13}
$S$ & $N_{turn}$ & $\ldots$ & $\tr$ & $k+1$ & $\tr$
& $k$ & $\tr$ & $\ldots$  & $\tr$
& $1$ & $\tr$ & $0$  \\ \cline{2-13}
 &  $\tse^{(\pm (k+1),\mp k)}$  & $\ldots$ & $\tr$ & $\tse^{(\pm (k+1),\mp k)}$ & $\tr$
& $\tse^{(\pm k,\mp (k-1))}$ & $\tr$ & $\ldots$  & $\tr$
& $\tse^{(\pm 1,\mp 0)}$ &   &    \\ \hline  \hline
outside & $y_{1c}$  & \multicolumn{11}{c|}{$y_{1c}(\vpo^{(\pm 1,\mp
0)})$} 
 \\ \cline{2-13}
$S$ & $t_{c}$  & \multicolumn{11}{c|}{$t_{c}(\vpo^{(\pm 1,\mp 0)})$} \\ \hline
\end{tabular}
\caption{Dynamical properties associated with the template map
of $\es^{(\pm (k+1),\mp k)}$ in the oscillatory (in $S$) and 
singular (outside $S$) domains. These properties characterize the 
dynamics in forward time
(see also Figure \ref{fg:basin_bB_n3m25g10} and Table \ref{tbl:all-mapp})
using the template map  ($\tr$).
}
\label{tbl:all-mapB}
\end{center}
\end{table}

\noindent
We now present the dynamical properties of arbitrary trajectories
in $B\sppm$ using the template map.
To do so, we partition $B\sppm$ into sub-basins limited by the segments
$\es^{(\pm (k+1),\mp k)}$.

\begin{Definition}{Sub-basins $\Bs^{\pm (k+1,k)}$}
   \label{def:all-Bs}
 We define a sub-basin to be a piece of the basin $B\sppm$ 
 limited by two adjacent turn segments 
 $\es^{(\pm (k+1),\mp k)}$ and $\es^{(\pm k,\mp (k-1))}$
 as follows (see Figure \ref{fg:basin_bB_n3m25g10}):
 \ba
   \Bs^{\pm (k+1,k)}
	&=&\{\vy|~\mbox{ region exclusively surrounded by}  \\
 & & \qquad \es^{(\pm (k+1),\mp k)},~ \vpo^{\pm k},~ 
		\bs^{\mp (k,(k-1))},~ \vpo^{\pm (k+1)}, \nonumber \\
 & & \qquad \es^{(\pm k,\mp (k-1))},~ \vpo^{\mp k },~
		\bs^{ \pm((k+1), k)},~ \vpo^{\mp (k-1)}  \}~.
	\nonumber
  \ea
  \end{Definition}

\begin{Remarks}{In comparison to  Remark \ref{rmk:all-bprop}:}
\label{rmk:all-Bprop}
\item
The semi-infinite unions of the sub-basins together with the
 semi-infinite unions of turn segments reconstruct the complete basin
(see Figure \ref{fg:basin_bB_n3m25g10}):
   \be
    B\sppm =\cup_{k=0}^{\infty} \Bs^{\pm (k+1,k)}
   ~\cup~\Bs^{\pm(0,\infty)}~\cup_{k=0}^{\infty} \es^{(\pm k+1,k)}~~,
   \ee
   where
  \ba 
   \Bs^{\pm(0,\infty)}
   &=&\{\vy|~\mbox{ region exclusively surrounded by } \\
    & & \qquad \qquad \bs^{+(0,\cdot)},~\vpo^{+0},~\bs^{+(0,1)},~\vpo^{+1},~
        \es^{(+1,-0)},~\vpo^{-0},~\bs^{-(0,\cdot)}  \}~~
	\nonumber 
  \ea
  where $\bs^{\pm(0,\cdot)}$ are defined by (\ref{eq:all-bsz}).
\item 
By construction, no trajectory makes any turn in $\Bs^{\pm (k+1,k)}$.
\end{Remarks}

\noindent
The complete description of the dynamical properties in $B\sppm$ 
can now be given completely.

 \begin{Corollary}{Dynamical properties of a trajectory 
 $\vy(t;\vyz,\tz)$ in $B\sppm$}
    \label{crlly:all-exittB}
Let us consider a trajectory $\vy(t;\vyz,\tz)$ in $B\sppm$ starting
 from  a point $\vyz$ inside the sub-basin
 $\vyz\in \Bs^{\pm(k+1,k)}$ at an arbitrary time $\tz$.
 It reaches the turn segment $\es^{(\pm k,\mp (k-1))}$ in forward time
 after a time interval $-\tsB^{\pm(k+1,k)}(>0)$ without making any
turn, where
\ba
\tauo^{\pm (k+1)}-\tauo^{\pm k}<\tsB^{\pm(k+1,k)}<0~,
\ea
because $\tauo^{\pm k}-\tauo^{\pm (k+1)}(>0)$ is the time of flight of
a trajectory along $\bs^{ \pm((k+1), k)}$ which borders $\Bs^{\pm(k+1,k)}$.
Accordingly, $\vy(t;\vyz,\tz)$  with $\vyz\in \Bs^{\pm(k+1,k)}$ 
has the same dynamical properties as $\es^{(\pm k,\mp (k-1))}$
(Table \ref{tbl:all-mapB}),
except that the exit time is extended from
$\tse^{(\pm k,\mp (k-1))}$ to $\tse^{(\pm k,\mp
(k-1))}+\tsB^{\pm(k+1,k)}$
(compare with Corollary \ref{crlly:all-exitt}).
\end{Corollary}

\subsection{Scaling laws}
   \label{sec:case-fractal}

\subsubsection{Dynamical properties on the $\yo$-axis}
   \label{sec:case-fractal-yo}
   
\noindent
Because the system is autonomous, each trajectory $\vy(t;\vyz,\tz)$ 
is determined uniquely by its initial condition $\vyz$.
We have shown above  that the template maps defined
on the $\yo$-axis completely describe the dynamical properties
of the dynamical system. It is thus convenient to
summarize these dynamical properties as a
function of the initial condition $\vyz=(y_{10},0)$ taken on the $\yo$-axis.
There properties are quantified by the
finite singular time $\tc$,
the total number of turn $N_{turn}$, and the
finite terminal value $y_{1c}$
(see Remarks \ref{rmk:all-po} and \ref{rmk:all-poe}).

\begin{figure}
\begin{center}
\includegraphics[height=18cm]{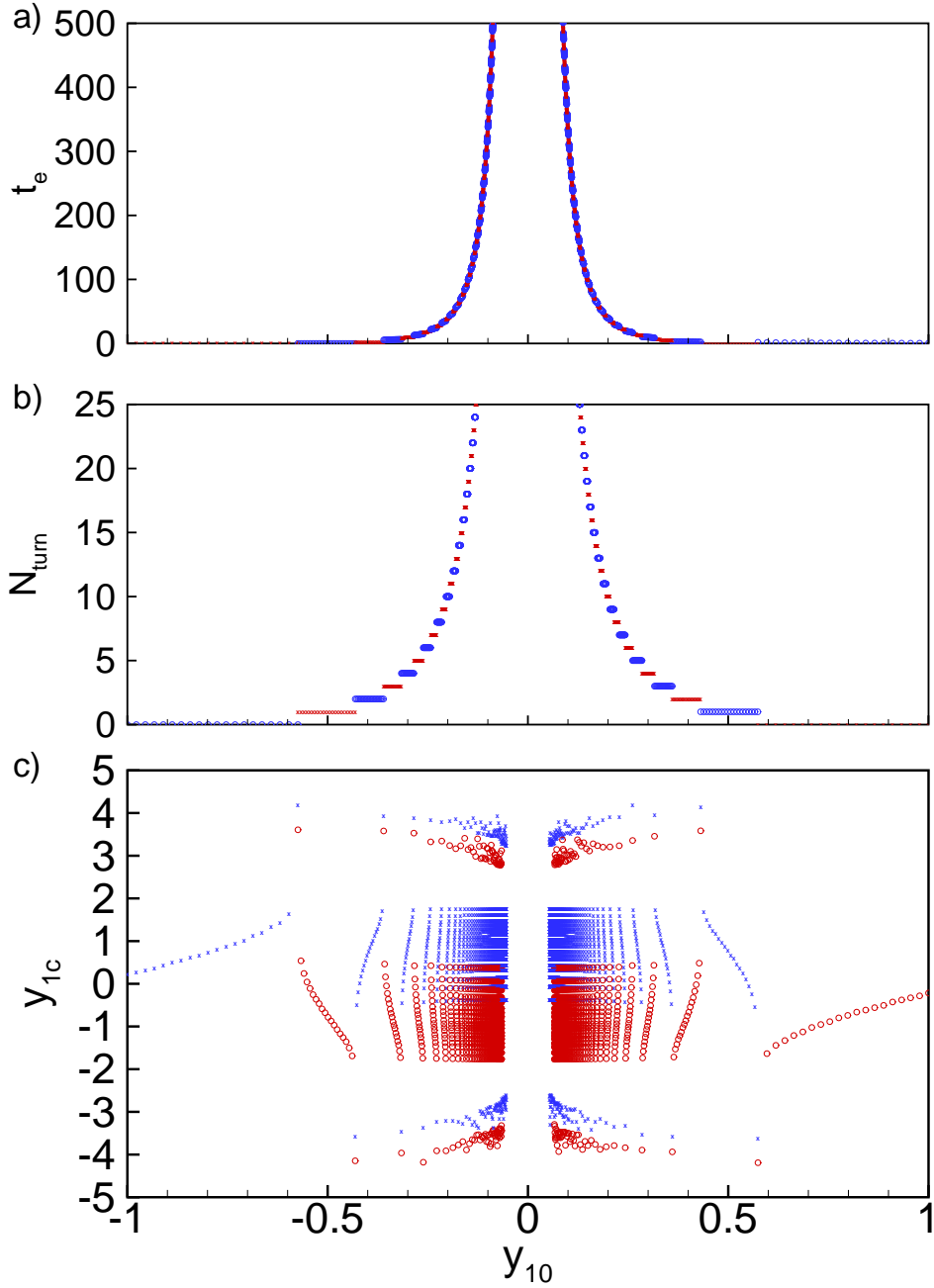}
\end{center}
\caption{Dependence of the key dynamical variables as a function 
of the initial condition
$\vyz=(y_{10},0)$ on the $\yo$=axis for
$(n,m)=(3,2.5)$ and $\gamma=10$: exit time $t^{\pm k}(\vyz)$ on $b^{\pm}$
into the non-oscillatory regime
beyond the intervals $\Delta e^{(\pm 1,\mp 0)}$ shown in 
figure \ref{fg:basin_bB_n3m25g10} (top panel);
$N_{turn}(\vyz)$ (middle panel); and
$y_{1c}(\vyz)$ (bottom panel).
In each panel, ``circle'' and  ``crosses'' symbols correspond to points in 
$\triangle e^{+(k+1),-k}$ and $\triangle e^{-k,+(k-1)}$, respectively.
Notice the alternate structure in panel c) reflecting the 
spiralling topology of the boundaries $b^+$ and $b^-$ shown in 
figure \ref{fg:basin_bB_n3m25g10}. In order to construct panel c), 
we have sampled each turn segment $\triangle e^{\pm k+1,\mp k}$
by 20 $y_{10}$ points. The two end points are
chosen to be less than 10$^{-8}$ away 
from $\vpo^{\pm k+1}$ and $\vpo^{\mp k}$.  The other 18 points are equally
spaced within $\triangle e^{\pm k+1,\mp k}$.
}
\label{fg:fractal_n3m25g10}
\end{figure}

Figure \ref{fg:fractal_n3m25g10} shows 
these dynamical properties as a function of $y_{10}$
for $(n,m)=(3,2.5)$ with
$\gamma=10$ and 1000, determined by direct numerical integration of the
equations of motion, using a fifth-order Runge-Kutta integration 
scheme with adjustable time step.
Each discontinuity of $N_{turn}$ and $y_{1c}$ as a function of $y_{10}$
occurs at a turn point $\vpo^{\pm k}$ which separates
two turn segments, one in $B\spp$ and the other in $B\spm$
(Figure \ref{fg:basin_bB_n3m25g10} and Remark \ref{rmk:all-es}).

$N_{turn}$ is directly associated with the oscillatory behavior of the
dynamics. Note that it exhibits a staircase structure, being constant
for any point in $\es^{(\pm (k+1),\mp k)}$  (Table \ref{tbl:all-mapB}).

In contrast, $y_{1c}$ is a property more directly associated with the singular
behavior and takes identical values for any point 
$\vyz\in\es^{\pm k,\mp(k-1)}$ which is mapped to 
$y_{1c}(\vpo^{(\pm 1,\mp 0)})=\langle \pm\infty, \mp\infty\rangle$ 
after $k$ turns (Corollary \ref{crlly:all-es-f})
with $y_{1c}(\vpo^{\pm k})=\pm\infty$.
Moreover, given a $y_{1c}$, each turn segment $\es^{(\pm (k+1),\mp k)}$ has a
unique point $\vyz^{(\pm (k+1),\mp k)}$ (Corollary \ref{crlly:all-es-f}).

The critical or singular time $\tc$ is function both of the oscillatory and the
singular terms:
\ba
\tc(\vyz)=\tc(\vpo^{(\pm 1,\mp 0)})-\tse^{(\pm (k+1),\mp k)}~,  \label{mfngalaa}
\ea
where $\vyz\in \es^{(\pm (k+1),\mp k)}$.
However, the explosive singular time scale $\tc(\vpo^{(\pm 1,\mp 0)})$
is generally much shorter than the slow oscillation time scale
$\tse^{(\pm (k+1),\mp k)}$ (see figure \ref{Figm1.5n3})).
Hence, the total time $\tc$ needed to reach the singularity
is dominated by $\tse^{(\pm (k+1),\mp k)}$.

\subsubsection{Definition and mechanism}
   \label{sec:case-fractal-def}

Our analysis up-to-now has demonstrated a hierarchical organization of the
spiraling trajectories diverging away from the origin in phase space, as shown
in figure \ref{fg:basin_bB_n3m25g10}. Figure
\ref{fg:fractal_n3m25g10} quantifies this hierarchical
organization by showing the dependence of the critical time $t_c$,
the number $N_{\rm turn}$ of rotations of
the spiraling trajectory in phase space and the final value $y_{1c}$,
as a function of the initial value of $y_{10} \equiv y_1(t_0)$ on the
$y_1$-axis. The two former quantities
diverge as power laws of $y_{10}$ with negative exponents as $y_{10} \to 0$.
The last quantity exhibits a ``local fractal'' structure around the origin
which reflects the nested spiral 
structure of the two basins $B\spp$ and $B\spm$ around the origin $S$,
and the fact that each turn segment $\es^{(\pm (k+1),\mp k)}$
shares the same singular dynamical properties as 
$\es^{(\pm 1,\mp 0)}$. Accordingly, $N_{\rm turn}$ is $k+1$ for
$y_{10} \in \es^{(\pm (k+1),\mp k)}$

\begin{figure}
\begin{center}
\includegraphics[width=18cm]{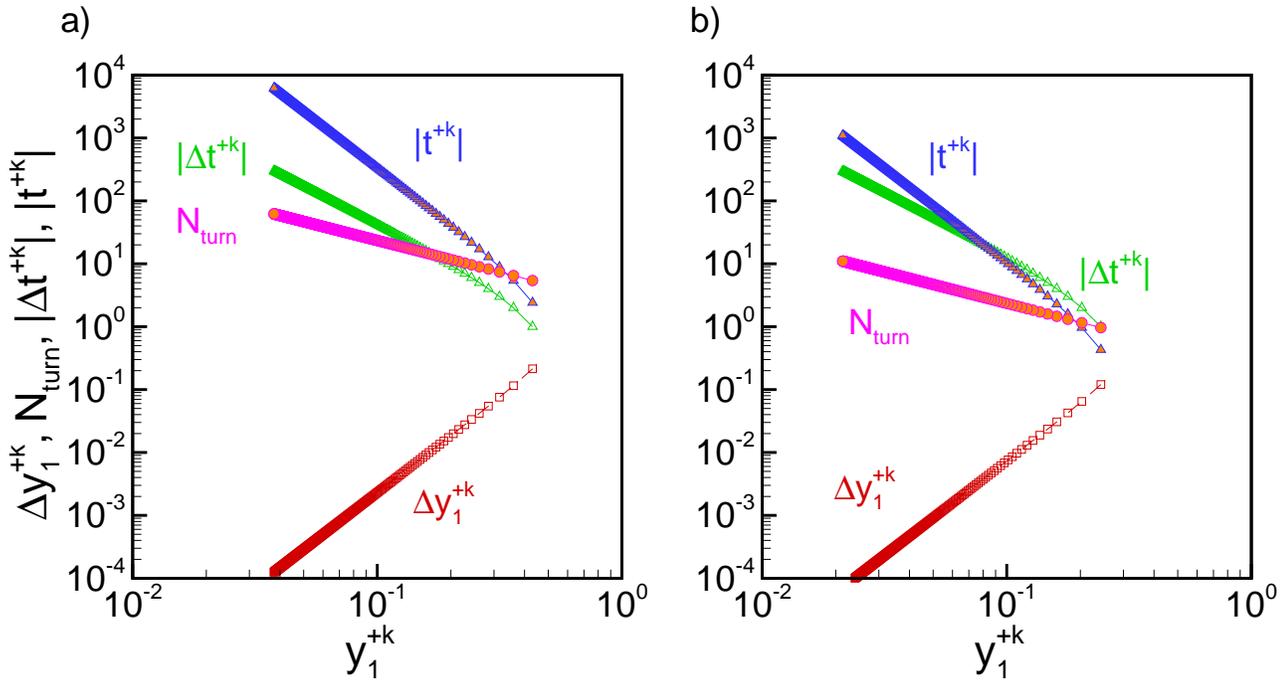}
\end{center}
\caption{Scaling laws associated with the fractal properties
as a function of initial condition at 
turn points $\vyz=(y_{10}^{+k},0)$ of $b\spp$ for $\yo>0$ on the
$\yo$-axis for $(n,m)=(3,2.5)$ as in Figure \ref{fg:phase_all}:
a) $\gamma=10$ and b) $\gamma=1000$.
The notations are: $\triangle \yo=|y_{10}^{+(k+2)}-y_{10}^{+ k}|$,
$\triangle t_e= t_e(\vpo^{+(k+2)})-t_e(\vpo^{+k})|$,
$t_e=t^{\pm k}(\vpo^{+k})$, and $N_{turn}=k$.}
\label{fg:scale_n3m25}
\end{figure}

Figure \ref{fg:scale_n3m25} make these
statements more quantitative by showing the log-log plots of $t_c(y_{10})$, of
$\Delta t_c(y_{10})$ (defined as the increment of $t_c$ over
one turn of the spiral starting from a given initial point),
of $N_{\rm turn}(y_{10})$ and of the increment $\Delta y_1(y_{10})$ 
over one turn of the spiral, as a function of 
$(y_{10})$. The observed straight lines qualify power laws. 
In order to get accurate and reliable
estimations of these dependences and of the exponents
defined below, we have integrated the dynamical
equations using a fifth-order Runge-Kutta integration 
scheme with adjustable time step.

Figure \ref{fg:power_n} shows that
the exponents (slopes of the log-log plots) are
essentially identical for $\gamma=10, 100$ and $1000$, indicating that
the scaling properties depend only on the exponents $(n,m)$
and are independent of $\gamma$.

These different power laws correspond to the linear behaviors shown in
figure \ref{fg:scale_n3m25} and can be represented as follows.
\ba
N_{\rm turn} &\sim& y_{10}^{-a}~,~~~~~~~~{\rm where}~ a>0~,  \label{rel1}\\
t_c &\sim& y_{10}^{-b}~,~~~~~~~~{\rm where}~ b>0~, \label{rel2}\\
\Delta y_1 &\sim& y_{10}^{c}~,~~~~~~~~{\rm where}~ c>0~, \label{rel3}\\
\Delta t_c &\sim& y_{10}^{-d}~,~~~~~~~~{\rm where}~ d>0~, \label{rel4}
\ea
where $y_{10} \equiv y_1(t_0)$.
Specifically, the definitions are (see figure \ref{fg:power_n}):
 $\triangle \yo=|y_{10}^{+ (k+2)}-y_{10}^{+ k}|$,
$\triangle\tc=|\tc(\vpo^{+ (k+2)})-\tc(\vpo^{+k})|$,
$\tc=\tc(\vpo^{+k})$, and $N_{turn}=k$.

The self-similar behavior and power laws occur because there is a 
countable infinite number
of $\Delta e^{\pm (k+1,k)}$ within an extremely slowly-divergent oscillatory
source region $S$ which reach the pre-exit basin segment
$\Delta e^{(\pm 1, \mp 0)}$ after $k$ mapping (therefore after about
$\tauo^{\pm (k+1)}$ exit time interval with $N_{\rm turn}=k+1$ forward turns).

Note that the self-similar behavior close to the origin is governed 
by the singular
boundaries $b\sppm$. The choice of the $\yo$-axis to
define the segments has been made for the simplicity associated with
counting number of forward turns. In theory, $\yo$-axis
can be replaced with any judiciously
chosen curve/lines. Here, we choose the $y_1$-axis to define the 
segments, so that
the self-similar properties are described as a function of
the initial position $y_{10}$ while its velocity $y_2$ is set to zero.

\subsubsection{Scaling relations from self-consistency}

Eliminating $y_{10}$ between (\ref{rel2}) and (\ref{rel4}) gives
\be
\Delta t_c \sim t_c^{d \over b}~.  \label{rel5}
\ee
Since $\Delta t_c$ is nothing but the difference
$\Delta t_c= t_c(N_{\rm turn}+1) - t_c(N_{\rm turn})$, (\ref{rel5})
gives the discrete difference equation on the function $t_c(N_{\rm turn})$
\be
{t_c(N_{\rm turn}+\Delta N_{\rm turn}) - t_c(N_{\rm turn}) \over \Delta N_{\rm turn}}
\propto t_c^{d \over b}~,  \label{jgjfja}
\ee
where $\Delta N_{\rm turn}= N_{\rm turn}+1 - N_{\rm turn}=1$. The left-hand-side
of (\ref{jgjfja}) provides a discrete difference representation of the 
derivative $dt_c/dN_{\rm turn}$. Equation (\ref{jgjfja}) can then be integrated
formally to give $N_{\rm turn} \sim t_c^{1-{d \over b}}$, which is
valid for $d<b$. Comparing with the
relation between $N_{\rm turn}$ and 
$t_c$ obtained by eliminating $y_{10}$ between (\ref{rel1}) and  (\ref{rel2}),
i.e., $N_{\rm turn} \sim t_c^{a \over b}$, we get the scaling relation
\be
a = b-d~.  \label{scal1}
\ee
Since $a>0$, the condition $d<b$ is automatically satisfied.

Similarly, $\Delta y_1 = y_{10}(N_{\rm turn}+1) - y_{10}(N_{\rm turn}) \propto
y_{10}^{c}$ according to definition (\ref{rel3}). This gives the differential
equation $dy_{10}/dN_{\rm turn} \sim y_{10}^c$, whose solution is $N_{\rm turn} 
\sim 1/y_{10}^{c-1}$, valid for $c>1$. Comparing with the definition (\ref{rel1}),
we get the second scaling relation
\be
a = c-1~.  \label{scal2}
\ee
Since $a>0$, the condition $c>1$ is automatically verified.

The scaling relations (\ref{scal2}) and (\ref{scal2}) are the only two
that can be extracted. This shows that the exponents defined in (\ref{rel1}-\ref{rel4}) are 
not independent: out of the four exponents $a, b, c, d$,
only two of them are independent.

\subsubsection{Determination of the critical exponents \label{scalitheo}}

To go further, we use the insight provided by sections \ref{sec:element}
to \ref{sec:case-global}.

Deep in the spiral structure described in the previous section
and depicted in figure \ref{fg:basin_bB_n3m25g10}, one full rotation
would close on itself in the absence of the trend term and would conserve
exactly the Hamiltonian $H$ defined in (\ref{eq:osc-model-H}). In this case,
we know that one full rotation takes a time
equal to the period $T(H)$ given by (\ref{eq:osc-global-T}). In the presence
of the trend term, one full rotation is not closed but the failure
to close is very small especially so very close to the origin. 
Actually, the failure of the trajectory to close is quantified by the variable
$\Delta y_1$ introduced in the previous section and used in (\ref{rel3}).

The approximation we are going to use is that the
value of the period $T(H)$ without the trend term gives the time
$\Delta t_c$ needed to make one full rotation (notwithstanding
the fact that it does not exactly close on itself). 
This is essentially an adiabatic approximation in which the Hamiltonian
$H$ and the period $T(H)$ are assumed to vary sufficiently slowly so that
the local period of rotation follows adiabatically the variation of the
Hamiltonian $H$. 

Equation (\ref{eq:osc-global-T}) gives $T(H) \propto H^{1-n \over 2(n+1)}$.
From the normalization (\ref{eq:osc-norm-var}), the amplitude maximum of
$y_1$ is proportional to $H^{1 \over n+1}$. Putting these two relationship
together gives 
\be
\Delta t_c \sim |y_{10}|^{1-n \over 2}~,   \label{dejbglal}
\ee
which, by comparison with (\ref{rel4}) gives
\be
d = {n-1 \over 2}~.  \label{eqexp1}
\ee

We need a second equation to determine completely the other exponents $a, b$ and $c$.
It is provided by $dT/dt$, expressed as $dT/dt=(dT/dH) \times (dH/dt)$, where
$dT/dH$ is obtained from (\ref{eq:osc-global-dTdH}) and $dH/dt$ is given by
(\ref{eq:all-phase-dHdt}). Estimating $dT/dH$ from (\ref{eq:osc-global-dTdH})
is consistent with the above approximation in which a full rotation
along the spiral takes the same time as the corresponding closed orbit
in absence of the trend term. Expressing $dH/dt$ using
(\ref{eq:all-phase-dHdt}) involves another approximation, which is
similar in spirit to a mean-field approximation corresponding to average
out the effect of the trend term over one full rotation. In so doing, we
average out the subtle positive and negative interferences
between the reversion and trend terms depicted in figure \ref{fg:all_F}.
Furthermore, consistent again with the above approximation in which a full rotation
along the spiral takes the same time as the corresponding closed orbit
in absence of the trend term, we replace $dT/dt$ by
$d \Delta t_c / d t_c$ and obtain
\be
{d \Delta t_c \over d t_c} \sim H^{-{3n+1 \over 2(n+1)}}~ |y_2|^{m+1}
\sim \Delta t_c^{3n+1 \over n-1}~ |y_1|^{(n+1)(m+1) \over 2}~,  \label{fnkala}
\ee
where the dependence in $\Delta t_c$ in the 
last expression of the right-hand-side is derived by replacing $H$ by its dependence 
as a function of $T$ (by inverting $T(H)$ given by (\ref{eq:osc-global-T})) 
and by identifying $T$ and 
$\Delta t_c$. In the last expression, we have replaced the dependence
in $y_2$ by the dependence in $y_1$ by using
the normalization (\ref{eq:osc-norm-var}), leading to 
$|y_2| \sim |y_1|^{n+1 \over 2}$.
Taking the derivative of (\ref{rel5}) with respect to $t_c$
provides another estimation of ${d \Delta t_c \over d t_c}$,
and replacing
$\Delta t_c$ in (\ref{fnkala}) by its dependence as a function of $y_{10}$
as defined by (\ref{rel4})
gives finally:
\be
a = {1 \over 2} (n+1)(m+1) - {1 \over 2} (3 n +1)~.  \label{eqexp2}
\ee

Figure \ref{fg:power_n} presents a
comparison of the 
theoretical predictions (\ref{scal1}), (\ref{scal2}), (\ref{eqexp1}), (\ref{eqexp2})
for the exponents $a, b, c, d$ defined by (\ref{rel1})-(\ref{rel4})
with an estimation obtained from the direct numerical simulation of
the dynamical equations. 
The lines are the theoretical predictions of the exponents $a, b, c, d$
for $m=2.5$ (solid line), $m=2.75$ (dotted line) and
$m=3$ (dashed line) as function of the exponent $n$.
The symbols correspond to the exponents obtained by numerical
simulation for $\gamma=10$ (square), $\gamma=100$ (diamond) and 
$\gamma=1000$ (crosses). The agreement is validated to within
numerical accuracy. We verify also the independence of the exponents 
$a, b, c, d$ with respect to
the amplitude $\gamma$ of the reversal term.

\subsubsection{Time-dependent expression of 
the envelop of $y_1(t; {\bf y}_0, t_0)$}

We have seen in section \ref{xsectionsaa} that, after exiting from
the spiral in phase space, the dynamics becomes completely controlled
by the trend term, while the reversal term responsible for the oscillations
becomes negligible. This leads to the asymptotic solution close to $t_c$ given
by expression (\ref{jfakka}), which we rewrite here
for the sake of comparison:
\be
y_1(t) \approx y_{1c} - A (t_c - t)^{m-2 \over m-1}~,~~~~~{\rm outside~the~oscillatory~regime}~,  
\label{jfakka2}
\ee
where $A$ is an amplitude. $y_1(t)$ has an infinite slope
but a finite value $y_{1c}$ at the critical time $t_c$ since $0 < (m-2)/(m-1) < 1$.
The dependence of this finite critical value $y_1(t_c) = y_{1c}$ as a function of
$y_{10}$ is shown in figure \ref{fg:fractal_n3m25g10}.

In the oscillatory regime, we can also obtain the growth of the amplitude
of $y_1(t)$ by
combining some of the previous scaling laws (\ref{rel1}-\ref{rel4}). 
Indeed, taking the ratio of (\ref{rel3}) and (\ref{rel4}) yields
$\Delta y_1/\Delta t_c \sim y_{10}^{c+d}$. Since $\Delta y_1$ corresponds to
the growth of the local amplitude $A_{y_1}$ of the oscillations of
$y_1(t) $ due to the trend term over one turn of the spiral
in phase space, this turn lasting $\Delta t_c$,
we identify this scaling law with the equation for the growth rate of 
the amplitude $A_{y_1}$ of $y_1$ in this oscillatory regime:
\be
{d A_{y_1} \over dt} \sim A_{y_1}^{c+d}~.
\label{oscigrowhthara}
\ee
Its solution is of the form
\be
A_{y_1}(t) = {B \over (t^*-t)^{1/b}}~,~~~~~{\rm within~the~oscillatory~regime} ~,
 \label{ghahfdjgas}
\ee
where $B$ is another amplitude.
We have used the scaling relations (\ref{scal1}) and (\ref{scal2}) leading 
to $c+d-1=b$. The time $t^*$ is a constant of integration such that 
$B/(t^*)^{1 \over b}=A_{y_1}(t_0)$, which can be interpreted as an
{\it apparent} or ``ghost'' critical time.
$t^*$ has no reason to be equal to $t_c$, in particular since the extrapolation
of (\ref{ghahfdjgas}) too close to $t^*$ would predict a divergence of $y_1(t)$.
The dynamical origin of the difference between $t^*$ and $t_c$ comes from the fact 
that $t^*$ is determined by the oscillatory regime while $t_c$
is the sum (\ref{mfngalaa}) of two contributions, one from
the oscillatory regime and the other from the singular regime.

The prediction (\ref{ghahfdjgas}) is verified accurately from our direct
numerical integration of the equations of motion, as shown in figure
\ref{figy1b_n3m25g10.eps}. To get this figure, we rewrite (\ref{ghahfdjgas}) as
\be
     t=t^{*}-\beta [A_{y_{1}}(t)]^{-b}~,    \label{gnfalal}
\ee
where $\beta=B^{-b}$.  Note that (\ref{gnfalal}) has the same
power law dependence on $y_1$ as (\ref{rel2}). The reason lies
in the fact that, for $y_{10}$ very close to the origin, the 
duration of the oscillatory regime is overwhelming that of the singular regime.
As a consequence, the two power laws (\ref{gnfalal}) and (\ref{rel2}) defined
in the asymptotic limit $y_{10} \to 0$ should and are the same.

We compare this expression with the numerical simulation
using $|y_{1}^{+k}|(t^{+k})$ for $A_{y_{1}}(t)$. The triangles are the 
data) $(t^{+k}, |y_{1}^{+k}|)$ which are fitted to (\ref{gnfalal})
to get $t^{*}$ and $\beta$. The exponent $b$ is fixed to its theoretical
value given by (\ref{scal1},\ref{eqexp1},\ref{eqexp2}).
As can be seen from the two panels, $t^{+k}$ as a function of $|y_{1}^{+k}|$
is a straight line as predicted by (\ref{gnfalal}) with a very good precision.
The first panel shows the whole calculated range. The second panel shows a 
magnification close to the exit point of the oscillatory regime. The different 
straight lines correspond to fits of the data with (\ref{gnfalal}) over
different intervals. We obtain respectively
\begin{itemize}
\item $t^{*}=0.3857$ and $\beta = 0.3381$ for the interval $k \in [1 \to 301]$;
\item $t^{*}=0.6665$ and $\beta = 0.3381$ for the interval $k \in [61 \to 301]$;
\item $t^{*}=1.9577$ and $\beta = 0.3380$ for the interval $k \in [241 \to 301]$;
\item $t^{*}=2.4330$ and $\beta = 0.3380$ for the interval $k \in [271 \to 301]$;
\item $t^{*}=3.5695$ and $\beta = 0.3379$ for the interval $k \in [286 \to 301]$.
\end{itemize}
These five fits performed increasingly deeper within the oscillatory regime
exhibit a good stability for the determination of the slope parameter
$\beta = 0.3380 \pm 0.0001$ but an alarmingly strong variation of the 
``ghost'' critical time $t^*$. Essentially, we must conclude that it is
impossible to determine $t^*$ with any reasonable accuracy. There are two
reasons for this. First, as the next section \ref{deviacrosover} shows, there
is a very slow shift or cross-over within the oscillatory regime to
the final power law singularity at $t_c$. This cross-over corresponds to 
$y_1(t)$ starting off from its initial value with an amplitude growing
according to (\ref{ghahfdjgas}) and then crossing over to (\ref{jfakka2}) close to $t_c$.
Second, while we hoped
that the asymptotic behavior (\ref{ghahfdjgas},\ref{gnfalal})
would become stable deeper within the oscillatory regime, this is counteracted
by larger distances of $|y_{1}^{+k}|(t^{+k})$ from the origin, which make the
determination of $t^*$ more unstable. These difficulties are similar to but
stronger than the well-known problems of the accurate determination of critical
transition parameters.

Combining (\ref{rel4}) with (\ref{ghahfdjgas}) gives the time dependence of the
local period $\Delta t_c$ of the oscillation in the oscillatory regime:
\be
\Delta t_c \sim (t^*-t)^{d \over b}~.   \label{hggqq}
\ee
Notice that the local period of the oscillation is not constant but shrinks
progressively to zero as time increases. This is 
qualitatively reminiscent to the log-periodic
oscillations discussed in the introduction \cite{reviewsor}. Quantitatively,
there is a difference which can be characterized as follows. 
A log-periodic oscillation associated with discrete scale invariance (DSI)
corresponds to characteristic time scales $t^{\pm k}$ (for instance, the times
of the local maxima of the oscillations) such that 
\be
t^{\pm(k+1)} - t^{\pm k} \sim 1/\lambda^k~,   \label{ghahvd}
\ee
where $k>0$ is an integer and $\lambda>1$ is a prefered scaling ratio of DSI.
Expression (\ref{ghahvd}) predicts that $t^*-t^{\pm k} \sim 1/\lambda^k$, i.e.,
goes to zero exponentially as a function of the index $k$. 
In contrast, expression (\ref{hggqq}) can be rewritten
\be
t^{\pm(k+1)} - t^{\pm k} \sim (t^*-t^{\pm k})^{d \over b}~.  \label{ghqllqa}
\ee
Expression (\ref{ghqllqa}) predicts that $t^*-t^{\pm k} \sim (k^*-k)^{1/(1-{d \over b})}$,
i.e., the local period goes to zero in a finite number of turns. The
log-periodic result is obtained as the limit $d/b \to 1^-$, which is reached
for $n \to \infty$ and $m \to 2$. We thus have a dynamical theory that provides
a mechanism for
quasi-log-periodic oscillations with, in addition, a finite number of them 
due to the cross-over to the non-oscillatory regime. A similar almost log-periodic
but nevertheless distinctly different frequency modulation has recently been
observed numerically on the average of the logistic map variable close to a 
tangent bifurcation associated with deterministic intermittency of type I 
\cite{powerlawperiodic}.
These ``power law periodicity'' are originated in the mechanism of reinjection
of the iterates on the channel of near-periodic behavior.

\subsubsection{Deviation from scaling \label{deviacrosover}}

\begin{figure}
\begin{center}
\includegraphics{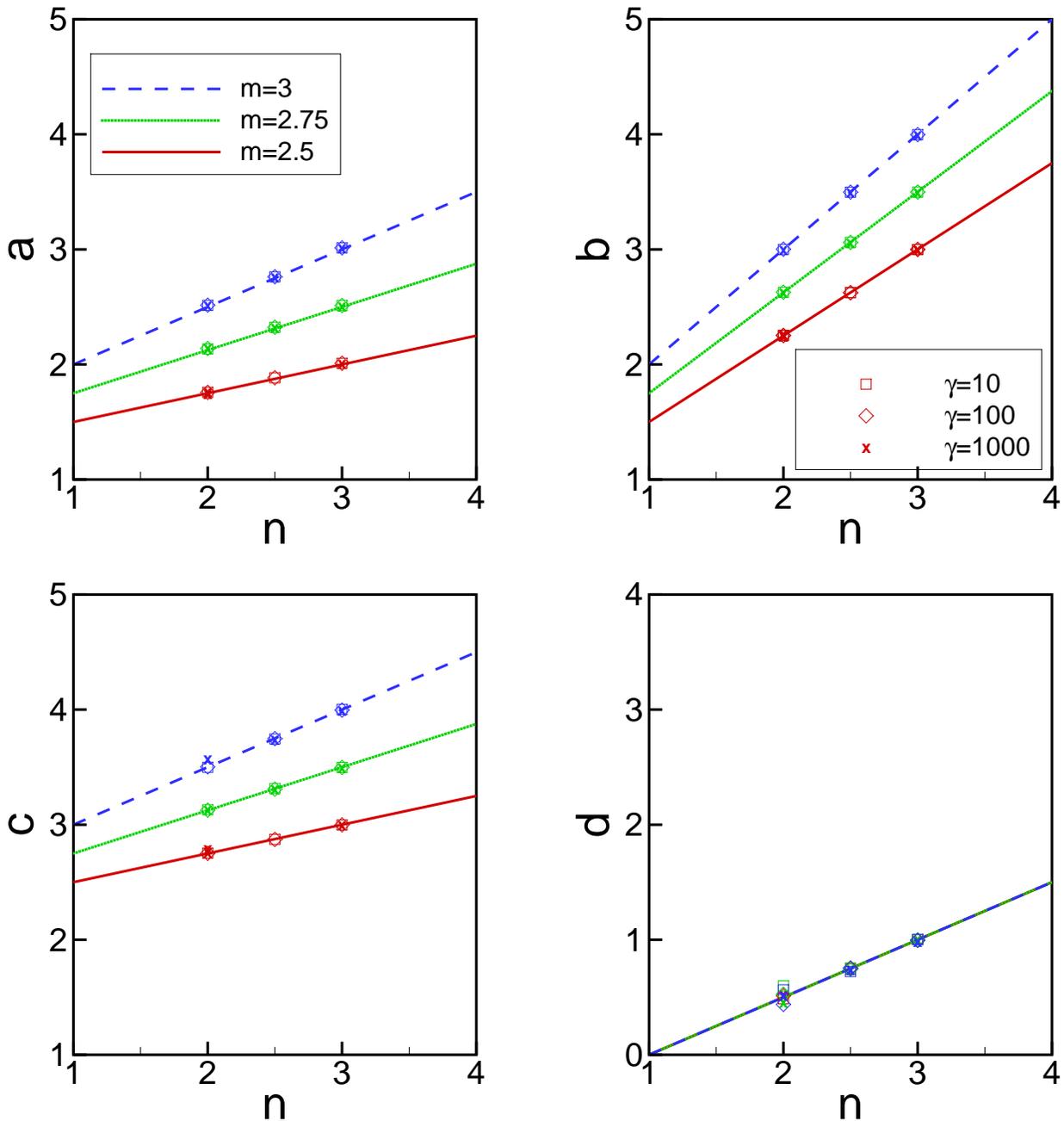}
\end{center}
\caption{Comparison of the 
theoretical predictions (\ref{scal1}), (\ref{scal2}), (\ref{eqexp1}), (\ref{eqexp2})
for the exponents $a, b, c, d$ defined by (\ref{rel1})-(\ref{rel4})
with estimations obtained from the direct numerical integration of
the dynamical equations using a fifth-order Runge-Kutta integration 
scheme with adjustable time step. 
The lines are the theoretical predictions as a function of $n$
for $m=2.5$ (solid line), $m=2.75$ (dotted line) and
$m=3$ (dashed line) as function of the exponent $n$.
The symbols correspond to the exponents obtained by numerical
simulation for $\gamma=10$ (square), $\gamma=100$ (diamond) and 
$\gamma=1000$ (crosses). 
}
\label{fg:power_n}
\end{figure}

\begin{figure}
\begin{center}
\includegraphics[width=16cm]{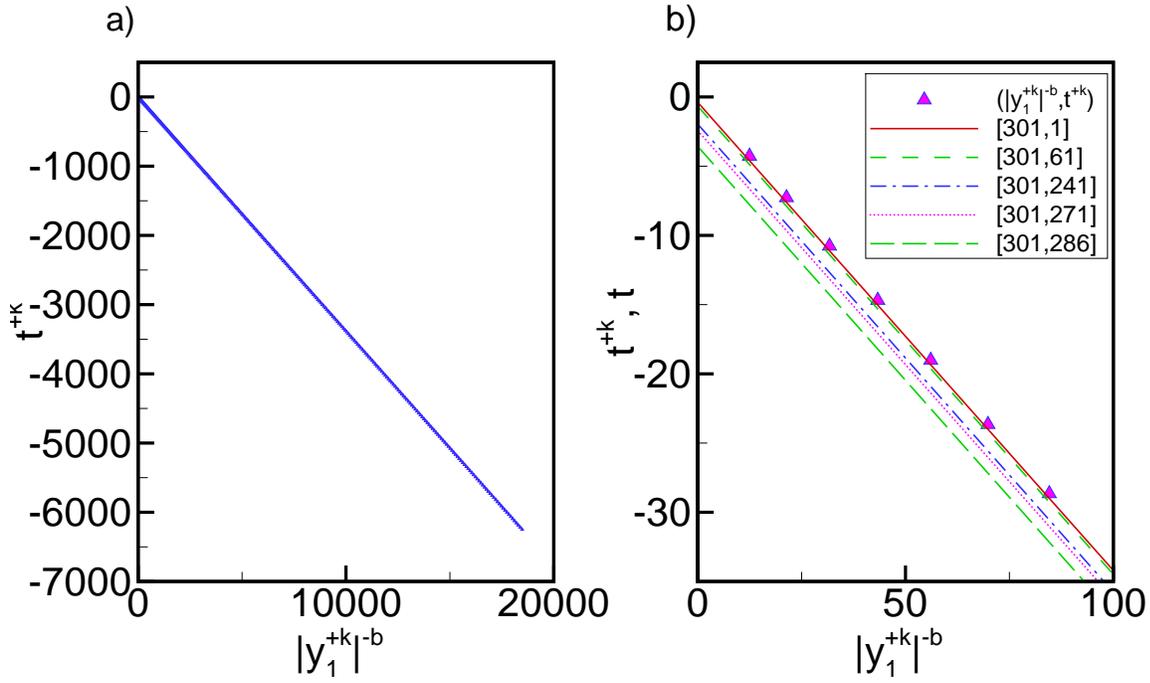}
\end{center}
\caption{Non-parametric test of the prediction (\ref{ghahfdjgas})  by
rewriting (\ref{ghahfdjgas}) as (\ref{gnfalal}) so as it qualifies by a
linear behavior. The time $t^{+k}$ is plotted (triangles) as a function of
$|y_{1}^{+k}|(t^{+k})$ which are proxies for $(A_{y_{1}}(t)$. The triangles 
are fitted to (\ref{gnfalal})
to get $t^{*}$ and $\beta$. The exponent $b$ is fixed to its theoretical
value given by (\ref{scal1},\ref{eqexp1},\ref{eqexp2}).
The first panel shows the whole calculated range. The second panel shows a 
magnification close to the exit point of the oscillatory regime. The different 
straight lines corresponds to fits of the data with (\ref{gnfalal}) over
different intervals, with
$t^{*}=0.3857$ and $\beta = 0.3381$ for the interval $k = 1 \to 301$;
$t^{*}=0.6665$ and $\beta = 0.3381$ for the interval $k = 61 \to 301$;
$t^{*}=1.9577$ and $\beta = 0.3380$ for the interval $k = 241 \to 301$;
$t^{*}=2.4330$ and $\beta = 0.3381$ for the interval $k = 271 \to 301$;
$t^{*}=3.5695$ and $\beta = 0.3379$ for the interval $k = 286 \to 301$.
}
\label{figy1b_n3m25g10.eps}
\end{figure}

Figure \ref{fg:scale_n3m25} and even more so figure \ref{fg:power_n}
have been constructed in the ``scaling'' regime in which the two approximations
used above hold strongly.
\begin{enumerate}

\item {\bf Adiabatic approximation}:
the value of the period $T(H)$ calculated without the trend term gives the time
$\Delta t_c$ needed to make one full rotation (notwithstanding
the fact that it does not exactly close on itself). Equivalently, the Hamiltonian
$H$ and the period $T(H)$ are assumed to vary sufficiently slowly so that
the local period of rotation follows adiabatically the variation of the
Hamiltonian $H$. 

\item {\bf Mean-field approximation}: we average
out the effect of the trend term over one full rotation. In so doing, we
average out the subtle positive and negative interferences
between the reversion and trend terms depicted in figure \ref{fg:all_F}.
This allowed us to estimate $dH/dt$ using
(\ref{eq:all-phase-dHdt}).

\end{enumerate}

The theoretical predictions (\ref{scal1}), (\ref{scal2}), (\ref{eqexp1}), (\ref{eqexp2})
for the exponents $a, b, c, d$ defined by (\ref{rel1})-(\ref{rel4})
obtained using these two approximations have been found to be in very good
agreement with direct numerical estimations, as shown in figure \ref{fg:power_n}.

\begin{figure}
\begin{center}
\includegraphics[height=16cm]{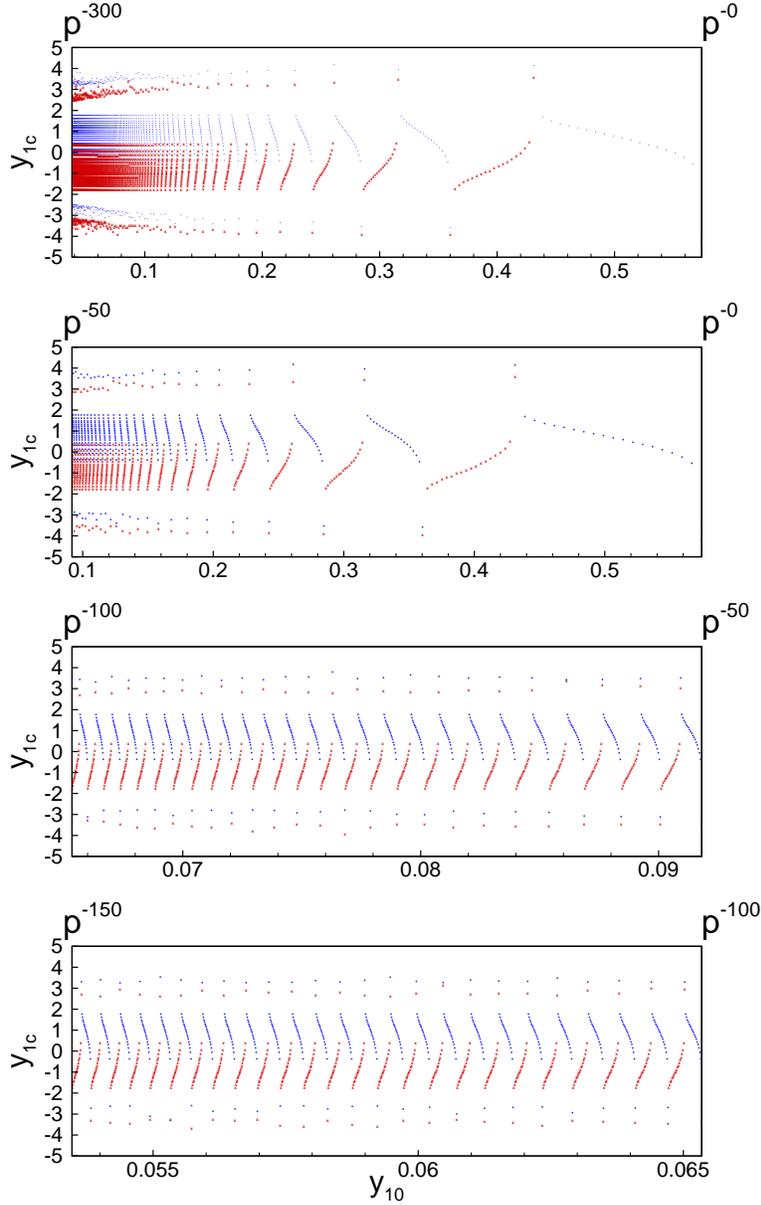}
\end{center}
\caption{Terminal critical value $y_1(t_c)=y_{1c}$ as a function
of initial value $y_{10}$. The top panel shows the oscillations
of $y_{1c}$ as a function of $y_{10}$ in the first 300 
turn segments $\es^{(\pm (k+1),\mp k)}$ (see definition \ref{def:all-es}).
Recall that $y_{1c}$ diverges at the boundaries corresponding to the intersection
of the two curves $b^{\pm}$ with the $y_1$-axis. Due to finite numerical
and graphical resolution, we can only observe a cusp-like behavior
associated with points approaching very close to these boundaries.
The other three panels gives magnifications of the top panel for the
first 50 turn segments $\es^{(\pm (k+1),\mp k)}$ (second panel),
from the 50th to the 100th turn segments $\es^{(\pm (k+1),\mp k)}$ 
(third panel) and from the 100th to the 150th turn segments $\es^{(\pm (k+1),\mp k)}$ 
(fourth panel). 
In each panel, ``circle'' and  ``crosses'' symbols correspond to points in 
$\triangle e^{+(k+1),-k}$ and $\triangle e^{-k,+(k-1)}$, respectively.
In order to construct these panels, 
we have sampled each turn segment $\triangle e^{\pm k+1,\mp k}$
by 20 $y_{10}$ points. The two end points are
chosen to be less than $10^{-8}$ away 
from $\vpo^{\pm k+1}$ and $\vpo^{\mp k}$.  The other 18 points are equally
spaced within $\triangle e^{\pm k+1,\mp k}$.
}
\label{fg:test50simil}
\end{figure}

However, this agreement is obtained only within a ``scaling'' regime, 
sufficiently close to the origin in phase space, i.e., such that 
the Hamiltonian $H$ of the oscillations grows sufficiently slowly.
To quantify the concept of a ``scaling'' regime, figure \ref{fg:test50simil}
represents the terminal critical value $y_1(t_c)=y_{1c}$ as a function
of initial value $y_1(t_0)=y_{10}$ in intervals such that 
perfect self-similarity can be checked readily. The top panel presents
a synopsis by showing the oscillations
of $y_{1c}$ as a function of $y_{10}$ for the first 300 
turn segments $\es^{(\pm (k+1),\mp k)}$ (see definition \ref{def:all-es}).
If self-similarity was true, the three following panels in figure
\ref{fg:test50simil} should be essentially identical because they show
exactly the same number of oscillations. However, it is clear that there is a 
systematic drift, which is fast at first (second panel for the
first 50 turn segments $\es^{(\pm (k+1),\mp k)}$ and then slows down
deep within the spiral structure for the third
and fourth panel (see figure \ref{fg:basin_bB_n3m25g10} for definitions).

\begin{figure}
\begin{center}
\includegraphics{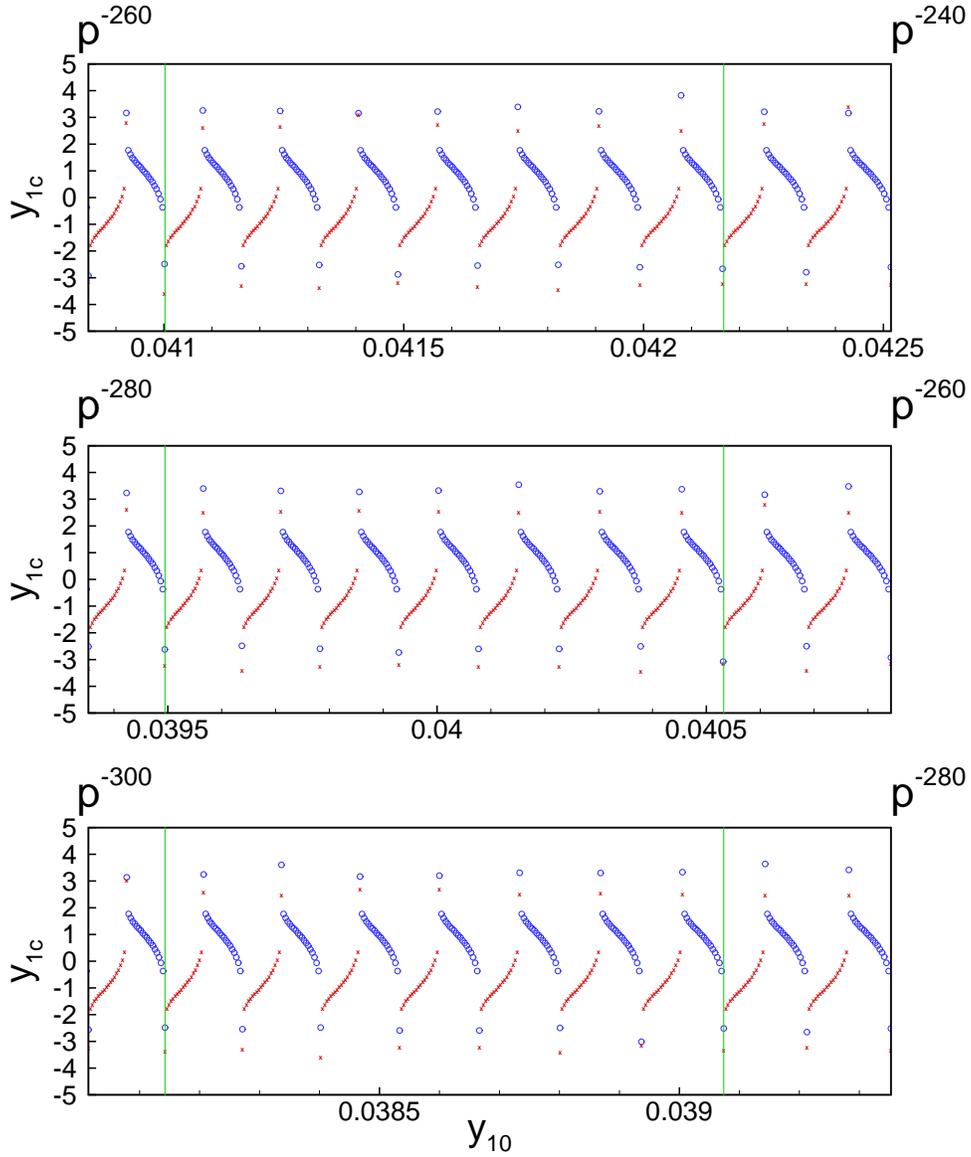}
\end{center}
\caption{Same as figure \protect\ref{fg:test50simil}, i.e., 
terminal critical value $y_1(t_c)=y_{1c}$ as a function
of initial value $y_{10}\equiv y_1(t_0)$ 
from the 240th to the 260th turn segments $\es^{(\pm (k+1),\mp k)}$ 
(top panel), from the 260th to the 280th turn segments $\es^{(\pm (k+1),\mp k)}$ 
(middle panel) and from the 280th to the 300th turn segments $\es^{(\pm (k+1),\mp k)}$
(bottom panel). The two vertical lines provide a guide to the eye to verify 
the almost perfect self-similarity.
}
\label{fg:test20simildeep}
\end{figure}

\begin{figure}
\begin{center}
\includegraphics[height=20cm,width=16.cm]{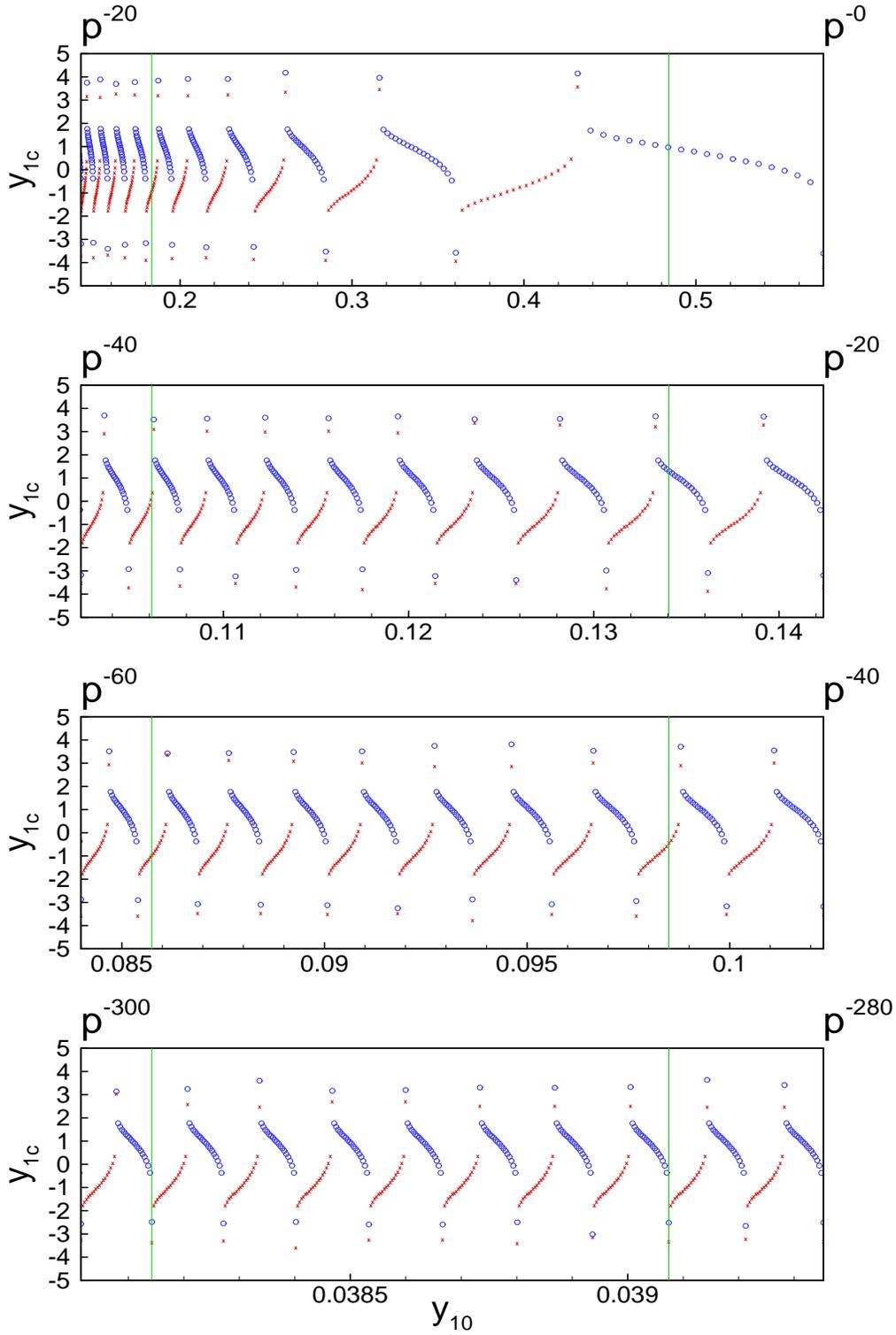}
\end{center}
\caption{Same as figure \protect\ref{fg:test20simildeep}
from the 0th to the 20th turn segments $\es^{(\pm (k+1),\mp k)}$
(first top panel), from the 20th to the 40th turn segments $\es^{(\pm (k+1),\mp k)}$
(second panel), from the 40th to the 60th turn segments $\es^{(\pm (k+1),\mp k)}$ 
(third panel) and from the 280th to the 300th turn segments $\es^{(\pm (k+1),\mp k)}$ 
(fourth bottom panel). The two vertical lines provide a guide to the eye to show
that self-similarity is not qualified.
}
\label{fg:test20similcross}
\end{figure}

Figure \ref{fg:test20simildeep} exhibits almost perfect self-similarity
by plotting the terminal critical value $y_1(t_c)=y_{1c}$ as a function
of initial value $y_1(t_0)=y_{10}$ 
from the 240th to the 260th turn segments $\es^{(\pm (k+1),\mp k)}$ 
(top panel), from the 260th to the 280th turn segments $\es^{(\pm (k+1),\mp k)}$ 
(middle panel) and from the 280th to the 300th turn segments $\es^{(\pm (k+1),\mp k)}$ 
(bottom panel). The two vertical lines provide a guide to the eye to verify 
the almost perfect self-similarity. This is the regime and beyond where 
the scaling laws reported in figures \ref{fg:scale_n3m25} and \ref{fg:power_n}
hold. The cross-over to this scaling regime is presented visually in figure 
\ref{fg:test20similcross} which compares exactly twenty oscillations of
the terminal critical value $y_1(t_c)=y_{1c}$ as a function
of initial value $y_1(t_0)=y_{10}$ in different intervals, from 
close to the exit point (first top panel) to deep within the spiral
structure (bottom panel). The two vertical lines point out the deviations
from the close-to-asymptotic regime of the bottom panel reached 
from the 280th to the 300th turn segments $\es^{(\pm (k+1),\mp k)}$
as the dynamics gets closer and closer to the exit point (from the third to the
first panel).

It is probably possible to improve upon the scaling theory offered in section \ref{scalitheo}
and go beyond the adiabatic mean-field approximation in order to describe the 
cross-over regime between the asymptotic scaling regime and the exit 
of the spiral structure. This should be done by quantifying the relative reinforcing and
opposing effects of the trend and reversal terms within each turn, as shown in 
figure \ref{fg:all_F}.

\section{Concluding remarks}

We have introduced a second-order ordinary differential equation describing
an oscillator exploding in finite time whose
dynamics results from the interplay between a nonlinear negative viscosity
and a nonlinear reversal term. This system provides a simplified description of
stock market prices, population dynamics and material rupture, in regimes
where the growth rates are an increasing function of the price, 
population or stress, respectively, in the presence of important negative
feedback effects.
Our approach using dynamical system theory has shown that the
trajectories can be understood in details from the spiraling structure of
two sets of specials curves in phase space linking the origin to
points at infinity.

The message of our work is threefold. First, it is important to take into 
account the delayed response of dynamical variables to past states, leading
technically to the presence of a second (or higher) order
differential equation. This inertia is essential for the generation of 
oscillations resulting from overshooting. Second, positive nonlinear
feedback is a general and ubiquitous mechanism for generating 
accelerated super-exponential growth ending in finite-time singularities.
Third, reversal, recovery or healing mechanisms with nonlinear threshold-like
behavior, together with inertia, ensure the existence of overshooting and thus of
oscillations 
controlled by the amplitude of the variable. Our 
two-dimensional nonlinear dynamical system opens the road
to a re-examination of many systems which contain these elements but which
have been linearized, thus missing the novel phenomenology 
unraveled here. Our study suggests that super-exponential growth 
modulated by amplitude-dependent oscillations may be a general feature
of complex systems, such as financial markets, population and heterogeneous
materials. We believe that this work provides a novel framework to 
model these systems and to discover new precursory indicators or patterns
of ``rupture''. This seems to offer a generalization to the reported
log-periodic critical oscillations previously reported
for these systems (see \cite{reviewsor} and references therein).

The study presented here can be enriched in many ways. For several applications,
three obvious missing ingredients are additive noise, stochastic reversal
to the fixed point $(y_1=0, y_2=0)$
and saturation effects with reinjection limiting the growth before reaching the singularity.
The addition of these terms 
will naturally lead to intermittent oscillatory
structures (deterministic, stochastic, or both resulting from the interplay between 
deterministic chaos and noise), 
similar to those documented as log-periodic power laws
in financial bubbles \cite{crash,antilogperiod2,nasdaq} as well
as in the population dynamics
\cite{JohSorgreat} and in rupture \cite{Anifrani,critruptcan,critrupt}.

\pagebreak

\pagebreak

\pagebreak

\pagebreak

\pagebreak

\pagebreak

\pagebreak

\pagebreak

\pagebreak

\pagebreak

\pagebreak

\pagebreak

\pagebreak

\pagebreak

\pagebreak

\pagebreak

\pagebreak

\pagebreak

\clearpage

\clearpage

\clearpage

\end{document}